# CARGO:
# Effective format-free compressed storage of genomic information

Łukasz Roguski[1,2]    Paolo Ribeca[3,1]

[1]Centro Nacional de Análisis Genómico, Barcelona, Spain
[2]Universitat Pompeu Fabra, Barcelona, Spain
[3]The Pirbright Institute, Woking, UK

*The recent super-exponential growth in the amount of sequencing data generated worldwide has put techniques for compressed storage into the focus. Most available solutions, however, are strictly tied to specific bioinformatics formats, sometimes inheriting from them suboptimal design choices; this hinders flexible and effective data sharing. Here we present CARGO (Compressed ARchiving for GenOmics), a high-level framework to automatically generate software systems optimized for the compressed storage of arbitrary types of large genomic data collections. Straightforward applications of our approach to FASTQ and SAM archives require a few lines of code, produce solutions that match and sometimes outperform specialized format-tailored compressors, and scale well to multi-TB datasets.*

# Introduction

**The state of the art**

A few typical strategies have been employed so far to implement compressed storage for genomic data: either compression tied to some specific datafile format [1][2][3][4] or compression implemented on the top of a pre-existing database engine or data framework [5][6][7]. Usually the first solution is relatively straightforward to implement and optimize, and sufficient for archival purposes; however, it is unsuitable for applications requiring more sophisticated or volatile derived data models, like "on-the-fly" intermediate schemes produced by data analysis workflows that do not necessarily correspond to an available long-term storage format. The second solution automatically inherits many pros of the framework it builds upon, mainly high-level features and seamless integration with other tools; however, it also usually constrains the user within the boundaries of a possibly cumbersome pre-defined architectural computational ecosystem, often resulting in solutions that are suboptimal in terms of either storage usage or computing resources consumption.

**A novel approach: CARGO**

Here we present a different approach, illustrated in **Figure 1**, that focuses on flexibility and immediate deployability in any local high-performance computing setup. It is based on a few key concepts:

1. We model each dataset as a free-text *header* plus a sequence of structured, and possibly sorted, *records*.

2. As when using full-fledged databases, each record is precisely defined in terms of an expressive domain-specific meta-language (**Figure 1A**, **Methods**). This allows to create rich data types, including arbitrarily nested vectors, records, unions, and special data types optimized for the storage of genomic information. From the record specification, in a way which is completely transparent to the user, our framework is able to automatically produce an optimized C++ specification for a compressor/decompressor implementation (**Figure 1D,E,F,G**, **Methods**). The data fields belonging to the record are analyzed and internally rearranged; fields having the same nature are brought together; and the record is turned into a collection of *streams* which contain homogeneous data and hence can be compressed more effectively, in the spirit of column-oriented databases (**Methods**). The meta-language allows the user to specify the record in great detail, including the way each field should be compressed. Our framework supports many stream compression methods (at the moment `gzip`, `bzip2`, PPMd and LZMA); new methods can be easily added as plug-ins. Parallel multi-threaded compression of the streams is transparently provided (**Methods**).

3. Describing the conceptual ontology of the data in the CARGO meta-language (for instance, a record data type suitable to store short-read alignment information as in **Figure 3, Methods**) is not enough to create a fully working (de)compressor. In order to do so, the user has to provide additional record parsing/unparsing methods written in C++ (**Supplementary Documentation sections 6,8**), that specify how a chunk of an input text file (for instance, a line in SAM [1] format) should actually be turned into the abstract record, and vice versa (**Figure 1B**). Optionally, some parameterized adaptor functions can also be specified, to perform on-the-fly transformations on the record content (for instance, quality down-sampling) before compression/after decompression (**Figure 1C**). Once all the needed inputs for a given format have been provided, they can be compiled by using the standard CARGO toolchain commands (**Supplementary Documentation sections 2,4,8**) to produce a compressor/decompressor program in binary form (**Figure 1D,H,I**). Support for several widely-used formats (mainly FASTQ and SAM), i.e. suitable input source files as in **Figure 1A,B,C** ready to be compiled, is provided in the standard CARGO distribution. In other words, optimized CARGO-based compressors for FASTQ and SAM are available out of the box, without any need for the user to develop code or learn the inner workings of CARGO.

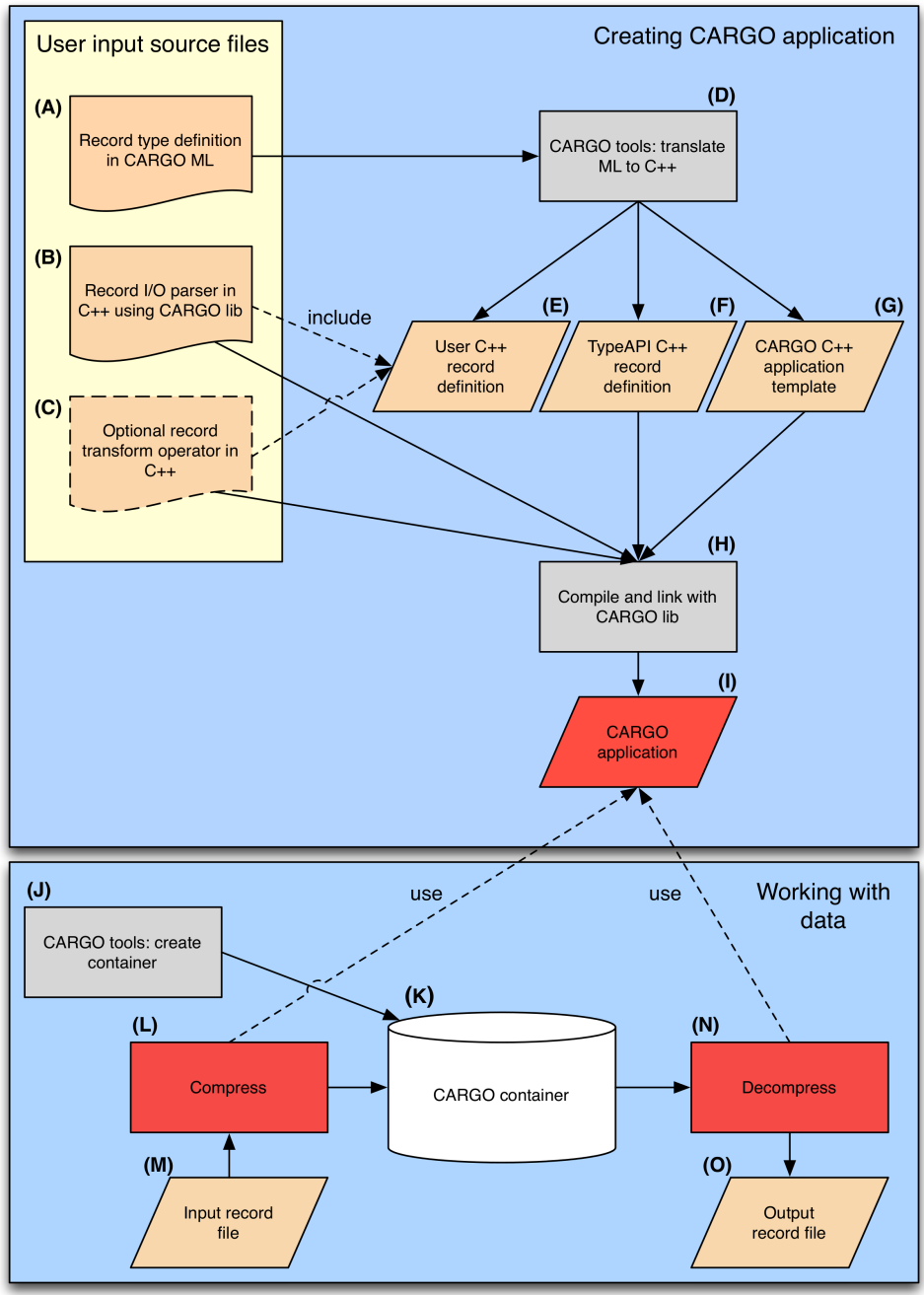

**Figure 1.** Storing big genomic data in compressed form within the CARGO framework. CARGO is a compressor compiler: all the user has to do in order to produce a family of compressors/decompressors is specifying a record data type, and how to parse/unparse it.

4. Once one or more (de)compressors are available, the user can allocate and populate a CARGO *container* (**Figure 1J,K,L,M,N,O**). A container is a large disk-allocated file designed for the efficient storage of many compressed CARGO streams (**Methods**). Different formats and different datasets can be written to, or read from, the same container through any of the (de)compressors produced as per the previous step. Albeit some features are not yet implemented, the CARGO tools provide conceptual support for a range of operations on containers, including querying for contained datasets, expansion, shrinkage, concurrent write/read of independent datasets, back-up of meta-data, recovery in case of corruption and so on.

5. Datasets can be pre-sorted according to a criterion specified by the user, and range queries on the key become then possible for each dataset. This is a generalization of the sort-by-position capability offered by indexed BAM [1] files.

## Methods

In this section we describe in more detail how the major features provided by CARGO are implemented.

**How CARGO meta-language record definitions are turned into streams**

In order to illustrate how CARGO meta-language is translated to C++ code, we will use a simple proof-of-concept FASTQ record definition (see **Figure 2A**). More realistic worked-out examples for both FASTQ and SAM format can be found in the **Supplementary Documentation**. The full definition of the CARGO meta-language in Backus-Naur form can also be found in the **Supplementary Documentation**.

According to the workflow depicted in **Figure 1**, the CARGO meta-language record definition (as in **Figure 2A**) will need to be processed with the CARGO tools. This operation will generate several files containing C++ code. The most relevant ones are:

1. A C++ record type definition (see **Figure 1E** and **Figure 2B**). It represents the handle by which the user can implement operations on the record such as parsing, transformations and key generation (for sorting and querying).

2. A C++ TypeAPI-based record description (see **Figure 1F** and **Figure 2C**). The TypeAPI is a core component of the CARGO framework, made of a large set of C++ template classes (see **Supplementary Documentation** for a full description). They encapsulate the stream I/O management logic by interfacing the C++ type definition with the container functionality. Hence the TypeAPI provides a middle layer between the high-level meta-language description and the low-level data stream representation.

3. Some C++ classes providing the skeleton of a record parser and, optionally, functionality to transform records and generate keys to store the data in sorted order (as in **Figure 1C**). In order to be able to produce a fully working

**A**

```
FastqRecord = {
  tag = string;
  seq = string;
  qua = string
}

@record FastqRecord
```

**B**

```
struct FastqRecord {
  std::string tag;
  std::string seq;
  std::string qua;
};
```

**C**

```
struct __Compound1 {
  typedef FastqRecord UserDataType;
  static const uint32 FieldCount = 3;
  typedef type::TStringType<streams::CompressionText,
                            streams::BlockSizeText> Type1;
  static typename Type1::UserDataType&
  Get1(UserDataType& data_) {
    return data_.tag;
  }
  static const typename Type1::UserDataType&
  Get1(const UserDataType& data_) {
    return data_.tag;
  }
  typedef type::TStringType<streams::CompressionText,
                            streams::BlockSizeText> Type2;
  static typename Type2::UserDataType&
  Get2(UserDataType& data_) {
    return data_.seq;
  }
  static const typename Type2::UserDataType&
  Get2(const UserDataType& data_) {
    return data_.seq;
  }
  typedef type::TStringType<streams::CompressionText,
                            streams::BlockSizeText> Type3;
  static typename Type3::UserDataType&
  Get3(UserDataType& data_) {
    return data_.qua;
  }
  static const typename Type3::UserDataType&
  Get3(const UserDataType& data_) {
    return data_.qua;
  }
};
typedef type::TStructType<__Compound1> FastqRecord_Type;
```

**Figure 2.** How a type definition in CARGO meta-language is translated into low-level C++ code. **Panel A**: A simple yet complete CARGO meta-language specification for a FASTQ record (as in **Figure 1A**). **Panel B**: The corresponding C++ record definition automatically generated (as in **Figure 1E**) by processing the meta-language definition with the CARGO tools (as in **Figure 1D**). **Panel C**: The C++ TypeAPI definition automatically generated (as in **Figure 1F**) by the CARGO tools from the meta-language definition.

compressor/decompressor, a few missing code pieces need to be manually filled out by the user.

All the C++ files mentioned above are automatically deduced from the provided CARGO meta-language type definition. In particular, both the C++ record type definition and the C++ TypeAPI-based record description can be emitted without any user intervention. This is possible thanks to the CARGO translator application (see **Supplementary Documentation** for a full description of the program and its options): it parses programs in CARGO meta-language, builds an internal representation of each type in the form of a set of annotated *abstract syntax trees* (ASTs; one per each @record directive present in the meta-code, as in **Figure 2A**) and turns each AST into valid C++ code. Such code relies upon the TypeAPI classes in order to interface with the CARGO library and so provide access functionality to compressed containers.

Even a very simple CARGO meta-code like that of **Figure 2A** is able to produce optimized compressors/decompressors that rival in performance with the current state of the art (see **Supplementary Data**). However, more complex file formats and more sophisticated data transformation strategies (like for instance those relying upon reference-based sequence compression) are likely to require more fine-tuned CARGO meta-language record definitions. A fully fledged record type definition, suitable to implement the high-performance SAM compressor presented in **Figure 5 methods 1,2 and 7**, is shown in **Figure 3**.

```
OptionalValue = [
  intValue = int^32;
  charValue = char;
  stringValue = string;
] :
  .intValue.Pack = Bzip2L4,
  .charValue.Pack = GzipL2,
  .stringValue.Pack = Bzip2L4;

OptionalField = {
  tag = string;
  value = OptionalValue;
} :
  .tag.Pack = PPMdL4;

MdOperationValue = [
  intValue = int^32;
  stringValue = string;
] :
  .intValue.Pack = Bzip2L4,
  .stringValue.Pack = LZMAL1;

MdOperation = {
  operation =
    enum [
      "None",
      "Match",
      "Substitution",
      "Insert",
      "Delete",
      "SoftClip",
      "HardClip"
    ];
  value = MdOperationValue;
} :
  .operation.Pack = Bzip2L4;
```

```
SamRecord = {
  qname = string;
  flag = uint^16;
  rname = string;
  pos = uint^32;
  mapq = uint^8;
  cigar = string;
  next = string;
  pnext = int^32;
  tlen = int^32;
  seq = string;
  qua = string;
  opt = OptionalField array;
  md = MdOperation array;
} :
  .qname.Pack = PPMdL4,
  .flag.Pack = PPMdL4,
  .rname.Pack = GzipL2,
  .pos.Pack = LZMAL1,
  .mapq.Pack = PPMdL1,
  .cigar.Pack = GzipL1,
  .next.Pack = GzipL2,
  .pnext.Pack = LZMAL1,
  .tlen.Pack = PPMdL4,
  .seq.Pack = LZMAL1,
  .qua.Pack = PPMdL1;

@record SamRecord
```

**Figure 3.** A more realistic SAM record type definition in CARGO meta-language, based on which our best reference-based SAM compressors are implemented. One can notice that different streams are compressed by different methods. This "optimal" configuration was selected based on a number of experiments carried out on different datasets.

**Internal organization of containers**

As illustrated in **Figure 4**, containers are made of three different storage areas:

- A *stream block area*, where the content of streams is stored. Each stream is conceptually a linked list of separated blocks. Blocks can be either big or small; the sizes of both small and big blocks for each container are configured once and for all when the container is created. Big blocks are used by default in order to minimize disk accesses and improve throughput, whereas small blocks are designed to deal with the last portions of the streams in order to minimize the amount of unused space in big blocks. Blocks also store checksum information for the contained data at regular intervals.

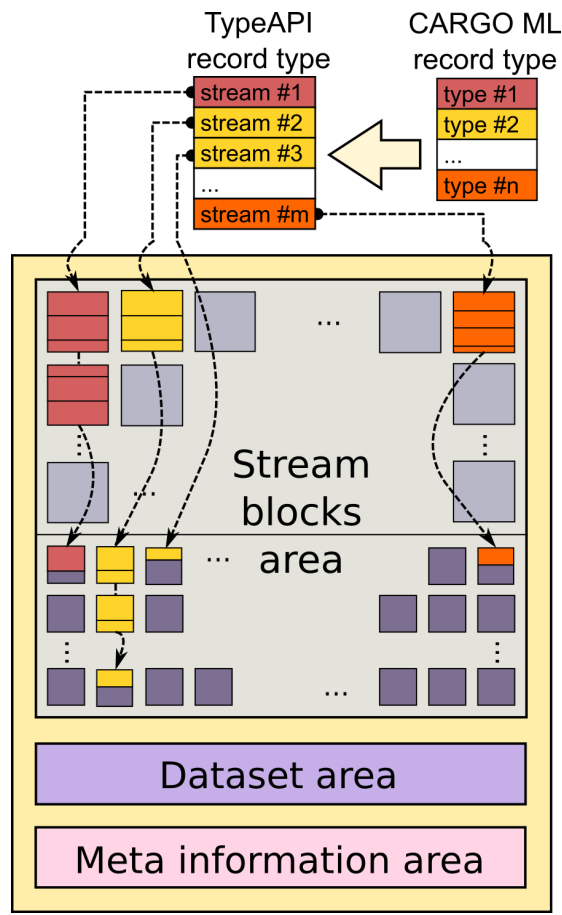

**Figure 4.** A conceptual representation of low-level information storage in CARGO. Once a meta-language record type has been turned into a C++ TypeAPI definition, the input data is automatically split into a collection of streams. Stream configuration is recorded in the *dataset area*. Streams are stored as linked lists of *blocks*. Blocks can be big or small, small blocks usually being used to store the terminal part of the stream in order to optimize space usage. Block state at any timepoint is stored in the *meta-information area*.

- A *meta-information area*. It contains a *block allocation table* with information about the occupancy state of each block (which can be either *free*, or *reserved* i.e. currently being written to but not yet finalized, or *occupied*).

- A *dataset area*. It contains a description of all datasets stored in the container: their type definition and their composition in terms of streams; for each stream, the compression method and a list of block IDs, the number of blocks and all the relevant metrics.

In order to provide for easier backup and recovery, in our current implementation each storage area corresponds to a separate file.

A key design feature of streams is that the blocks composing them do not need to be consecutive. As a result, many different interleaved streams can coexist in the same container, and, provided that a suitable locking mechanism is put in place, many processes can read and write to the same container concurrently.

Overall our container design is very flexible, in that it allows for an easy implementation of many high-level features. For instance, when the amount of needed space is not known in advance a container can be generated with an arbitrary size; however, once all the datasets of interest are stored in it, the container can be shrunk in order to remove unused blocks and facilitate data archival/exchange. Although not yet currently implemented, other possible features would be block encryption and resilience to error. The latter would work in a way similar to what is offered by filesystem recovery programs: it would rely on the presence of checksum information within the blocks in order to detect possible corruptions due to hardware errors occurring in the physical layer of the storage, and would allow the reconstruction of the whole dataset except for the records directly hinging on the affected block.

As each stream is saved as an independent list of blocks, after a record had been parsed and optionally transformed using a user-specified adaptor function, each block can be compressed independently (the same holds in the reverse order for decompression). Considering that record parsing and transformation are usually not the data-processing bottlenecks, this scheme offers big room for implementing transparent parallelization within the CARGO library. In our current implementation we follow a standard producer/consumer paradigm and dispatch a small slice of the input records to each processing thread. After streams have been parallelized in this way, I/O throughput usually becomes the performance bottleneck. This is why selecting big block sizes can improve performance, especially on the network-distributed filesystems usually found in high-performance computation environments.

In addition to the custom record transformation functionality, another important feature offered by CARGO is the possibility of specifying an arbitrary custom sorting function defined in terms of the record, and then performing range queries on such a generated key. In this case streams will contain additional information stored in the dataset area, consisting of keys sampled at regular intervals. This allows us to implement locating searches by simple bisection methods. Once blocks within the query range have been identified, subsequent data extraction can be easily parallelized.

# Results

In the **Supplementary Documentation section 8** we demonstrate in a step-by-step tutorial how families of compressors/decompressors from/to formats widely used in the bioinformatics of high-throughput sequencing, like FASTQ [10] and SAM, can be implemented with CARGO. In fact, several implementations of increasing complexity and sophistication are included in the standard CARGO distribution. Our full results are presented in the **Supplementary Data**; here we focus on two benchmarks testing recompression of SAM files, although similar results hold true for FASTQ files. In general our framework provides unprecedented flexibility, excellent performance, and good scalability.

**Re-compression of single SAM files**

**Figure 5** illustrates our results when re-compressing a 82 GB SAM file with different methods (dataset HG01880 from the 1000 Genome Project [11], see **Supplementary Data** for a complete description of all methods). Perhaps the most striking feature of our approach is that in general very little code is required to achieve results that are comparable to, or better than, what state-of-the art compressors can obtain. In fact about 30 lines of CARGO meta-code supplemented with less than 90 lines of C++ code on the top of our framework are sufficient to implement a SAM compressor achieving compression levels similar to those offered by the recently published DeeZ [4] and being several times faster at both compression and decompression (**Figure 5**, **methods 4 and 12** versus **methods 6, 8 and 9**). With some more code (70 lines of CARGO meta-code and about 2,000 lines of C++ code) one can implement a fully-fledged SAM compressor offering advanced features like reference-based sequence compression and optional lossy Illumina 8-level quality re-binning; both compression levels and (de)compression speed are on par with those provided by the latest version of HTSLib-based [9] CRAM [8] implementation (**Figure 5**, **method 2 versus 3** and **method 7 versus 10**). In addition, by changing a few lines in the meta-code description we can easily generate a family of different compressors that are suitable to different scenarios (like: slower compression but faster decompression; or both slower compression and decompression but smaller final size). Our best lossy quality re-sampling scheme uses the same compression setups as CRAM, additionally stripping out read names and optional SAM fields as in [8]; it produces archives that are about two times smaller than the corresponding CRAM ones, and is able to compress/decompress about twice as fast as CRAM (**Figure 5**, **method 1 versus 3**). Of note, although performing the same function our family of compressors and those developed so far to operate on fixed-format files differ in a number of fundamental philosophical points. In our case, good performance stems as a straightforward by-product from simple design choices (like separate streams and automatic user-transparent multitasking, see **Methods**) rather than from highly optimized ad-hoc code. In addition while, say, a SAM file compressed to CRAM needs a specific tool for the semantics of its content to be successfully recovered, the same data compressed within our framework does not. For instance, we fully

## Re-compression of SAM files, (A)

**HG01880, 82 GB, unsorted, different compression scenarios**

| Scenario | | Method ID & name | Compression | | | Speed (MB/s) | | Lines of code | |
|---|---|---|---|---|---|---|---|---|---|
| | | | Size (MB) | Fraction | Fold | Compr. | Decompr. | Meta | C++ |
| Lossy quality binning | 1 | CARGO-Ref-Q8-Max | 3,852 | 4.7% | 21.25 | 147.1 | 224.6 | 67 | ~2.2K |
| | 2 | CARGO-Ref-Q8 | 5,233 | 6.4% | 15.64 | 79.0 | 122.3 | 67 | ~2.2K |
| | 3 | SCRAMBLE-CRAM-Q8 [8][9] | 6,077 | 7.4% | 13.47 | 110.4 | 117.9 | | |
| | 4 | CARGO-Std-Q8 | 6,468 | 7.9% | 12.66 | 37.6 | 119.3 | 28 | 86 |
| | 5 | CARGO-Ext-Q8 | 6,894 | 8.4% | 11.87 | 47.6 | 129.9 | 42 | 157 |
| | 6 | DEEZ-Q8 [4] | 7,767 | 9.5% | 10.54 | 25.6 | 27.9 | | |
| Lossless | 7 | CARGO-Ref | 9,623 | 11.8% | 8.51 | 74.8 | 115.1 | 67 | ~2.2K |
| | 8 | DEEZ-Samcomp | 10,120 | 12.4% | 8.09 | 21.8 | 22.4 | | |
| | 9 | DEEZ-Normal | 10,596 | 12.9% | 7.73 | 25.1 | 27.1 | | |
| | 10 | SCRAMBLE-CRAM | 10,698 | 13.1% | 7.65 | 100.9 | 115.6 | | |
| | 11 | SAMTOOLS-CRAM [1][9] | 10,712 | 13.1% | 7.64 | 24.2 | 13.2 | | |
| | 12 | CARGO-Std | 10,869 | 13.3% | 7.53 | 36.4 | 111.1 | 28 | 58 |
| | 13 | CARGO-Ext | 11,284 | 13.8% | 7.26 | 46.1 | 115.9 | 42 | 129 |
| | 14 | BZIP2-Best | 14,271 | 17.4% | 5.74 | 52.2 | 111.3 | | |
| | 15 | BZIP2-Fast | 15,091 | 18.5% | 5.42 | 63.4 | 121.7 | | |
| | 16 | GZIP-Best | 16,540 | 20.2% | 4.95 | 95.6 | 113.3 | | |
| | 17 | SCRAMBLE-BAM [8][9] | 18,418 | 22.5% | 4.44 | 136.4 | 105.1 | | |
| | 18 | SAMTOOLS-BAM [1][9] | 18,420 | 22.5% | 4.44 | 70.4 | 86.2 | | |
| | 19 | GZIP-Fast | 20,056 | 24.5% | 4.08 | 263.4 | 101.7 | | |

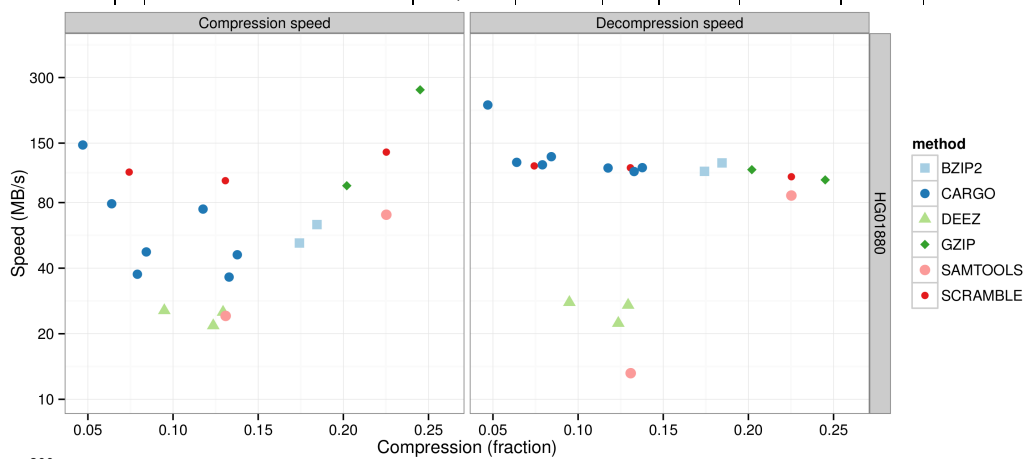

**Figure 5.** Several methods for re-compressing SAM files benchmarked on a 82 GB file containing DNA-sequencing data. In the names of lossy methods, "Q8" means that qualities have been downsampled to 8 possible values following the Illumina binning scheme; "Max" means that read names and optional SAM fields have been discarded as well (see the **Supplementary Data** for a precise definition of each method). "Fraction" is the ratio between the sizes of compressed and original file; "Fold" is the ratio between the sizes of original and compressed file. For each CARGO method, the number of source code lines needed on the top of the CARGO library is also shown.

parse the SAM record including its optional fields, and every component of the data structure is subsequently available to (de)compressing programs on the C++ side of the CARGO implementation. It would be perfectly possible to output a different format (for instance, one including less or additional fields) starting from the same database.

**Scaling up to multi-TB datasets**

In addition, our approach can easily scale up to multi-TB datasets. **Figure 6** illustrates the results of an experiment where we compress a large collection of SAM volumes from the 1000 Genome Project (17 TB uncompressed total, see **Supplementary Data** section 2 for the complete list of archives) into (1) a single CARGO container using our second most effective compression scheme, same as in **Figure 5 method 2** (2) a single BAM archive recompressed with SCRAMBLE, same as in **Figure 5 method 17** (3) a single CRAM archive recompressed with SCRAMBLE, same as in **Figure 5 method 3**. The CARGO scheme was selected because it is directly comparable to BAM or CRAM, provided that suitable parameters are selected for those methods (see **Supplementary Data**). CARGO streams were stored sorted by genomic position, in order to make the archive searchable and mimic the search-by-position capabilities offered by the BAM/CRAM formats. In line with the results presented in **Figure 5**, we achieved a compression rate that is almost twice as much as that of BAM's, and significantly better than the one provided by CRAM. In addition, the time needed to query a CARGO container is either slightly worse, or comparable to, the one needed to query the much larger BAM container; and it is much better than the querying time for a CRAM container (see **Supplementary Data** section 1.2). Finally, the data throughput obtained by CARGO is several times higher than what SAMtools implementations of either BAM or CRAM format can offer at the moment (see **Supplementary Data** sections 1.1 and 1.2), making CARGO the ideal storage tool for high-performance downstream applications.

# Outlook

In general, what our approach can achieve goes far beyond the compression of formats like FASTQ or SAM, offering many advantages with respect to most solutions currently available in the field of genomics. First, in the spirit of database format design and different from rigid, complicated, ambiguously-defined file-based formats like SAM, our data representation is based on an *abstract description of the content* defined in terms of a specialized domain-specific language: this results in flexibility, in fast reaction to new genomic technologies or specific data analysis frameworks requiring amendments to the data structure (whereas solutions like SAM would require a re-definition of the format and its re-implementation in all downstream tools), in meaningful data interchange based on the semantics of the record description, in efficient automatic optimization and multi-threading of CARGO applications. Second, our framework is high-level and modular, thus fostering *parsimonious implementations* that require very little amounts of CARGO meta-language and C++ code to produce optimized compressed storage systems for complex data structures: the user does not need to install, or delve into the complicated technical details of, a general-purpose database system/data framework. Third, our simple and concise scheme allows for a *versatile approach to storage*: in order to identify the combination that best matches the requirements of an application, the user can easily try out different data structures and combinations of compression methods for each data field; in

## Re-compression of SAM files, (B)

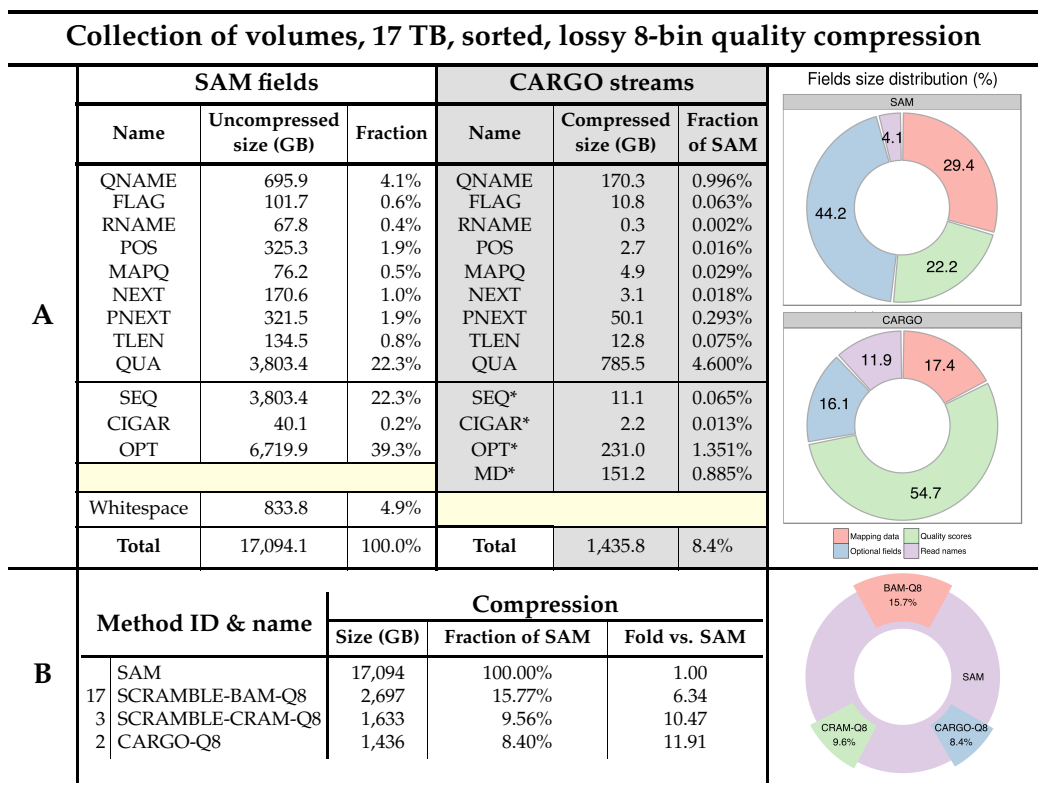

**Collection of volumes, 17 TB, sorted, lossy 8-bin quality compression**

|   | SAM fields ||| CARGO streams |||
|---|---|---|---|---|---|---|
|   | Name | Uncompressed size (GB) | Fraction | Name | Compressed size (GB) | Fraction of SAM |
| **A** | QNAME | 695.9 | 4.1% | QNAME | 170.3 | 0.996% |
|   | FLAG | 101.7 | 0.6% | FLAG | 10.8 | 0.063% |
|   | RNAME | 67.8 | 0.4% | RNAME | 0.3 | 0.002% |
|   | POS | 325.3 | 1.9% | POS | 2.7 | 0.016% |
|   | MAPQ | 76.2 | 0.5% | MAPQ | 4.9 | 0.029% |
|   | NEXT | 170.6 | 1.0% | NEXT | 3.1 | 0.018% |
|   | PNEXT | 321.5 | 1.9% | PNEXT | 50.1 | 0.293% |
|   | TLEN | 134.5 | 0.8% | TLEN | 12.8 | 0.075% |
|   | QUA | 3,803.4 | 22.3% | QUA | 785.5 | 4.600% |
|   | SEQ | 3,803.4 | 22.3% | SEQ* | 11.1 | 0.065% |
|   | CIGAR | 40.1 | 0.2% | CIGAR* | 2.2 | 0.013% |
|   | OPT | 6,719.9 | 39.3% | OPT* | 231.0 | 1.351% |
|   |   |   |   | MD* | 151.2 | 0.885% |
|   | Whitespace | 833.8 | 4.9% |   |   |   |
|   | **Total** | **17,094.1** | **100.0%** | **Total** | **1,435.8** | **8.4%** |

|   | Method ID & name | Compression |||
|---|---|---|---|---|
|   |   | Size (GB) | Fraction of SAM | Fold vs. SAM |
| **B** | SAM | 17,094 | 100.00% | 1.00 |
| 17 | SCRAMBLE-BAM-Q8 | 2,697 | 15.77% | 6.34 |
| 3 | SCRAMBLE-CRAM-Q8 | 1,633 | 9.56% | 10.47 |
| 2 | CARGO-Q8 | 1,436 | 8.40% | 11.91 |

**Figure 6.** Three methods for re-compressing SAM files benchmarked on a collection of files containing 17,094 GB of DNA-sequencing data. "Fraction" is the ratio between the sizes of compressed and original file; "Fold" is the ratio between the sizes of original and compressed file. **Panel A**: Breakdown of SAM file size by SAM field (see [1] for the precise definition) and of CARGO container size by CARGO streams (see **Supplementary Data** for a precise definition — when a CARGO stream does not exactly correspond to a SAM field, it is marked with an asterisk). **Panel B**: Benchmark results at a glance.

addition, the stock CARGO engine can be easily extended by incorporating new compression methods and/or support for accelerated hardware where available. Finally, with its *container-based approach* the system offers a simple way to store analysis intermediates without cluttering the storage with too many files, which is usually very expensive on PB-scale filesystems: this helps improving overall performance on computing clusters. As our results stem from a few lines of code, however, they are just a glimpse of what CARGO can offer to the field of high-throughput genomics.

# CARGO Supplementary

*Release 0.7rc-internal*

**Lukasz Roguski, Paolo Ribeca**

May 26, 2015

CHAPTER

ONE

SUPPLEMENTARY DATA

## 1.1 SAM format benchmarks

In this section we describe in detail how our SAM format compression ratio and throughput benchmarks were performed. We compare the results obtained by several *CARGO* implementations, each one having a different degree of sophistication and complexity, to those of several state-of-the-art SAM format-specific compressors.

### 1.1.1 Data set

The test data sets consist of mapped sequences of *H. Sapiens* individuals, which were downloaded from the 1000 Genomes Project in BAM [1] format. We decompressed files and converted them to the SAM format using *SAMtools* version 1.1. Some test scenarios use reference-based compression techniques and thus require the original FASTA reference file which the sequencing reads have been mapped to – in the case of the 1000 Genomes Project the *GRCh37* Human Genome assembly was used, which we also downloaded from the Project's repository (see *Appendix A - datasets*).

### 1.1.2 Reference compressors

The binaries for reference compressors were downloaded from their official websites or compiled from source with default options set in build scripts.

All applications were tested using 8 processing threads (whenever multi-threading is supported by the application).

#### GZIP

We used version 2.3.1 of *pigz*, a parallel implementation of the *gzip* compression tool (available at http://zlib.net/pigz/). The application was tested in 2 configurations: *GZIP-FAST* providing a good compression ratio together with fast performance, and *GZIP-BEST* providing the highest compression ratio.

#### BZIP2

We used version 1.1.8 of *pbzip2*, a parallel implementation of the *bzip2* compression tool (available at http://compression.ca/pbzip2/). The application was tested in 2 configurations: *BZIP2-FAST* providing a good compression ratio together with fast performance, and *BZIP2-BEST* providing the highest compression ratio.

**SAMTOOLS**

We used *SAMtools* version 1.1 (available at http://www.htslib.org/); it implements both BAM and CRAM [2] formats. Similar to the newest versions of *SAMtools*, it is based on *HTSLib*. We tested 2 configurations of the application: *SAMTOOLS-BAM* implementing the standard SAM-to-BAM format compression, and *SAMTOOLS-CRAM* implementing reference-based compression and storing the data in CRAM format.

**SCRAMBLE**

We used version 1.13.7 of *sCRAMble* [3], a SAM/BAM/CRAM format conversion toolkit (available at http://sourceforge.net/projects/staden/files/io_lib/). We tested 3 configurations of the application: *SCRAMBLE-BAM* implementing the standard SAM-to-BAM format compression, *SCRAMBLE-CRAM* implementing reference-based compression, and *SCRAMBLE-CRAM-Q8* implementing both reference-based compression alongside with the Illumina Q-scores reduction [4] scheme. The first configuration outputs a BAM file, the two latter ones CRAM.

**DEEZ**

We tested 3 configurations of *DeeZ* [5] (available at http://sfu-compbio.github.io/deez/). *DEEZ-NORMAL* uses default compression parameters, *DEEZ-SAMCOMP* uses a *sam_comp*-compatible [6] compression method and *DEEZ-Q8* uses the default compression method alongside with the Illumina Q-scores reduction scheme.

### 1.1.3 CARGO

We also tested a number of different *CARGO* implementations of the SAM format. The source code for them is available in the standard *CARGO* distribution in the directory `cargo/examples/sam/` and the pre-compiled binaries can be found at `cargo/examples/bin/`. To build and test the executables for all the implementations one needs the following tools: `cargo_translate` (generates C++ files from record definition in *CARGO* meta-language) and `cargo_tool` (allows container management). They are available in the directory `cargo/tools/`.

The examples were all compiled from source and tested using as runtime parameters 8 processing threads and a 64 MB block for the input file buffer.

A more detailed description about how the examples were implemented, how they can be compiled step-by-step and which command-line parameters should be used with them is available in the **Supplementary Documentation**.

**Container configuration**

Before each test, a temporary container was created, setting the available compressible storage space up to a size of 12.8 GB (for details about the creation of the container see *Appendix B - tools invocation*).

**Size measurements**

The space occupancy results reported for *CARGO* are those provided by `cargo_tool` about the cumulative size of the compressed dataset. As containers are usually arbitrarily big and can contain more than a single dataset, this might be considered as an indirect measurement of the size of the dataset. However, containers can be shrunk to eliminate free blocks and thus reclaim unused space. In the current implementation, the difference between the actual size of the compressed dataset and the size of the container having been shrunk (which depends on the container block size selected by the user and the amount of block padding inside the container due to data granularity) will usually be negligible with respect to the large size of the compressed dataset, as corroborated by many shrinkage tests we conducted. Being this the case, we finally decided to skip the shrinkage step altogether, and directly report the size of the compressed database instead.

*CARGO* methods

**CARGO-SAM-STD**

*CARGO-SAM-STD* is a simple, proof-of-concept SAM format file compressor, where most of the SAM fields are represented either as a string or an integer and compressed using different schemes – the source code for this example is available in the directory `cargo/examples/sam-std/` of the standard *CARGO* distribution, and a pre-compiled binary in `cargo/examples/bin/cargo_samrecord_toolkit-std`. This solution was tested in 2 configurations: lossless *CARGO-SAM-STD*, and lossy *CARGO-SAM-STD-Q8* implementing the Illumina Q-scores reduction scheme.

**CARGO-SAM-EXT**

*CARGO-SAM-EXT* is an extended version of the *CARGO-SAM-STD* example, which, in addition to the methods described in *CARGO-SAM-STD*, also performs tokenization of the optional fields instead of compressing them together as one long string. The source code for this example is available in `cargo/examples/sam-ext/`, and the pre-compiled binary can be found at `cargo/examples/bin/cargo_samrecord_toolkit-ext`. This solution was tested in 2 configurations: lossless *CARGO-SAM-EXT*, and lossy *CARGO-SAM-EXT-Q8* implementing the Illumina Q-scores reduction scheme.

**CARGO-SAM-REF**

*CARGO-SAM-REF* is a more advanced lossless SAM format compressor, which in addition to the methods used in *CARGO-SAM-EXT* also implements:

- more complex tokenization of the SAM optional fields,
- internal alignment description in terms of a special combination of `SEQ` SAM field, `CIGAR` field and `MD` optional field,
- reference-based sequence compression,
- transformations of several SAM numerical fields including `TLEN` and `PNEXT`.

Additionally, we have two more lossy schemes: *CARGO-SAM-REF-Q8* implements the Illumina Q-scores reduction transformation on the top of the previous lossless manipulations, while *CARGO-SAM-REF-Q8-MAX* also discards both the `QNAME` field and the optional fields `OPT` (keeping intact only the alignment mismatch information defined by the `MD` tag).

The source codes for this example are available in the directory `cargo/examples/sam/sam-ref/` of the *CARGO* distribution, while the pre-compiled binares, one for each different test case, can be found as `cargo/examples/bin/cargo_samrecord_toolkit_{ ref, ref-q8, ref-q8-max }`.

### 1.1.4 Test setup

The experiments were performed on a server machine equipped with four 8-core AMD OpteronTM 6136 2.4GHz CPUs, 128 GB of RAM and a RAID-5 disk matrix containing 6 non-solid state HDDs.

### 1.1.5 Results

In this section we present the results of our benchmarks on the compression of SAM format. In the tables below the **Ratio** column is computed as *original SAM size / compressed size*, while **C/size** is the size of the compressed dataset in MB. The **C/speed** and **D/speed** fields represent compression and decompression speed in MB/s, while **C/time** and **D/time** are the total compression/decompression processing times in seconds.

Additionally, in Figure *Compression of SAM format: results of throughput vs. ratio benchmarks* we visualize how our *CARGO*-generated solutions and other SAM format-specific compressors perform in terms of both (de)compression speed and ratio.

Table 1.1: Compressing file HG00306 from the 1000 Genomes Project with several methods. Initial SAM file size: 68.3 GB

| Method | Ratio | C/size | C/speed | D/speed | C/time | D/time |
|---|---|---|---|---|---|---|
| CARGO-REF-Q8-MAX | 16.99 | 4017 | 109 | 204 | 626 | 335 |
| CARGO-REF-Q8 | 13.13 | 5199 | 74 | 121 | 921 | 562 |
| SCRAMBLE-CRAM-Q8 | 12.17 | 5608 | 118 | 201 | 579 | 340 |
| CARGO-STD-Q8 | 11.26 | 6061 | 38 | 130 | 1788 | 527 |
| CARGO-EXT-Q8 | 10.52 | 6490 | 44 | 126 | 1547 | 541 |
| DEEZ-Q8 | 10.44 | 6537 | 24 | 29 | 2822 | 2328 |
| CARGO-REF | 7.85 | 8699 | 70 | 132 | 973 | 518 |
| DEEZ-SAMCOMP | 7.6 | 8976 | 22 | 22 | 3123 | 3078 |
| SCRAMBLE-CRAM | 7.33 | 9318 | 113 | 180 | 603 | 379 |
| DEEZ-NORMAL | 7.29 | 9369 | 26 | 28 | 2649 | 2467 |
| SAMTOOLS-CRAM | 7.29 | 9364 | 27 | 10 | 2522 | 6601 |
| CARGO-STD | 7.13 | 9579 | 37 | 125 | 1853 | 544 |
| CARGO-EXT | 6.83 | 9990 | 43 | 123 | 1600 | 554 |
| BZIP2-BEST | 5.57 | 12246 | 58 | 189 | 1186 | 360 |
| BZIP2-FAST | 5.23 | 13062 | 68 | 219 | 1008 | 311 |
| GZIP-BEST | 4.82 | 14155 | 93 | 162 | 733 | 421 |
| SAMTOOLS-BAM | 4.31 | 15830 | 72 | 86 | 943 | 792 |
| SCRAMBLE-BAM | 4.31 | 15829 | 172 | 149 | 397 | 457 |
| GZIP-FAST | 3.96 | 17243 | 311 | 151 | 219 | 452 |

Table 1.2: Compressing file HG01880 from the 1000 Genomes Project with several methods. Initial SAM file size: 81.9 GB

| Method | Ratio | C/size | C/speed | D/speed | C/time | D/time |
|---|---|---|---|---|---|---|
| CARGO-REF-Q8-MAX | 21.25 | 3852 | 147 | 225 | 556 | 364 |
| CARGO-REF-Q8 | 15.64 | 5233 | 79 | 122 | 1035 | 669 |
| SCRAMBLE-CRAM-Q8 | 13.47 | 6077 | 110 | 118 | 741 | 694 |
| CARGO-STD-Q8 | 12.66 | 6468 | 38 | 119 | 2178 | 686 |
| CARGO-EXT-Q8 | 11.87 | 6894 | 48 | 130 | 1720 | 630 |
| DEEZ-Q8 | 10.54 | 7767 | 26 | 28 | 3201 | 2937 |
| CARGO-REF | 8.51 | 9623 | 75 | 115 | 1094 | 711 |
| DEEZ-SAMCOMP | 8.09 | 10120 | 22 | 22 | 3756 | 3662 |
| DEEZ-NORMAL | 7.73 | 10596 | 25 | 27 | 3257 | 3024 |
| SCRAMBLE-CRAM | 7.65 | 10698 | 101 | 116 | 811 | 708 |
| SAMTOOLS-CRAM | 7.64 | 10712 | 24 | 13 | 3389 | 6210 |
| CARGO-STD | 7.53 | 10869 | 36 | 111 | 2250 | 737 |
| CARGO-EXT | 7.26 | 11284 | 46 | 116 | 1776 | 706 |
| BZIP2-BEST | 5.74 | 14271 | 52 | 111 | 1568 | 735 |
| BZIP2-FAST | 5.42 | 15091 | 63 | 122 | 1291 | 672 |
| GZIP-BEST | 4.95 | 16540 | 96 | 113 | 856 | 722 |
| SCRAMBLE-BAM | 4.44 | 18418 | 136 | 105 | 600 | 778 |
| SAMTOOLS-BAM | 4.44 | 18420 | 70 | 86 | 1163 | 949 |
| GZIP-FAST | 4.08 | 20056 | 263 | 102 | 310 | 805 |

Table 1.3: Compressing file HG03780 from the 1000 Genomes Project with several methods. Initial SAM file size: 75.5 GB

| Method | Ratio | C/size | C/speed | D/speed | C/time | D/time |
|---|---|---|---|---|---|---|
| CARGO-REF-Q8-MAX | 16.37 | 4611 | 169 | 252 | 447 | 299 |
| CARGO-REF-Q8 | 13.15 | 5738 | 88 | 140 | 857 | 537 |
| SCRAMBLE-CRAM-Q8 | 11.93 | 6329 | 112 | 117 | 674 | 645 |
| CARGO-STD-Q8 | 10.67 | 7071 | 39 | 125 | 1948 | 606 |
| DEEZ-Q8 | 10.45 | 7220 | 27 | 28 | 2818 | 2657 |
| CARGO-EXT-Q8 | 10.21 | 7392 | 47 | 132 | 1615 | 572 |
| CARGO-REF | 7.40 | 10202 | 78 | 125 | 962 | 605 |
| SCRAMBLE-CRAM | 7.22 | 10451 | 113 | 121 | 670 | 626 |
| SAMTOOLS-CRAM | 7.20 | 10486 | 25 | 12 | 3046 | 6083 |
| DEEZ-SAMCOMP | 7.00 | 10786 | 20 | 21 | 3735 | 3569 |
| DEEZ-NORMAL | 6.76 | 11168 | 24 | 26 | 3121 | 2866 |
| CARGO-STD | 6.53 | 11550 | 37 | 111 | 2037 | 678 |
| CARGO-EXT | 6.37 | 11855 | 44 | 110 | 1697 | 683 |
| BZIP2-BEST | 5.10 | 14799 | 54 | 123 | 1395 | 614 |
| BZIP2-FAST | 4.86 | 15540 | 65 | 111 | 1169 | 681 |
| GZIP-BEST | 4.57 | 16499 | 104 | 120 | 725 | 631 |
| SCRAMBLE-BAM | 4.12 | 18321 | 139 | 98 | 545 | 767 |
| SAMTOOLS-BAM | 4.12 | 18322 | 71 | 85 | 1061 | 889 |
| GZIP-FAST | 3.84 | 19676 | 208 | 103 | 362 | 730 |

## 1.2 Queryable large-scale SAM format benchmarks

In this section we describe in detail how our large-scale queryable SAM format benchmarks were performed. We compare the results of a typical *CARGO* implementation with those obtained by the *SAMtools* implementations of both BAM and CRAM formats (the syntax for command line invocation of all considered tools can be found in *Appendix B - tools invocation*). We report on the achieved compression rates, the size of the underlying compressed streams and the range querying times of the compressed datasets.

### 1.2.1 Data sets

The test data set consist of 2 collections of several BAM files. The first collection (to which we will refer as *small volume* from now on) consists of 8 BAM files for a total compressed size of 209 GB (880 GB of decompressed SAM); the second collection is made of 157 BAM files for a total compressed size of 4.1 TB (17.1 TB of decompressed SAM). All BAM files were publicly available from download on the 1000 Genomes Project repository, and should be regarded as an essentially random selection of the available data (see: *Appendix A - datasets* for the complete list of the files in each collection).

### 1.2.2 Data preparation

First all BAM files in each collection were merged into a single large and sorted BAM volume (labelled as "BAM") using the *SAMtools* toolkit (`merge` subcommand). In the next step, the volumes were re-compressed

- into another BAM archive (labelled as "BAM-Q8") with *sCRAMble*, applying the Illumina Q-scores reduction scheme
- into a CRAM archive (labeled as "CRAM") with *sCRAMble*, applying the Illumina Q-scores reduction scheme

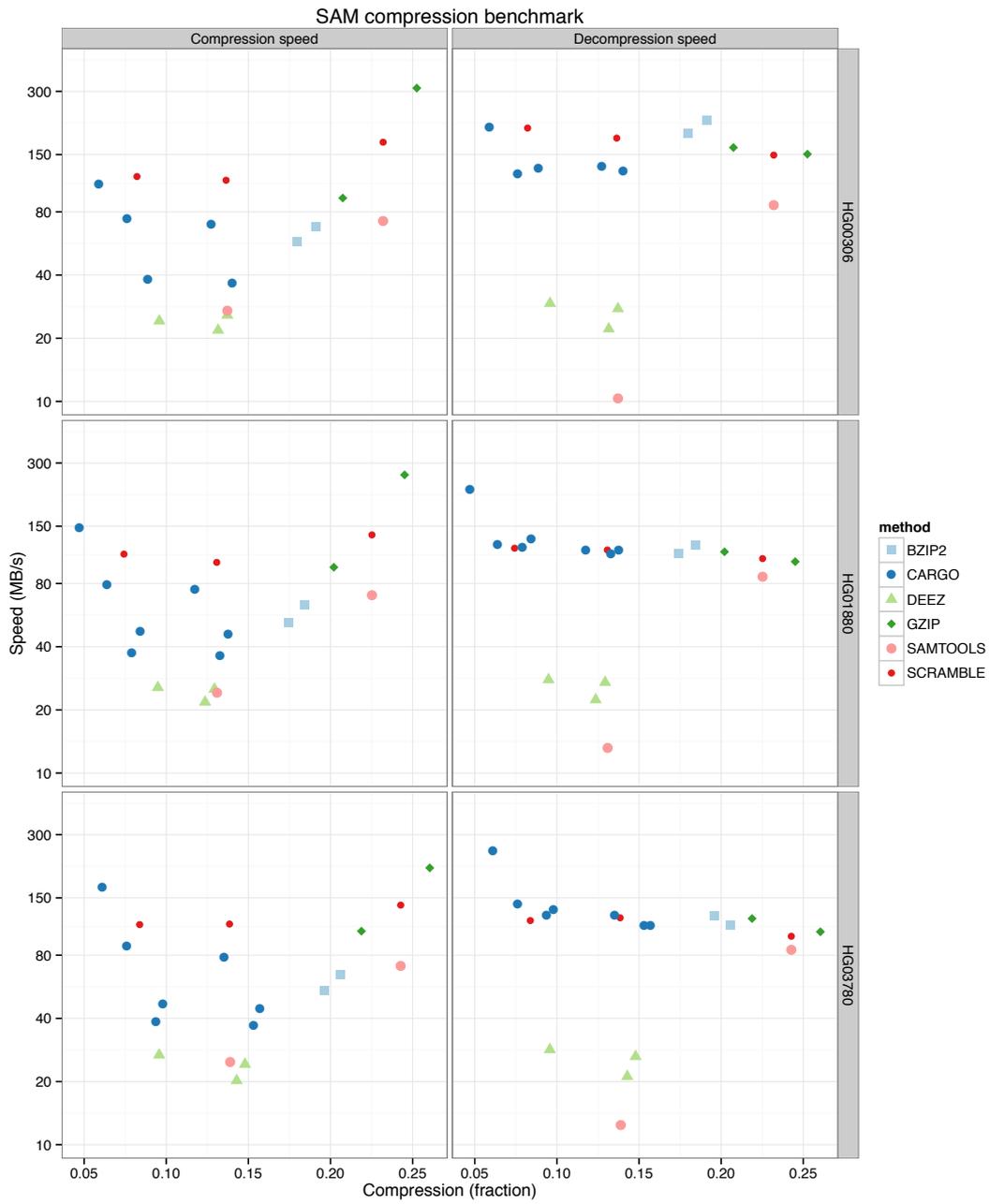

Fig. 1.1: Compression of SAM format: results of throughput vs. ratio benchmarks

- into a *CARGO* archive, using implementation *CARGO-SAM-REF-Q8*.

In all cases, the records in the archives were stored in sorted order, indexed by chromosome and position.

### 1.2.3 *CARGO*

For this test we used a *CARGO-SAM-REF-Q8* method – that is, the SAM format compressor implementing reference-based compression and the Illumina Q-scores reduction scheme that was benchmarked in the previous section. The application source code is available in the standard *CARGO* distribution in the directory `cargo/examples/sam/sam-ref`, and the pre-compiled binaries can be found at `cargo/examples/bin/cargo_samrecord_toolkit-ref-q8`.

#### Container configuration

For each data collection, we created a container with a total available space being 10% of the total uncompressed input data to be stored – respectively up to 1.72 TB (*large volume*) and 86.4 GB (*small volume*). This is enough to accommodate the compressed datasets produced by all *CARGO* methods.

#### Size measurements

The space occupancy results reported for *CARGO* are those provided by `cargo_tool` about the cumulative size of the compressed dataset. As containers are usually arbitrarily big and can contain more than a single dataset, this might be considered as an indirect measurement of the size of the dataset. However, containers can be shrunk to eliminate free blocks and thus reclaim unused space (see *Appendix B - tools invocation*). In the current implementation, the difference between the actual size of the compressed dataset and the size of the container having been shrunk (which depends on the container block size selected by the user and the amount of block padding inside the container due to data granularity) will usually be negligible with respect to the large size of the compressed dataset, as corroborated by many shrinkage tests we conducted. Being this the case, we finally decided to skip the shrinkage step altogether, and directly report the size of the compressed database instead.

### 1.2.4 Test setup

The tests were performed on the CNAG cluster. It is made of more than 100 compute nodes each one having two Intel Xeon Quad Core 2.93 GHz processors with 48 GB of RAM. It has about 3 PB of network-distributed hard-drive storage mounted as a *Lustre* parallel file system (http://lustre.org/). Inter-node communication is performed via a dedicated Infiniband network, whereas the *Lustre* filesystem is connected to the cluster via a number of standard Gigabit Ethernet connections. In practice, filesystem access through the network turns out to be the computational bottleneck for many applications.

### 1.2.5 Results

#### Volume compression results

In this section we present the results of our large-scale benchmarks on the compression of SAM format. In the tables below the **SAM ratio** column is computed as *original SAM size / compressed size*, while **SAM fraction (%)** is the ratio in per cents between the *compressed size* and the *original SAM size*, that is 100 * *compressed size / original SAM size* (ratio and fraction are two common measures used to quantify compression). Analogous definitions hold true for **BAM ratio**, **BAM frac.(%)** and **CRAM frac.(%)**.

Table 1.4: Summary of SAM compression for the small volume

| Format | size (GB) | SAM ratio | BAM ratio | SAM frac.(%) | BAM frac.(%) | CRAM frac.(%) |
|---|---|---|---|---|---|---|
| SAM | 880.93 | — | — | — | — | — |
| BAM | 209.63 | 4.20 | — | 23.80 | — | — |
| BAM-Q8 | 145.00 | 6.08 | 1.45 | 16.46 | 69.17 | — |
| CRAM | 85.30 | 10.33 | 2.46 | 9.68 | 40.69 | — |
| CARGO | 72.91 | 12.08 | 2.88 | 8.28 | 34.78 | 85.48 |

Table 1.5: Summary of SAM compression for the large volume

| Format | size (TB) | SAM ratio | BAM ratio | SAM frac.(%) | BAM frac.(%) | CRAM frac.(%) |
|---|---|---|---|---|---|---|
| SAM | 17.09 | — | — | — | — | — |
| BAM | 4.03 | 4.24 | — | 23.58 | — | — |
| BAM-Q8 | 2.70 | 6.34 | 1.49 | 15.77 | 66.91 | — |
| CRAM | 1.63 | 10.47 | 2.47 | 9.56 | 40.53 | — |
| CARGO | 1.44 | 11.91 | 2.81 | 8.40 | 35.63 | 87.90 |

**SAM fields compression results**

In this section we break down the global compression results shown in the previous section (see *Volume compression results*) and show in detail how the *CARGO* method considered in our benchmark is able to compress various sets of SAM fields considered as separated entities. More in detail, in the following table we report information for the following (sets of) SAM fields:

- *Read names*: the contents of the `QNAME` SAM field
- *Quality scores*: the contents of the `QUAL` SAM field
- *Optional fields*: the contents of all the SAM optional fields excluding the `MD` tag (which in our implementation is compressed together with the *Mapping data* below)
- *Mapping data*: the content of all the remaining fields.

In the tables below **SAM fraction (%)** is the ratio in per cents between the *size of the information contained in the SAM field* and the *total size of the SAM file*, that is 100 * *size of the information contained in the SAM field* / *total size of the SAM file*; a similar definition holds true for **CARGO frac.(%)**. **CARGO ratio** is computed as *size of the information contained in the SAM field / compressed size of the corresponding 'CARGO streams collection*. All sizes are in GB.

Table 1.6: Summary of SAM fields compression for the small volume

| Fields | SAM size | SAM frac.(%) | CARGO size | CARGO frac.(%) | CARGO ratio |
|---|---|---|---|---|---|
| Read names | 35.5 | 4.0 | 7.7 | 10.2 | 4.59 |
| Mapping data | 260.5 | 29.6 | 17.3 | 22.8 | 15.07 |
| Quality scores | 197.3 | 22.4 | 39.8 | 52.4 | 4.96 |
| Optional fields | 387.6 | 44.0 | 11.1 | 14.7 | 34.77 |

Table 1.7: Summary of SAM fields compression for the large volume

| Fields | SAM size | SAM frac.(%) | CARGO size | CARGO frac.(%) | CARGO ratio |
|---|---|---|---|---|---|
| Read names | 695.9 | 4.1 | 170.3 | 11.2 | 4.09 |
| Mapping data | 5041.1 | 29.5 | 328.7 | 21.7 | 15.34 |
| Quality scores | 3803.4 | 22.2 | 785.5 | 51.8 | 4.84 |
| Optional fields | 7553.8 | 44.2 | 231.0 | 15.2 | 32.70 |

**SAM volumes query results**

In this section we perform range queries on each considered container (the one compressed with BAM, the one compressed with CRAM, and the one compressed with *CARGO*) and collect time measurements for each query. As *SAMtools* is single-threaded, only one result is reported when *SAMtools* is used to query either the BAM or the CRAM container (in columns **BAM** and **CRAM**, respectively). On the other hand, as *CARGO* can employ a different number of threads to perform the query, several results are reported for it (query performed with 1 thread in column **CARGO-T1**, query performed with 2 threads in column **CARGO-T2**, and so on). Column **Query size** contains the size of the output of the query, in MB. Each table row corresponds to a different query, as described in column **Range/type**. The query will span a `chr:pos_begin` – `chr:pos_end` range where both `chr` and `pos_begin` have been randomly selected; the range size is 1k nt for the first two rows of each table, 10k nt for rows 3-4, 100k nt for rows 5-6 and 1M nt for rows 7-8. Queries can be either `cold` (when a new range is sampled for the first time) or `hot` (when a query on the same range is repeated and slices of data representing the queried region are likely to be already present in the cache of the filesystem); both timings are reported in order to evaluate the latency of the distributed network file system. Each table entry is an average on 10 runs.

Table 1.8: Summary of range-querying timings for the small volume

| Range/type | Query size | BAM | CRAM | CARGO-T1 | CARGO-T2 | CARGO-T4 | CARGO-T8 |
|---|---|---|---|---|---|---|---|
| 1k-cold | 0.2 | 0.3 | 49.5 | 1.2 | 1.2 | 1.2 | 1.2 |
| 1k-hot | 0.2 | 0.2 | 48.9 | 1.2 | 1.2 | 1.2 | 1.2 |
| 10k-cold | 2.3 | 0.5 | 50.3 | 1.9 | 1.8 | 2.1 | 1.8 |
| 10k-hot | 2.3 | 0.5 | 46.2 | 1.3 | 1.3 | 1.3 | 1.3 |
| 100k-cold | 27.2 | 0.8 | 50.3 | 2.2 | 1.7 | 1.4 | 1.4 |
| 100k-hot | 27.2 | 0.9 | 49.0 | 2.2 | 1.9 | 1.4 | 1.4 |
| 1M-cold | 262.7 | 2.8 | 47.0 | 14.4 | 5.6 | 3.5 | 3.0 |
| 1M-hot | 262.7 | 2.4 | 46.0 | 10.5 | 5.6 | 3.6 | 3.6 |

Table 1.9: Summary of range-querying timings for the large volume. SAMtools was unable to query CRAM volume

| Range/type | Query size | BAM | CARGO-T1 | CARGO-T2 | CARGO-T4 | CARGO-T8 |
|---|---|---|---|---|---|---|
| 1k-cold | 7.2 | 0.8 | 6.6 | 4.8 | 4.9 | 4.6 |
| 1k-hot | 7.2 | 0.5 | 3.2 | 3.1 | 3.1 | 3.0 |
| 10k-cold | 57.1 | 2.2 | 20.0 | 10.9 | 10.8 | 10.0 |
| 10k-hot | 57.1 | 1.2 | 5.4 | 3.9 | 3.9 | 4.1 |
| 100k-cold | 548.4 | 7.8 | 39.4 | 19.5 | 14.9 | 13.2 |
| 100k-hot | 548.4 | 5.2 | 23.2 | 13.1 | 8.0 | 6.9 |
| 1M-cold | 5607.8 | 45.1 | 181.8 | 92.6 | 48.1 | 36.2 |
| 1M-hot | 5607.8 | 42.2 | 176.4 | 91.9 | 48.0 | 39.6 |

## 1.3 FASTQ format benchmarks

In this section we describe in detail how our FASTQ format compression ratio and throughput benchmarks were performed. We compare the results obtained by a very simple proof-of-concept *CARGO* implementation to those of several state-of-the-art FASTQ format-specific compressors.

### 1.3.1 Data sets

The test data sets consist of FASTQ files produced by different sequencing platforms: Illumina (*SRR608906_2*), Ion Torrent (*ERR039503*) and SOLiD (*SRR445256*). They were downloaded from the Short Reads Archive (SRA) database (see: *Appendix A - datasets*). Unfortunately the files for Ion Torrent and SOLiD are relatively small (6 and 5 GB uncompressed, respectively); however, in line with what is usually done in similar benchmarks and in order to achieve a better representation of available technologies, we decided to include them anyway.

### 1.3.2 Reference solutions

The binaries for reference compressors were downloaded from their official websites or compiled from source with default options set in build scripts.

All applications were tested using 8 processing threads (whenever multi-threading is supported by the application).

All applications were tested in 2 lossless compression configurations: *-FAST* providing a good compression ratio together with fast performance, and *-MAX* providing the highest compression ratio.

#### GZIP

We used version 2.3.1 of `pigz`, a parallel implementation of the standard `gzip` compression tool (available at http://zlib.net/pigz/).

#### BZIP2

We used version 1.1.8 of *pbzip2*, a parallel implementation of the *bzip2* compression tool (available at http://compression.ca/pbzip2/).

#### DSRC

We used version 2.0 of *DSRC* [7] (available at http://sun.aei.polsl.pl/dsrc/).

#### FQZCOMP

We used version 4.6 of *FQZcomp* [6] (available at http://sourceforge.net/projects/fqzcomp/).

#### QUIP

We used version 1.1.7 of *Quip* [8] (available at http://homes.cs.washington.edu/~dcjones/quip/).

### 1.3.3 *CARGO*

We also tested a relatively simple-minded *CARGO* implementations of the FASTQ format. The source code is available in the standard *CARGO* distribution in the directory `cargo/examples/fastq/fastq-multi`; together with it, a specialized build script `build.sh` to generate the test binaries. To build and test all the executables (on per set of compression methods chosen for the FASTQ fields, see *CARGO methods*) one needs the following tools: `cargo_translate` (generates C++ files from record definition in *CARGO* meta-language) and `cargo_tool` (allows container management). They are available in the directory `cargo/tools/`.

The examples were all compiled from source and tested using as runtime parameters 8 processing threads and an 8 MB (in the case of the *-FAST compressors, see *CARGO methods*) or 64 MB (in the case of the *-BEST compressors) block for the input file buffer.

A more detailed description about how the examples were implemented, how they can be compiled step-by-step and which command-line parameters should be used with them is available in the **Supplementary Documentation**.

**Container configuration**

Before each test, a temporary container was created, setting the available compressible storage space up to a size of 8.7 GB (for details about the creation of the container see *Appendix B - tools invocation*).

### 1.3.4 Size measurements

The space occupancy results reported for *CARGO* are those provided by `cargo_tool` about the cumulative size of the compressed dataset. As containers are usually arbitrarily big and can contain more than a single dataset, this might be considered as an indirect measurement of the size of the dataset. However, containers can be shrunk to eliminate free blocks and thus reclaim unused space. In the current implementation, the difference between the actual size of the compressed dataset and the size of the container having been shrunk (which depends on the container block size selected by the user and the amount of block padding inside the container due to data granularity) will usually be negligible with respect to the large size of the compressed dataset, as corroborated by many shrinkage tests we conducted. Being this the case, we finally decided to skip the shrinkage step altogether, and directly report the size of the compressed database instead.

**CARGO methods**

The names of the tested *CARGO* methods follow the pattern:

```
CARGO-<comp_method>-<option>
```

where `comp_method` specifies the compression method name (one of *GZIP*, *BZIP*, *PPMD*, *LZMA* and *OPT*) and `option` specifies the compression method option (either *FAST*, a fast compression setup offering a lower compression ratio, or *MAX*, a slower compression setup offering the maximum compression ratio). The *FAST* method uses compression level 1 for each algorithm, whereas *BEST* uses compression level 4 (see **Supplementary Documentation** for a precise definition of compression levels and methods). The *OPT* compression methods uses different combinations of compression methods for each of the FASTQ fields. In addition, compressors with the *FAST* option use 8 MB, whereas those with *MAX* and *OPT* options use 64 MB as the input file buffer block size.

### 1.3.5 Test setup

The experiments were performed on a server machine equipped with four 8-core AMD OpteronTM 6136 2.4GHz CPUs, 128 GB of RAM and a RAID-5 disk matrix containing 6 non-solid state HDDs.

### 1.3.6 Results

In this section we present the results of our benchmarks on the compression of FASTQ format. In the tables below the **Ratio** column is computed as *original FASTQ size / compressed size*, while **C/size** is the size of the compressed dataset in MB. The **C/speed** and **D/speed** fields represent compression and decompression speed in MB/s, while **C/time** and **D/time** are the total compression/decompression processing times in seconds.

Additionally, in Figure *Compression of FASTQ format: results of throughput vs. ratio benchmarks* we visualize how our *CARGO*-generated solutions and other FASTQ format-specific compressors perform in terms of both (de)compression speed and ratio.

Table 1.10: Compressing file SRR608906_2 (sample Illumina reads from the SRA) with several methods. Initial FASTQ file size: 12.2 GB

| Method | Ratio | C/size | C/speed | D/speed | C/time | D/time |
|---|---|---|---|---|---|---|
| FQZCOMP-BEST | 4.98 | 2437 | 7 | 8 | 1664 | 1578 |
| QUIP-BEST | 4.66 | 2607 | 13 | 13 | 911 | 926 |
| QUIP-FAST | 4.61 | 2633 | 21 | 16 | 580 | 759 |
| CARGO-PPMD-MAX | 4.52 | 2688 | 50 | 44 | 243 | 274 |
| CARGO-OPT-PPP | 4.52 | 2689 | 76 | 67 | 160 | 181 |
| DSRC2-MAX | 4.39 | 2770 | 66 | 69 | 183 | 175 |
| FQZCOMP-FAST | 4.38 | 2774 | 48 | 34 | 253 | 357 |
| CARGO-OPT-PLL | 4.33 | 2805 | 12 | 96 | 1037 | 127 |
| CARGO-LZMA-MAX | 4.31 | 2819 | 9 | 216 | 1326 | 56 |
| CARGO-PPMD-FAST | 4.3 | 2828 | 128 | 107 | 95 | 113 |
| CARGO-OPT-GPP | 4.21 | 2886 | 76 | 117 | 160 | 103 |
| CARGO-LZMA-FAST | 4.16 | 2923 | 13 | 186 | 914 | 65 |
| CARGO-BZIP2-MAX | 4.11 | 2953 | 47 | 134 | 259 | 90 |
| CARGO-BZIP2-FAST | 4.04 | 3009 | 61 | 182 | 197 | 66 |
| DSRC2-FAST | 3.92 | 3098 | 267 | 370 | 45 | 32 |
| CARGO-GZIP-MAX | 3.7 | 3280 | 19 | 491 | 646 | 24 |
| BZIP2-BEST | 3.53 | 3446 | 47 | 100 | 257 | 120 |
| BZIP2-FAST | 3.36 | 3615 | 57 | 185 | 212 | 65 |
| CARGO-GZIP-FAST | 3.26 | 3726 | 287 | 395 | 42 | 30 |
| GZIP-BEST | 2.89 | 4198 | 28 | 132 | 430 | 91 |
| GZIP-FAST | 2.53 | 4802 | 200 | 118 | 60 | 102 |

Table 1.11: Compressing file ERR039503 (sample Ion Torrent reads from the SRA) with several methods. Initial FASTQ file size: 6.0 GB

| Method | Ratio | C/size | C/speed | D/speed | C/time | D/time |
|---|---|---|---|---|---|---|
| FQZCOMP-BEST | 5.58 | 1069 | 7 | 7 | 900 | 906 |
| DSRC2-MAX | 5.15 | 1157 | 81 | 76 | 73 | 78 |
| FQZCOMP-FAST | 5.14 | 1160 | 47 | 35 | 127 | 171 |
| CARGO-PPMD-MAX | 4.91 | 1214 | 71 | 62 | 84 | 96 |
| CARGO-OPT-PPP | 4.87 | 1225 | 98 | 94 | 60 | 63 |
| QUIP-FAST | 4.8 | 1241 | 18 | 16 | 332 | 381 |
| QUIP-BEST | 4.8 | 1241 | 7 | 9 | 813 | 655 |
| CARGO-OPT-GPP | 4.66 | 1278 | 83 | 108 | 71 | 55 |
| CARGO-PPMD-FAST | 4.56 | 1306 | 134 | 109 | 44 | 54 |
| CARGO-OPT-PLL | 4.42 | 1347 | 8 | 200 | 718 | 29 |
| CARGO-LZMA-MAX | 4.39 | 1356 | 8 | 245 | 793 | 24 |
| CARGO-BZIP2-MAX | 4.38 | 1360 | 62 | 133 | 95 | 44 |
| DSRC2-FAST | 4.34 | 1373 | 336 | 452 | 17 | 13 |
| CARGO-BZIP2-FAST | 4.2 | 1421 | 68 | 185 | 87 | 32 |
| CARGO-LZMA-FAST | 4.18 | 1425 | 11 | 250 | 528 | 23 |
| BZIP2-BEST | 4.09 | 1457 | 59 | 163 | 101 | 36 |
| BZIP2-FAST | 3.85 | 1548 | 57 | 239 | 104 | 24 |
| CARGO-GZIP-MAX | 3.83 | 1555 | 11 | 533 | 524 | 11 |
| GZIP-BEST | 3.26 | 1828 | 16 | 140 | 372 | 42 |
| CARGO-GZIP-FAST | 3.17 | 1883 | 86 | 559 | 69 | 10 |
| GZIP-FAST | 2.75 | 2165 | 189 | 115 | 31 | 51 |

Table 1.12: Compressing file SRR445256 (sample SOLiD reads from the SRA) with several methods. Initial FASTQ file size: 5.0 GB

| Method | Ratio | C/size | C/speed | D/speed | C/time | D/time |
| --- | --- | --- | --- | --- | --- | --- |
| FQZCOMP-BEST | 5.54 | 847 | 13 | 12 | 361 | 382 |
| FQZCOMP-FAST | 5.06 | 928 | 53 | 38 | 89 | 122 |
| DSRC2-MAX | 5.05 | 929 | 66 | 70 | 70 | 67 |
| DSRC2-FAST | 4.49 | 1046 | 242 | 417 | 19 | 11 |
| CARGO-LZMA-MAX | 4.46 | 1054 | 8 | 191 | 587 | 24 |
| CARGO-OPT-PLL | 4.36 | 1076 | 15 | 131 | 303 | 35 |
| CARGO-OPT-PPP | 4.28 | 1096 | 74 | 65 | 63 | 71 |
| CARGO-PPMD-MAX | 4.26 | 1102 | 41 | 37 | 115 | 126 |
| CARGO-LZMA-FAST | 4.13 | 1137 | 18 | 204 | 262 | 23 |
| CARGO-BZIP2-MAX | 4.13 | 1136 | 34 | 133 | 137 | 35 |
| CARGO-OPT-GPP | 4.04 | 1163 | 71 | 80 | 65 | 58 |
| CARGO-BZIP2-FAST | 4 | 1174 | 49 | 176 | 96 | 26 |
| CARGO-PPMD-FAST | 3.94 | 1190 | 126 | 103 | 37 | 45 |
| BZIP2-BEST | 3.9 | 1203 | 37 | 185 | 127 | 25 |
| CARGO-GZIP-MAX | 3.75 | 1250 | 35 | 459 | 135 | 10 |
| BZIP2-FAST | 3.71 | 1267 | 50 | 253 | 94 | 18 |
| CARGO-GZIP-FAST | 3.33 | 1410 | 65 | 503 | 72 | 9 |
| GZIP-BEST | 3.25 | 1445 | 62 | 155 | 75 | 30 |
| GZIP-FAST | 2.8 | 1679 | 283 | 129 | 16 | 36 |

## 1.4 References

[1] BAM format specification. http://samtools.github.io/hts-specs/SAMv1.pdf

[2] CRAM format specification. https://samtools.github.io/hts-specs/CRAMv2.1.pdf

[3] Bonfield, JK (2014) *The Scramble conversion tool*. Bioinformatics vol. 30 no. 19

[4] Illumina (2012), *Reducing Whole-Genome Data Storage Footprint*. Technical report. http://www.illumina.com/documents/products/whitepapers/whitepaper_datacompression.pdf

[5] Hach I, Faraz S, Cenk S (2014) *DeeZ: reference-based compression by local assembly*. Nature Methods vol. 11

[6] Bonfield JK, Mahoney MV (2013) *Compression of FASTQ and SAM Format Sequencing Data*. PLoS ONE vol. 8 no. 3

[7] Roguski L, Deorowicz S (2014) *DSRC 2: Industry-oriented compression of FASTQ files*, Bioinformatics vol 30. no 15

[8] Jones DC, Ruzzo WL, Peng X. Katze MG (2012) *Compression of next-generation sequencing reads aided by highly efficient de novo assembly*. Nucleic Acids Research

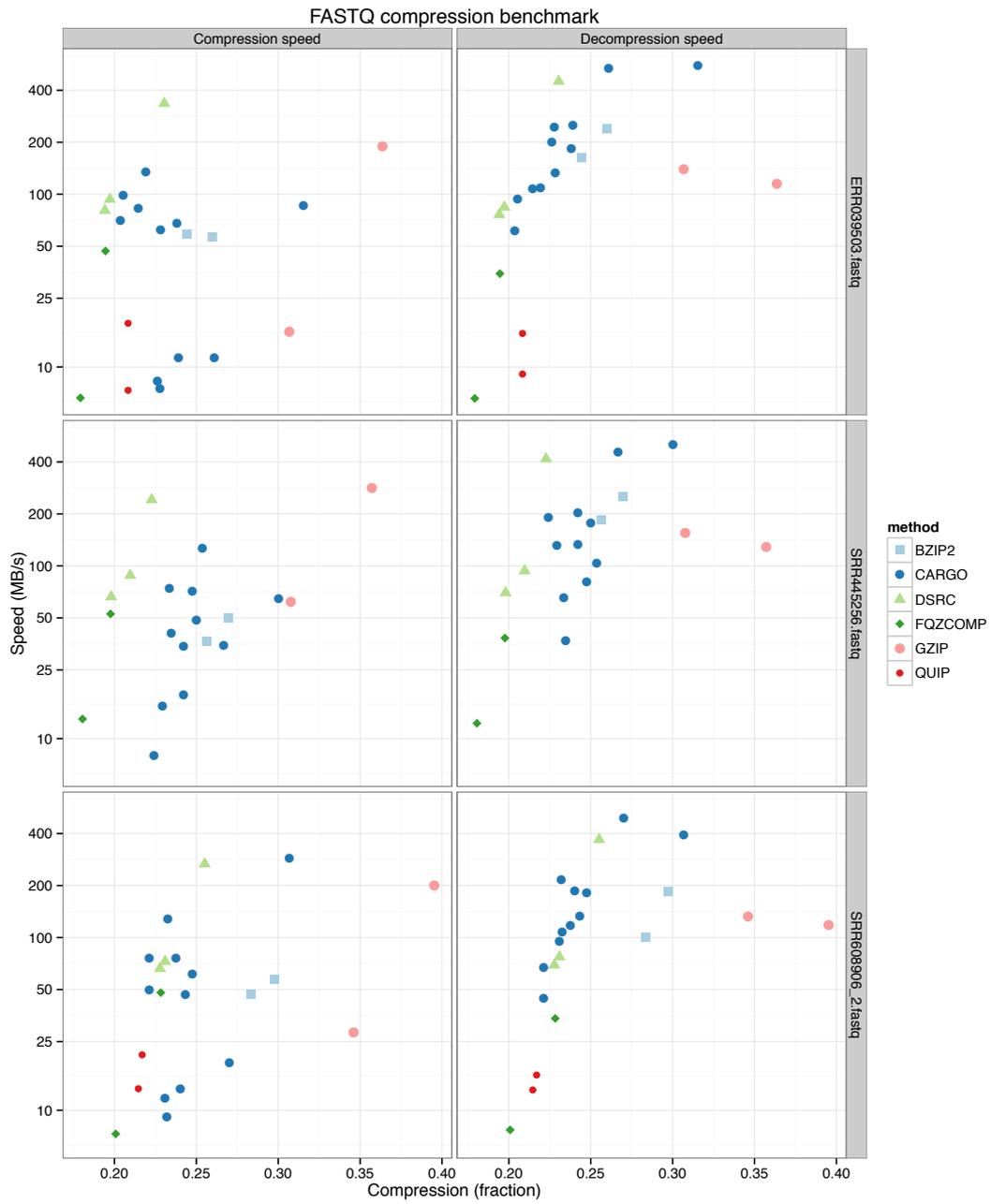

Fig. 1.2: Compression of FASTQ format: results of throughput vs. ratio benchmarks

# CHAPTER
# TWO

# APPENDIX A - DATASETS

## 2.1 SAM format benchmarks

Each file in our test data set consists of a number of aligned sequencing reads that come from a different *H. Sapiens* individual; all reads were produced within the 1000 Genomes project. The files in compressed BAM format were downloaded from the project's public `ftp` repository

```
ftp://ftp.ncbi.nlm.nih.gov/1000genomes/ftp/data/
```

Name and size information for all test files in BAM and SAM format (the latter obtained by decompressing the BAM files with *SAMtools* version 1.1) is presented in the table below.

Table 2.1: SAM files data set used in current benchmark

| File name | BAM size (GB) | SAM size (GB) |
|---|---|---|
| HG00306.mapped.ILLUMINA.bwa.FIN.low_coverage.20120522.sam | 15.8 | 68.3 |
| HG01880.mapped.ILLUMINA.bwa.ACB.low_coverage.20120522.sam | 18.4 | 81.9 |
| HG03780.mapped.ILLUMINA.bwa.ITU.low_coverage.20121211.sam | 18.3 | 75.5 |

Some of our test scenarios use reference-based compression techniques and thus require the original reference genome to which the sequencing reads have been aligned. In the case of the 1000 Genomes Project the *GRCh37* Human Genome assembly was used. The corresponding `gzip`-compressed FASTA file `hs37d5.fa.gz` was also downloaded from the 1000 Genomes project's public `ftp` repository,

```
ftp://ftp.1000genomes.ebi.ac.uk/vol1/ftp/technical/reference/phase2_reference_assembly_sequence/
```

## 2.2 Queryable large-scale SAM format benchmarks

The total data for the test consists of 157 BAM files, for a total (compressed) size of 4071 GB. The first collection of files tested (the *small volume*) consists of only 8 BAM files merged into one sorted BAM file; information for such files is provided in the first table, see *Small volume BAM files*. On the other hand, the second collection (the *large volume*) was prepared from all the 157 BAM files downloaded (all the relevant information about them can be found in the second table, *Large volume BAM files*). Also in this case the files were obtained from the 1000 Genomes project.

Table 2.2: Small volume BAM files

| File name | Size (GB) |
|---|---|
| HG00096.mapped.ILLUMINA.bwa.GBR.low_coverage.20120522.bam | 15.6 |
| HG00097.mapped.ILLUMINA.bwa.GBR.low_coverage.20130415.bam | 32.0 |
| HG00140.mapped.ILLUMINA.bwa.GBR.low_coverage.20130415.bam | 38.1 |
| HG00141.mapped.ILLUMINA.bwa.GBR.low_coverage.20130415.bam | 44.2 |
| HG00142.mapped.ILLUMINA.bwa.GBR.low_coverage.20120522.bam | 14.1 |
| HG00143.mapped.ILLUMINA.bwa.GBR.low_coverage.20121211.bam | 27.3 |
| HG00145.mapped.ILLUMINA.bwa.GBR.low_coverage.20120522.bam | 20.2 |
| HG00146.mapped.ILLUMINA.bwa.GBR.low_coverage.20120522.bam | 17.3 |
| **Total** | **208.9** |

Table 2.3: Large volume BAM files

| File name | Size (GB) |
|---|---|
| HG00096.mapped.ILLUMINA.bwa.GBR.low_coverage.20120522.bam | 15.6 |
| HG00097.mapped.ILLUMINA.bwa.GBR.low_coverage.20130415.bam | 32.0 |
| HG00099.mapped.ILLUMINA.bwa.GBR.low_coverage.20130415.bam | 26.0 |
| HG00100.mapped.ILLUMINA.bwa.GBR.low_coverage.20130415.bam | 44.4 |
| HG00101.mapped.ILLUMINA.bwa.GBR.low_coverage.20130415.bam | 24.0 |
| HG00102.mapped.ILLUMINA.bwa.GBR.low_coverage.20130415.bam | 23.4 |
| HG00103.mapped.ILLUMINA.bwa.GBR.low_coverage.20120522.bam | 17.1 |
| HG00105.mapped.ILLUMINA.bwa.GBR.low_coverage.20130415.bam | 23.7 |
| HG00106.mapped.ILLUMINA.bwa.GBR.low_coverage.20121211.bam | 25.8 |
| HG00107.mapped.ILLUMINA.bwa.GBR.low_coverage.20130415.bam | 25.3 |
| HG00108.mapped.ILLUMINA.bwa.GBR.low_coverage.20120522.bam | 17.9 |
| HG00109.mapped.ILLUMINA.bwa.GBR.low_coverage.20130415.bam | 21.4 |
| HG00110.mapped.ILLUMINA.bwa.GBR.low_coverage.20130415.bam | 32.4 |
| HG00111.mapped.ILLUMINA.bwa.GBR.low_coverage.20120522.bam | 19.1 |
| HG00112.mapped.ILLUMINA.bwa.GBR.low_coverage.20120522.bam | 14.5 |
| HG00113.mapped.ILLUMINA.bwa.GBR.low_coverage.20130415.bam | 31.1 |
| HG00114.mapped.ILLUMINA.bwa.GBR.low_coverage.20120522.bam | 13.6 |
| HG00115.mapped.ILLUMINA.bwa.GBR.low_coverage.20130415.bam | 28.4 |
| HG00116.mapped.ILLUMINA.bwa.GBR.low_coverage.20120522.bam | 19.6 |
| HG00117.mapped.ILLUMINA.bwa.GBR.low_coverage.20120522.bam | 20.7 |
| HG00118.mapped.ILLUMINA.bwa.GBR.low_coverage.20121211.bam | 39.7 |
| HG00119.mapped.ILLUMINA.bwa.GBR.low_coverage.20120522.bam | 17.1 |
| HG00120.mapped.ILLUMINA.bwa.GBR.low_coverage.20120522.bam | 17.5 |
| HG00121.mapped.ILLUMINA.bwa.GBR.low_coverage.20130415.bam | 22.2 |
| HG00122.mapped.ILLUMINA.bwa.GBR.low_coverage.20121211.bam | 24.6 |
| HG00123.mapped.ILLUMINA.bwa.GBR.low_coverage.20120522.bam | 31.0 |
| HG00124.mapped.ILLUMINA.bwa.GBR.low_coverage.20120522.bam | 39.6 |
| HG00125.mapped.ILLUMINA.bwa.GBR.low_coverage.20120522.bam | 13.2 |
| HG00126.mapped.ILLUMINA.bwa.GBR.low_coverage.20121211.bam | 26.0 |
| HG00127.mapped.ILLUMINA.bwa.GBR.low_coverage.20120522.bam | 15.8 |
| HG00128.mapped.ILLUMINA.bwa.GBR.low_coverage.20130415.bam | 23.5 |
| HG00129.mapped.ILLUMINA.bwa.GBR.low_coverage.20130415.bam | 23.7 |
| HG00130.mapped.ILLUMINA.bwa.GBR.low_coverage.20130415.bam | 15.7 |
| HG00131.mapped.ILLUMINA.bwa.GBR.low_coverage.20120522.bam | 18.3 |
| HG00132.mapped.ILLUMINA.bwa.GBR.low_coverage.20130415.bam | 25.8 |
|  | |

Table 2.3 – continued from previous page

| File name | Size (GB) |
|---|---|
| HG00133.mapped.ILLUMINA.bwa.GBR.low_coverage.20120522.bam | 33.0 |
| HG00136.mapped.ILLUMINA.bwa.GBR.low_coverage.20120522.bam | 12.6 |
| HG00137.mapped.ILLUMINA.bwa.GBR.low_coverage.20120522.bam | 12.7 |
| HG00138.mapped.ILLUMINA.bwa.GBR.low_coverage.20120522.bam | 18.3 |
| HG00139.mapped.ILLUMINA.bwa.GBR.low_coverage.20130415.bam | 39.5 |
| HG00140.mapped.ILLUMINA.bwa.GBR.low_coverage.20130415.bam | 38.1 |
| HG00141.mapped.ILLUMINA.bwa.GBR.low_coverage.20130415.bam | 44.2 |
| HG00142.mapped.ILLUMINA.bwa.GBR.low_coverage.20120522.bam | 14.1 |
| HG00143.mapped.ILLUMINA.bwa.GBR.low_coverage.20121211.bam | 27.3 |
| HG00145.mapped.ILLUMINA.bwa.GBR.low_coverage.20120522.bam | 20.2 |
| HG00146.mapped.ILLUMINA.bwa.GBR.low_coverage.20120522.bam | 17.3 |
| HG00148.mapped.ILLUMINA.bwa.GBR.low_coverage.20121211.bam | 34.2 |
| HG00160.mapped.ILLUMINA.bwa.GBR.low_coverage.20120522.bam | 22.0 |
| HG00178.mapped.ILLUMINA.bwa.FIN.low_coverage.20130415.bam | 30.0 |
| HG00189.mapped.ILLUMINA.bwa.FIN.low_coverage.20120522.bam | 13.5 |
| HG00235.mapped.ILLUMINA.bwa.GBR.low_coverage.20130415.bam | 26.9 |
| HG00245.mapped.ILLUMINA.bwa.GBR.low_coverage.20120522.bam | 20.7 |
| HG00255.mapped.ILLUMINA.bwa.GBR.low_coverage.20130415.bam | 24.0 |
| HG00260.mapped.ILLUMINA.bwa.GBR.low_coverage.20130415.bam | 34.9 |
| HG00275.mapped.ILLUMINA.bwa.FIN.low_coverage.20120522.bam | 16.8 |
| HG00282.mapped.ILLUMINA.bwa.FIN.low_coverage.20120522.bam | 17.4 |
| HG00290.mapped.ILLUMINA.bwa.FIN.low_coverage.20130415.bam | 24.4 |
| HG00306.mapped.ILLUMINA.bwa.FIN.low_coverage.20120522.bam | 15.8 |
| HG00315.mapped.ILLUMINA.bwa.FIN.low_coverage.20120522.bam | 12.4 |
| HG00324.mapped.ILLUMINA.bwa.FIN.low_coverage.20120522.bam | 18.8 |
| HG00332.mapped.ILLUMINA.bwa.FIN.low_coverage.20130415.bam | 22.2 |
| HG00341.mapped.ILLUMINA.bwa.FIN.low_coverage.20130415.bam | 26.5 |
| HG00353.mapped.ILLUMINA.bwa.FIN.low_coverage.20130415.bam | 39.2 |
| HG00361.mapped.ILLUMINA.bwa.FIN.low_coverage.20120522.bam | 19.7 |
| HG00371.mapped.ILLUMINA.bwa.FIN.low_coverage.20130415.bam | 24.3 |
| HG00379.mapped.ILLUMINA.bwa.FIN.low_coverage.20130415.bam | 23.3 |
| HG00410.mapped.ILLUMINA.bwa.CHS.low_coverage.20121211.bam | 35.0 |
| HG00422.mapped.ILLUMINA.bwa.CHS.low_coverage.20130415.bam | 38.3 |
| HG00448.mapped.ILLUMINA.bwa.CHS.low_coverage.20130415.bam | 32.6 |
| HG00458.mapped.ILLUMINA.bwa.CHS.low_coverage.20130415.bam | 28.8 |
| HG00530.mapped.ILLUMINA.bwa.CHS.low_coverage.20120522.bam | 16.1 |
| HG00534.mapped.ILLUMINA.bwa.CHS.low_coverage.20120522.bam | 17.3 |
| HG00557.mapped.ILLUMINA.bwa.CHS.low_coverage.20130415.bam | 34.8 |
| HG00566.mapped.ILLUMINA.bwa.CHS.low_coverage.20120522.bam | 14.3 |
| HG00610.mapped.ILLUMINA.bwa.CHS.low_coverage.20120522.bam | 16.9 |
| HG00638.mapped.ILLUMINA.bwa.PUR.low_coverage.20120522.bam | 16.1 |
| HG00674.mapped.ILLUMINA.bwa.CHS.low_coverage.20121211.bam | 31.0 |
| HG00702.mapped.ILLUMINA.bwa.CHS.low_coverage.20120522.bam | 15.1 |
| HG00732.mapped.ILLUMINA.bwa.PUR.low_coverage.20130422.bam | 105.8 |
| HG01029.mapped.ILLUMINA.bwa.CDX.low_coverage.20130415.bam | 24.4 |
| HG01075.mapped.ILLUMINA.bwa.PUR.low_coverage.20120522.bam | 58.0 |
| HG01089.mapped.ILLUMINA.bwa.PUR.low_coverage.20121211.bam | 16.3 |
| HG01104.mapped.ILLUMINA.bwa.PUR.low_coverage.20130415.bam | 24.4 |
| HG01272.mapped.ILLUMINA.bwa.CLM.low_coverage.20130415.bam | 21.4 |
|  | |

Table 2.3 – continued from previous page

| File name | Size (GB) |
|---|---|
| HG01441.mapped.ILLUMINA.bwa.CLM.low_coverage.20120522.bam | 21.4 |
| HG01443.mapped.ILLUMINA.bwa.CLM.low_coverage.20121211.bam | 18.0 |
| HG01589.mapped.ILLUMINA.bwa.PJL.low_coverage.20130415.bam | 17.1 |
| HG01767.mapped.ILLUMINA.bwa.IBS.low_coverage.20130415.bam | 26.7 |
| HG01861.mapped.ILLUMINA.bwa.KHV.low_coverage.20130415.bam | 23.3 |
| HG01880.mapped.ILLUMINA.bwa.ACB.low_coverage.20120522.bam | 18.4 |
| HG02006.mapped.ILLUMINA.bwa.PEL.low_coverage.20130415.bam | 25.9 |
| HG02082.mapped.ILLUMINA.bwa.KHV.low_coverage.20130415.bam | 28.7 |
| HG02184.mapped.ILLUMINA.bwa.CDX.low_coverage.20120522.bam | 15.0 |
| HG02387.mapped.ILLUMINA.bwa.CDX.low_coverage.20120522.bam | 18.3 |
| HG02425.mapped.ILLUMINA.bwa.PEL.low_coverage.20130415.bam | 20.4 |
| HG02541.mapped.ILLUMINA.bwa.ACB.low_coverage.20130415.bam | 22.6 |
| HG02611.mapped.ILLUMINA.bwa.GWD.low_coverage.20121211.bam | 39.0 |
| HG02716.mapped.ILLUMINA.bwa.GWD.low_coverage.20121211.bam | 33.4 |
| HG02840.mapped.ILLUMINA.bwa.GWD.low_coverage.20121211.bam | 20.1 |
| HG03054.mapped.ILLUMINA.bwa.MSL.low_coverage.20130415.bam | 30.5 |
| HG03189.mapped.ILLUMINA.bwa.ESN.low_coverage.20130415.bam | 21.8 |
| HG03445.mapped.ILLUMINA.bwa.MSL.low_coverage.20121211.bam | 26.1 |
| HG03604.mapped.ILLUMINA.bwa.BEB.low_coverage.20130415.bam | 20.1 |
| HG03625.mapped.ILLUMINA.bwa.PJL.low_coverage.20130415.bam | 30.7 |
| HG03731.mapped.ILLUMINA.bwa.ITU.low_coverage.20130415.bam | 36.9 |
| HG03780.mapped.ILLUMINA.bwa.ITU.low_coverage.20121211.bam | 18.3 |
| HG03808.mapped.ILLUMINA.bwa.BEB.low_coverage.20121211.bam | 18.9 |
| HG03905.mapped.ILLUMINA.bwa.BEB.low_coverage.20121211.bam | 17.1 |
| HG04038.mapped.ILLUMINA.bwa.STU.low_coverage.20130415.bam | 16.6 |
| HG04185.mapped.ILLUMINA.bwa.BEB.low_coverage.20130415.bam | 14.8 |
| HG04239.mapped.ILLUMINA.bwa.ITU.low_coverage.20130415.bam | 25.1 |
| NA06984.mapped.ILLUMINA.bwa.CEU.low_coverage.20120522.bam | 29.0 |
| NA06985.mapped.ILLUMINA.bwa.CEU.low_coverage.20120522.bam | 57.7 |
| NA06986.mapped.ILLUMINA.bwa.CEU.low_coverage.20130415.bam | 50.7 |
| NA06989.mapped.ILLUMINA.bwa.CEU.low_coverage.20120522.bam | 17.6 |
| NA06994.mapped.ILLUMINA.bwa.CEU.low_coverage.20120522.bam | 18.4 |
| NA07000.mapped.ILLUMINA.bwa.CEU.low_coverage.20130415.bam | 35.2 |
| NA07037.mapped.ILLUMINA.bwa.CEU.low_coverage.20130502.bam | 26.1 |
| NA07048.mapped.ILLUMINA.bwa.CEU.low_coverage.20120522.bam | 19.0 |
| NA07051.mapped.ILLUMINA.bwa.CEU.low_coverage.20120522.bam | 12.5 |
| NA07056.mapped.ILLUMINA.bwa.CEU.low_coverage.20130415.bam | 18.7 |
| NA07347.mapped.ILLUMINA.bwa.CEU.low_coverage.20130415.bam | 46.0 |
| NA07357.mapped.ILLUMINA.bwa.CEU.low_coverage.20130415.bam | 21.1 |
| NA10847.mapped.ILLUMINA.bwa.CEU.low_coverage.20130502.bam | 34.1 |
| NA10851.mapped.ILLUMINA.bwa.CEU.low_coverage.20130415.bam | 18.7 |
| NA11829.mapped.ILLUMINA.bwa.CEU.low_coverage.20130415.bam | 46.5 |
| NA11830.mapped.ILLUMINA.bwa.CEU.low_coverage.20120522.bam | 11.7 |
| NA11831.mapped.ILLUMINA.bwa.CEU.low_coverage.20120522.bam | 22.9 |
| NA11832.mapped.ILLUMINA.bwa.CEU.low_coverage.20120522.bam | 62.1 |
| NA11840.mapped.ILLUMINA.bwa.CEU.low_coverage.20120522.bam | 30.8 |
| NA11843.mapped.ILLUMINA.bwa.CEU.low_coverage.20120522.bam | 19.6 |
| NA11881.mapped.ILLUMINA.bwa.CEU.low_coverage.20120522.bam | 26.9 |
| NA11892.mapped.ILLUMINA.bwa.CEU.low_coverage.20130415.bam | 28.3 |

Table 2.3 – continued from previous page

| File name | Size (GB) |
|---|---|
| NA11893.mapped.ILLUMINA.bwa.CEU.low_coverage.20130415.bam | 24.0 |
| NA11894.mapped.ILLUMINA.bwa.CEU.low_coverage.20130415.bam | 19.3 |
| NA11918.mapped.ILLUMINA.bwa.CEU.low_coverage.20130415.bam | 33.6 |
| NA11919.mapped.ILLUMINA.bwa.CEU.low_coverage.20130415.bam | 47.5 |
| NA11920.mapped.ILLUMINA.bwa.CEU.low_coverage.20130415.bam | 42.9 |
| NA11930.mapped.ILLUMINA.bwa.CEU.low_coverage.20130415.bam | 26.7 |
| NA11931.mapped.ILLUMINA.bwa.CEU.low_coverage.20130415.bam | 18.8 |
| NA11932.mapped.ILLUMINA.bwa.CEU.low_coverage.20130415.bam | 22.1 |
| NA11933.mapped.ILLUMINA.bwa.CEU.low_coverage.20130415.bam | 21.9 |
| NA11992.mapped.ILLUMINA.bwa.CEU.low_coverage.20120522.bam | 30.4 |
| NA11994.mapped.ILLUMINA.bwa.CEU.low_coverage.20120522.bam | 56.2 |
| NA11995.mapped.ILLUMINA.bwa.CEU.low_coverage.20120522.bam | 19.4 |
| NA12003.mapped.ILLUMINA.bwa.CEU.low_coverage.20120522.bam | 21.4 |
| NA12004.mapped.ILLUMINA.bwa.CEU.low_coverage.20121211.bam | 36.3 |
| NA12005.mapped.ILLUMINA.bwa.CEU.low_coverage.20120522.bam | 45.6 |
| NA18548.mapped.ILLUMINA.bwa.CHB.low_coverage.20130415.bam | 23.3 |
| NA18747.mapped.ILLUMINA.bwa.CHB.low_coverage.20130415.bam | 29.9 |
| NA18881.mapped.ILLUMINA.bwa.YRI.low_coverage.20130415.bam | 19.7 |
| NA19005.mapped.ILLUMINA.bwa.JPT.low_coverage.20120522.bam | 27.5 |
| NA19735.mapped.ILLUMINA.bwa.MXL.low_coverage.20130415.bam | 22.5 |
| NA20509.mapped.ILLUMINA.bwa.TSI.low_coverage.20130415.bam | 23.0 |
| NA20773.mapped.ILLUMINA.bwa.TSI.low_coverage.20130415.bam | 18.0 |
| NA20895.mapped.ILLUMINA.bwa.GIH.low_coverage.20120522.bam | 24.6 |
| NA21115.mapped.ILLUMINA.bwa.GIH.low_coverage.20130415.bam | 18.0 |
| **Total** | **4071.6** |

## 2.3 FASTQ format benchmarks

In this case the data set for each test we performed consists of a single FASTQ file containing reads generated by a specific sequencing platform (we ran tests on Illumina, Ion Torrent and SOLiD reads). The test files were downloaded in `gzip` compressed format from the Short Read Archive (SRA) public `ftp` repository. In more detail, the Illumina and SOLiD data sets were downloaded from

```
http://ftp.sra.ebi.ac.uk/vol1/fastq/
```

while Ion Torrent reads were obtained from:

```
ftp://ftp-trace.ncbi.nlm.nih.gov/sra/sra-instant/reads/ByRun/sra/
```

Information about name, size, originating organism, sequencing protocol and read lengths is presented for each test file in the following table.

Table 2.4: FASTQ files used in the benchmark

| File name | Format | Organism | Strategy | Size (MB) | Seq. length |
|---|---|---|---|---|---|
| SRR608906_2.fastq | Illumina | A. mexicanus | WGS | 12151 | 100 |
| ERR039503.fastq | Ion Torrent | H. sapiens | WGS | 5962 | 5 - 2716 |
| SRR445256.fastq | SOLiD | B. anthracis | RNA-Seq | 4696 | 51 |

# APPENDIX B - TOOLS INVOCATION

## 3.1 SAM format benchmarks

### 3.1.1 Reference compressors

Whenever the compressor supports multi-threading, it was tested using 8 processing threads.

**GZIP**

- *GZIP-FAST*

    compress:
    ```
    pigz --fast --processes 8 --stdout IN.sam > OUT.gz
    ```
    decompress:
    ```
    pigz -d --processes 8 --stdout IN.gz > OUT.sam
    ```

- *GZIP-BEST*

    compress:
    ```
    pigz --best --processes 8 --stdout IN.sam > OUT.gz
    ```
    decompress:
    ```
    pigz -d --processes 8 --stdout IN.gz > OUT.sam
    ```

**BZIP2**

- *BZIP2-FAST*

    compress:
    ```
    pbzip2 -1 -p8 --stdout IN.sam > OUT.gz
    ```
    decompress:
    ```
    pbzip2 -d --p8 --stdout IN.gz > OUT.sam
    ```

- *BZIP2-BEST*

    compress:

```
pbzip2 -9 -p8 --stdout IN.sam > OUT.gz
```

decompress:

```
pbzip2 -d -p8 --stdout IN.gz > OUT.sam
```

**SAMTOOLS**

- *SAMTOOLS-BAM*

    compress:

    ```
    samtools view -b -@ 8 IN.sam > OUT.bam
    ```

    decompress:

    ```
    samtools view -h -@ 8 IN.bam > OUT.sam
    ```

- *SAMTOOLS-CRAM*

    compress:

    ```
    samtools view -C -T REF.fasta -@ 8 IN.sam > OUT.cram
    ```

    decompress:

    ```
    samtools view -h -T REF.fasta -@ 8 IN.sam > OUT.cram
    ```

**SCRAMBLE**

- *SCRAMBLE-BAM*

    compress:

    ```
    scramble -I sam -O bam -m -t 8 IN.sam > OUT.bam
    ```

    decompress:

    ```
    scramble -I bam -O sam -m -t 8 IN.bam > OUT.sam
    ```

- *SCRAMBLE-CRAM*

    compress:

    ```
    scramble -I sam -O cram -m -r REF.fasta -t 8 IN.sam > OUT.cram
    ```

    decompress:

    ```
    scramble -I cram -O sam -m -r REF.fasta -t 8 IN.cram > OUT.sam
    ```

- *SCRAMBLE-CRAM-Q8*

    compress:

    ```
    scramble -I sam -O cram -m -r REF.fasta -B -t 8 IN.sam > OUT.cram
    ```

    decompress:

    ```
    scramble -I cram -O sam -m -r REF.fasta -t 8 IN.cram > OUT.sam
    ```

**DEEZ**

- *DEEZ-NORMAL*

    compress:

    ```
    deez -t 8 -r REF.fasta IN.sam -c > OUT.dz
    ```

    decompress:

    ```
    deez -t 8 -r REF.fasta IN.dz -c > OUT.sam
    ```

- *DEEZ-SAMCOMP*

    compress:

    ```
    deez -t 8 -r REF.fasta -q1 IN.sam -c > OUT.dz
    ```

    decompress:

    ```
    deez -t 8 -r REF.fasta IN.dz -c > OUT.sam
    ```

- *DEEZ-Q8*

    compress:

    ```
    deez -t 8 -r REF.fasta -l30 IN.sam -c > OUT.dz
    ```

    decompress:

    ```
    deez -t 8 -r REF.fasta IN.dz -c > OUT.sam
    ```

### 3.1.2 *CARGO* container configuration

Before each test, we created a temporary container using 2048 ( `32 * 64` ) large blocks of `8` MiB each and 1024 ( `16 * 64` ) small blocks of `512` KiB each, containing enough empty space to store compressed data up to a size of

```
2048 * 8 MiB + 1024 * 512 KiB = 16 GiB (17180393472 B)
```

where

```
1 KiB = 1024 B
1 MiB = 1024 KiB = 1024 * 1024 B
...
```

The following command line was used in order to create such a container:

```
cargo_tool --create-container --container-file=CONTAINER \
    --large-block-size=8 --large-block-count=32 \
    --small-block-size=512 --small-block-count=16
```

The space available in this container was sufficient for all the tested SAM files. More information about the container architecture concept, the command-line description for `cargo_tool` and examples of use can be found in the **Supplementary Documentation**.

### 3.1.3 *CARGO* methods

Each *CARGO* method was tested using 8 processing threads and an input buffer size of 64 MiB.

### CARGO-SAM-STD

- *CARGO-SAM-STD*:

    compress:

    ```
    cargo_samrecord_toolkit-std c -c CONTAINER -n DATASET -i IN.sam -t 8 -b 64
    ```

    decompress:

    ```
    cargo_samrecord_toolkit-std d -c CONTAINER -n DATASET -o OUT.sam -t 8
    ```

- *CARGO-SAM-STD-Q8*

    compress:

    ```
    cargo_samrecord_toolkit-std c -c CONTAINER -n DATASET -i IN.sam -a -t 8 -b 64
    ```

    decompress:

    ```
    cargo_samrecord_toolkit-std d -c CONTAINER -n DATASET -o OUT.sam -t 8
    ```

### CARGO-SAM-EXT

- *CARGO-SAM-EXT*

    compress:

    ```
    cargo_samrecord_toolkit-ext c -c CONTAINER -n DATASET -t 8 -b 64 \
        -i IN.sam
    ```

    decompress:

    ```
    cargo_samrecord_toolkit-ext d -c CONTAINER -n DATASET -t 8 \
        -o OUT.sam
    ```

- *CARGO-SAM-EXT-Q8*

    compress:

    ```
    cargo_samrecord_toolkit-ext c -c CONTAINER -n DATASET -a -t 8 -b 64 \
        -i IN.sam
    ```

    decompress:

    ```
    cargo_samrecord_toolkit-ext d -c CONTAINER -n DATASET -t 8 \
        -o OUT.sam
    ```

### CARGO-SAM-REF

Before running the actual tests, we turned the *GRCh37* build of the reference genome for *H.sapiens* in FASTA format (see: **Appendix A**) into a compressed and searchable file (see: **CARGO documentation**). This binary representation, that we named *BFF (Binary Fasta File)* allows for fast subsequent sequence queries into the reference. The BFF file was created by running

```
cargo_samrecord_toolkit-ref r -i REF.fasta -o REF.bff
```

Having generated the index for the reference file, the following command lines were used to perform the tests:

- *CARGO-SAM-REF*

compress:

```
cargo_samrecord_toolkit-ref c -c CONTAINER -n DATASET -t 8 -b 64 -a \
    -i IN.sam
```

decompress:

```
cargo_samrecord_toolkit-ref d -c CONTAINER -n DATASET -t 8 -a \
    -f REF.bff -o OUT.sam
```

- *CARGO-SAM-REF-Q8*

    compress:

```
cargo_samrecord_toolkit-ref-q8 c -c CONTAINER -n DATASET -t 8 -b 64 -a \
    -i IN.sam
```

    decompress:

```
cargo_samrecord_toolkit-ref-q8 d -c CONTAINER -n DATASET -t 8 -a \
    -f REF.bff -o OUT.sam
```

- *CARGO-SAM-REF-Q8-MAX*

    compress:

```
cargo_samrecord_toolkit-ref-q8-max c -c CONTAINER -n DATASET -t 8 -b 64 -a \
    -i IN.sam
```

    decompress:

```
cargo_samrecord_toolkit-ref-q8-max d -c CONTAINER -n DATASET -t 8 -a \
    -f REF.bff -o OUT.sam
```

## 3.2 Queryable large-scale SAM format benchmarks

### 3.2.1 Volumes preparation

To prepare the data for both the *small* and the *large volume*, we used *SAMtools* and *sCRAMble* (with multi-threaded decompression support) and the *CARGO-SAM-REF-Q8* method, as explained in *Queryable large-scale SAM format benchmarks*.

In the first step, we merged the (already sorted) BAM files into a single large sorted BAM file:

```
samtools merge -@ 8 SAM_VOLUME.bam IN1.bam IN2.bam ...
```

Having the `SAM_VOLUME.bam` generated, in the next step we set to apply the Illumina Q-scores reduction scheme, saving the output as `SAM_VOLUME_Q8.bam`:

```
scramble -I bam -O bam -t 8 -B SAM_VOLUME.bam SAM_VOLUME_Q8.bam
```

In parallel, we performed a BAM-to-CRAM format conversion with applied Illumina Q-scores reduction:

```
scramble -I bam -O cram -r REF.fa -t 8 -B SAM_VOLUME.bam SAM_VOLUME_Q8.cram
```

Finally, we compressed the sorted `SAM_VOLUME.bam` into a dataset named `SAM`, and stored it into the *CARGO* container `CARGO_VOLUME`:

```
scramble -I bam -O sam -t 6 | cargo_samrecord_toolkit-ref-q8-max c -c CARGO_VOLUME \
    -n SAM -t 6 -b 64 -a -g -z
```

**CARGO Supplementary, Release 0.7rc-internal**

Having the volumes generated in BAM and CRAM formats, we indexed them in order to be able to perform subsequent range queries with *SAMtools*:

```
samtools index SAM_VOLUME.bam
samtools index SAM_VOLUME_Q8.bam
samtools index SAM_VOLUME_Q8.cram
```

In the case of *CARGO* methods, the generated record blocks inside the container were transparently indexed during compression thanks to option `-g`.

### *CARGO* containers configuration

Before running the tests, we created containers able to accommodate the compressed data produced by the selected *CARGO-SAM-REF-Q8* compressor. In order to do so, we decided to set the container size to the 10% of the uncompressed input data. Hence we created an empty container of ~86 GB for the *small volume* test, and an empty container of 1.72 TB for the *large volume* test.

#### Small volume

We created a temporary container using 10240 (`160 * 64`) large blocks of `8` MiB each and 1024 (`16 * 64`) small blocks of `512` KiB each, containing enough empty space to store compressed data up to a size of

```
10240 * 8 MiB + 1024 * 512 KiB = 78 GiB (86436216832 B)
```

The following command line was used in order to create such a container:

```
cargo_tool --create-container --container-file=CARGO_VOLUME \
    --large-block-size=8 --large-block-count=150 \
    --small-block-size=512 --small-block-count=16
```

#### Large volume

We created a temporary container using 204800 (`3200 * 64`) large blocks of `8` MiB each and 1024 (`16 * 64`) small blocks of `512` KiB each, containing enough empty space to store compressed data up to a size of

```
204800 * 8 MiB + 1024 * 512 KiB = 1.56 TiB (1718523789312 B)
```

The following command line was used in order to create such a container:

```
cargo_tool --create-container --container-file=CARGO_VOLUME \
    --large-block-size=8 --large-block-count=3000 \
    --small-block-size=512 --small-block-count=16
```

More information about the container architecture concept, the command-line description for `cargo_tool` and examples of use can be found in the **Supplementary Documentation**.

### 3.2.2 Optional volumes shrinkage

In order to optimize space usage, a container can be optionally shrank to fit its size to the size of the contained data. This operation removes free and unoccupied blocks and pads the number of the available blocks to the nearest multiple of *64* (for more information see the **Supplementary Documentation**). To shrink the volume one should say

```
cargo_tool --shrink-container --container-file=CARGO_VOLUME
```

In our test cases the *small volume* was shrunk to 73.1 GB, corresponding to the following blocks configuration:

```
8704 * 8 MiB + 192 * 512 KiB = 68.1 GiB (73115107328 B)
```

whereas the *large volume* was shrunk to 1.44 TB, corresponding to the following blocks configuration:

```
171200 * 8 MiB + 192 * 512 KiB = 1.31 GiB (1436230352896 B)
```

Apart from a tiny overhead, those sizes essentially coincide with the corrsponding total sizes of the underlying compressed streams (72.9 GB and 1.44 TB, respectively).

Of note, `cargo_tool` provides a command to display the information about the container size configuration including how many blocks were allocated:

```
cargo_tool --print-blocks --container-file=CARGO_VOLUME
```

### 3.2.3 Querying

- BAM:

```
samtools view SAM_VOLUME_Q8.bam KEY > OUT.sam
```

- CRAM:

```
samtools view SAM_VOLUME_Q8.cram -T REF.fa KEY > OUT.sam
```

- CARGO:

```
cargo_samrecord_toolkit-ref-q8 e -c CARGO_VOLUME -n SAM \
    -f REF.bff -t TH -k KEY -a -g -z > OUT.sam
```

where `TH` specifies the number of decompressing threads used when querying for data and `KEY` specifies the query range, as follows:

- BAM / CRAM:

```
chrom:pos_begin-pos_end
```

- CARGO:

```
chrom:pos_begin::chrom:pos_end
```

For example, to extract from `SAM_VOLUME.bam` all the SAM records belonging to chromosome 2 in the range from position 20,100,000 to position 20,200,000 with *SAMtools* one should say

```
samtools view SAM_VOLUME.bam 2:20,100,000-20,200,000 > OUT.sam
```

while the equivalent *CARGO* command (extracting from the `SAM` dataset contained in `CARGO_VOLUME` and using 2 extraction threads) would be

```
cargo_samrecord_toolkit-ref-q8 e -c CARGO_VOLUME -n SAM \
    -f REF.bff -t 2 -k 2:20100000::2:20200000 -a -g -z > OUT.sam
```

## 3.3 FASTQ format benchmarks

### 3.3.1 Reference compressors

**GZIP**

- *GZIP-FAST*

    compress:
```
pigz --fast --processes 8 --stdout IN.fastq > OUT.gz
```
    decompress:
```
pigz -d --processes 8 --stdout IN.gz > OUT.fastq
```

- *GZIP-BEST*

    compress:
```
pigz --best --processes 8 --stdout IN.fastq > OUT.gz
```
    decompress:
```
pigz -d --processes 8 --stdout IN.gz > OUT.fastq
```

**BZIP2**

- *BZIP2-FAST*

    compress:
```
pbzip2 -1 -p8 --stdout IN.fastq > OUT.gz
```
    decompress:
```
pbzip2 -d --p8 --stdout IN.gz > OUT.fastq
```

- *BZIP2-BEST*

    compress:
```
pbzip2 -9 -p8 --stdout IN.fastq > OUT.gz
```
    decompress:
```
pbzip2 -d -p8 --stdout IN.gz > OUT.fastq
```

**DSRC**

- *DSRC-FAST*

    compress:
```
dsrc c -m0 -t 8 IN.fastq OUT.dsrc
```
    decompress:

```
dsrc d -t 8 IN.dsrc OUT.fastq
```

- *DSRC-MED*

    compress:

```
dsrc c -m1 -t 8 IN.fastq OUT.dsrc
```

    decompress:

```
dsrc d -t 8 IN.dsrc OUT.fastq
```

- *DSRC-BEST*

    compress:

```
dsrc c -m2 -t 8 IN.fastq OUT.dsrc
```

    decompress:

```
dsrc d -t 8 IN.dsrc OUT.fastq
```

### FQZCOMP

- *FQZCOMP-FAST*

    compress:

```
fqz_comp c -n1 -s1 -q1 IN.fastq OUT.fqz
```

    decompress:

```
fqz_comp -d IN.fqz OUT.fastq
```

- *FQZCOMP-BEST*

    compress:

```
fqz_comp -n2 -q3 -s8+ -b IN.fastq OUT.fqz
```

    decompress:

```
fqz_comp -d IN.fqz OUT.fastq
```

### QUIP

- *QUIP-FAST*

    compress:

```
quip -c IN.fastq > OUT.qp
```

    decompress:

```
quip -d -c IN.qp OUT.fastq
```

- *QUIP-BEST*

    compress:

```
        quip -c IN.fastq OUT.qp
```

decompress:

```
        quip -d -c IN.qp OUT.fastq
```

### 3.3.2 *CARGO* methods

**Container configuration**

Before each test, we created a temporary container using 1024 (`16 * 64`) large blocks of `8` MiB each and 2048 (`32 * 64`) small blocks of `256` KiB each, containing enough empty space to store compressed data up to a size of

```
1024 * 8 MiB + 2048 * 256 KiB = 8.0 GiB (8590458880 B)
```

The following command line was used in order to create such a container:

```
cargo_tool --create-container --container-file=CONTAINER \
    --large-block-size=8 --large-block-count=8 \
    --small-block-size=256 --small-block-count=32
```

The space available in this container was sufficient for all the tested FASTQ files. More information about the container architecture concept, the command-line description for `cargo_tool` and examples of use can be found in the **Supplementary Documentation**.

**Running *CARGO***

In order to compress a FASTQ file into a *CARGO* container one should say

```
cargo_fastqrecord_toolkit-* c -c CONTAINER -n DATASET -t 8 \
    -b BUFFER_SIZE -i IN.fastq
```

where `BUFFER_SIZE` specifies the size of the input block buffer and `DATASET` the dataset name under which the FASTQ data will be saved. In addition, in our tests we used an `8` MiB buffer for all the *-FAST* methods and a `64` MiB buffer for all the *-BEST* methods.

One can decompress a FASTQ file from a *CARGO* container with the command

```
cargo_fastqrecord_toolkit-* d -c CONTAINER -n DATASET -t 8 \
    -o OUT.fastq
```

where `cargo_fastqrecord_toolkit-*` is any FASTQ compressor/decompressor corresponding to one of the implemented *CARGO* methods.

# CARGO Documentation

*Release 0.7rc-internal*

Lukasz Roguski, Paolo Ribeca

May 27, 2015

CONTENTS

# CHAPTER
# ONE

# INTRODUCTION

## 1.1 What is *CARGO*?

*CARGO – Compressed ARchiving for GenOmics* – is a set of tools and a library providing building blocks for the creation of applications to store, compress and manipulate large-scale genomic data. The main goal of *CARGO* is to supply universal and format-independent storage methods, whereby the record data type can be easily described by the user in terms of a special meta-language, high-performance compressing/decompressing tools can be easily generated from the record data type with little effort, and the tools thus produced can be used in order to store compressed genomic datasets in big containers.

## 1.2 Main features

The main features of *CARGO* are:

- Efficient storage of genomics data in compressed form
- Data aggregated into configurable containers of giga– and terabytes in size, which can hold multiple datasets having different formats
- Record format defined by the user in special meta-language allowing to describe any file format used in genomic applications. For some of those (at the moment FASTQ and SAM) support is provided out-of-the-box (see *Quickstart* and *Examples*)
- Automatic high-performance and multi-threaded processing of the data
- Possibility of implementing range searches on the top of an arbitrary order defined by the user
- Data parsing and transformation methods explicitly specified by user
- Multiple compression methods to be selected by the user depending on the characteristics of the input data.

## 1.3 General workflow

To create from scratch a simple compressor for a specified genomic file format the user only needs to:

1. Define the record data type in the high-level *CARGO* meta-language
2. Translate the record definition into a set of C++ files with the *CARGO* tools
3. Write a simple record parser in C++ using the record data type automatically generated during the previous step
4. Compile the automatically generated application template using the automatically generated *Makefile*
5. Create a container or use the existing one having enough available free space

6. Store / retrieve the data.

Once created, a compressor for a specific file format can be reused multiple times; similarly, a container can be used to store multiple genomic files possibly having different formats.

## 1.4 Contact and support

Official web site: http://algorithms.cnag.cat/cargo/

Authors:

- Lukasz Roguski : `lucas [dot] roguski [at] gmail [dot] com`
- Paolo Ribeca : `paolo [dot] ribeca [at] gmail [dot] com`

# CHAPTER
# TWO

# QUICKSTART

## 2.1 Download

The *CARGO* tools, the *CARGO* C++ library, the examples and the pre-compiled compressors for several popular genomic formats can be downloaded directly from the official web site.

## 2.2 Project structure

The *CARGO* project main directory is structured in the following way:

```
cargo
*-- contrib
|   *-- lib
*-- examples
|   *-- bin
|   *-- fastq
|   |   *-- fastq-simple
|   |   *-- fastq-multi
|   *-- sam
|       *-- common
|       *-- sam-ref
|       *-- sam-std
*-- include
|   *-- cargo
|       *-- core
|       *-- type
*-- lib
*-- tools
```

Those directories contain:

- `include`: the *CARGO* C++ library headers necesary for building the applications
- `lib`: precompiled *libcargo.a* library, which is used when linking compiled *CARGO* applications
- `contrib`: third-party source codes and libraries, which include: PPMd and LZMA compression libraries
- `tools`: precompiled *CARGO* tools:
    - `cargo_translate`: *CARGO* metalanguage-to-C++ type translator
    - `cargo_tool`: container management tool
- `examples`: examples of compressor applications for several popular genomics formats:
    - `fast-simple`: a simple *FASTQ* format lossless compressor

- `fast-multi`: a family of simple *FASTQ* format lossless compressors
- `sam-std`: a simple SAM format lossless and lossy compressor
- `sam-ext`: an advanced SAM format lossless and lossy compressor
- `sam-ref`: a more advanced SAM format compressor providing: reference-based sequence compression, optimized reads alignment representation, several transformations of numerical fields, and lossy Illumina Q-scores reduction scheme.

## 2.3 Simple FASTQ format compressor

**Tip:** Application source files for this tutorial are available in the directory `cargo/examples/fastq/fastq-simple` of the official *CARGO* distribution. Should one prefer to skip introduction, record data type translation, parser coding and application building steps, a pre-compiled ready-to-use *FASTQ* format compressor is available in the directory `cargo/examples/bin`. The sample FASTQ file compression use-case is presented starting from *Creating container* subchapter below.

### 2.3.1 Prerequisites

Before compiling the application, the path to *CARGO* distribution directory needs to be set in the build environment in order to access *CARGO* C++ header files and libraries:

```
export CARGO_PATH=/path/to/cargo/directory/
```

As *CARGO* relies on several publicly-available compression libraries, the zlib (`libz`) and bzip2 (`libbz2`) libraries need to be present in the system for linking.

Compiling *CARGO* applications will also require a compiler with *C++11* standard support (for multi–threading support) – by default the *gcc* compiler version *4.8* or above should be used.

### 2.3.2 FASTQ format

The FASTQ format is an ASCII text-based format useful to store biological sequences together with their quality score values. A sample record looks like the following:

```
1  @SRR001666.1 071112_SLXA-EAS1_s_7:5:1:817:345 length=36
2  GGGTGATGGCCGCTGCCGATGGCGTCAAATCCCACC
3  +
4  IIIIIIIIIIIIIIIIIIIIIIIIIIIIII9IG9IC
```

while a general record contains:

- *read id* – an identifier of the read starting after the `@` symbol,
- *sequence* – a sequence of nucleotides encoded using `AGCTN` letters,
- *plus* – a control line, optionally containing a repetition of the read identifier,
- *quality* – a Phred sequencing quality score of the sequence.

### 2.3.3 Record type definition

In general, a *FASTQ* record can be seen as a triplet, consisting of 3 fields: *read id*, *sequence* and *read quality* (discarding the redundant information in the *plus* field). The *read id* field is usually a collection of tokens separated by a set of delimiters incl. `-_ ,.;:/#`; *sequence* field is – in majority of the cases - a list of nucleotide bases `AGCTN` and the *quality* is a list of Phred numeric values encoded as ASCII characters. However, for a simplicity of this example, all the record fields will be represented as a `string` type. Such record definition in *CARGO* metalanguage is as follows:

```
1  FastqRecord = {
2    tag = string;
3    seq = string;
4    qua = string
5  }
6
7  @record FastqRecord
```

**Note:** More details about the *CARGO* metalanguage syntax with available data types can be found in *The CARGO meta-language* chapter.

### 2.3.4 Translating

Having the FASTQ record definition saved in `FastqRecord.cargo` file, next step is to translate the definition from *CARGO* metalanguage to C++ code by running the `cargo_translate` tool available in the `cargo/tools` subdirectory:

```
cargo_translate -i FastqRecord.cargo
```

After the translation, a set of files will be generated:

- `FastqRecord.h`: C++ definition of the `FastqRecord` user record type
- `FastqRecord_Parser.h`: C++ parser template for the `FastqRecord` C++ user record - will need to be completed by the user
- `FastqRecord_Type.h`: C++ *TypeAPI*-based record type specification for the `FastqRecord` record type (for subsequent internal use, it does not need to be opened or modified by the user)
- `FastqRecord_main.cpp`: template file containing compressor/decompressor applications writing/reading a stream of `FastqRecord` records to/from containers
- `FastqRecord_Makefile.mk`: *Makefile* template file to build such applications.

The translated C++ record definition, which will be used later when implementing parsing methods, is:

```cpp
1  struct FastqRecord {
2    std::string tag;
3    std::string seq;
4    std::string qua;
5  };
```

**Note:** More details regarding the translated *FASTQ* record definition into C++ code can be found in *FASTQ example* subchapter.

### 2.3.5 Writing a FASTQ records parser

In the next step, the missing record parsing functions will be implemented – the C++ functions are in class `FastqRecord_Parser` (`FastqRecord_Parser.h`) and include:

- `void SkipToEndOfRecord(io::MemoryStream& )` – skips the characters in the memory stream until the end of the current record (if any),
- `void SkipToEndOfHeader(io::MemoryStream& )` – skips the characters in the memory stream until the end of the file header (none in case of FASTQ),
- `bool ReadNextRecord(io::MemoryStream& stream_, FastqRecord& record_)` – reads the next record from the memory stream and fills the `FastqRecord` structure member fields with the parsed data; returns `true` on success,
- `bool WriteNextRecord(io::MemoryStream& stream_, FastqRecord& record_)` – parses the `FastqRecord` structure member fields into a textual *FASTQ* format to, saving the output to the memory stream.

The complete code snippet for `FastqRecord_Parser` class is as follows:

```cpp
using namespace type;

class FastqRecord_Parser
{
public:
    // no header in the FASTQ file format – member function left empty
    static void SkipToEndOfHeader(io::MemoryStream& stream_) { }

    static void SkipToEndOfRecord(io::MemoryStream& stream_)
    {
        // skip to the new line
        FieldParser::SkipNextField(stream_, '\n');

        byte nextByte = 0;
        while (FieldParser::PeekNextByte(stream_, nextByte))
        {
            // beginning of next record – found read id line
            if (nextByte == '@')
                return;
            FieldParser::SkipNextField(stream_, '\n');
        }
    }

    static bool ReadNextRecord(io::MemoryStream& stream_, FastqRecord& record_)
    {
        bool result = FieldParser::ReadNextField(stream_, record_.tag, '\n');
        result &= FieldParser::ReadNextField(stream_, record_.seq, '\n');
        FieldParser::SkipNextField(stream_, '\n');              // skip plus line
        result &= FieldParser::ReadNextField(stream_, record_.qua, '\n');
        return result;
    }

    static bool WriteNextRecord(io::MemoryStream& stream_, FastqRecord& record_)
    {
        bool result = FieldParser::WriteNextField(stream_, record_.tag, '\n');
        result &= FieldParser::WriteNextField(stream_, record_.seq, '\n');
        result &= FieldParser::WriteNextField(stream_, "+", '\n');
        result &= FieldParser::WriteNextField(stream_, record_.qua, '\n');
        return result;
```

```
40        }
41 };
```

### 2.3.6 Building

**Important:** Before building the compressor the *Prerequisites* need to be met on the development machine.

To build the simple *FASTQ* compressor a generated makefile `FastqRecord_Makefile.mk` is available to be used with gnu `make`:

```
make -f FastqRecord_Makefile.mk
```

As a result, a `cargo_fastqrecord` executable will be created.

### 2.3.7 Creating container

As the sequencing data is stored in *CARGO* containers (independently of the records formats), an existing one needs to be used or a new one created using `cargo_tool` utility from `cargo/tools` directory. Creating a container `fastq_container` with a sample configuration of `1024` (16 multiplied by 64) large blocks of 4 MiB in size and `4069` (64 multiplied by 64) small blocks of `256` KiB in size in straightforward:

```
cargo_tool --container-file=fastq_container --create-container \
        --large-block-count=16 --large-block-size=4 \
        --small-block-count=64 --small-block-size=256
```

As a result, 3 files will be generated, which define a single container:

- `fastq_container.cargo-meta` – holds the container's meta information,
- `fastq_container.cargo-stream` – contains the data streams,
- `fastq_container.cargo-dataset` – holds the stored datasets information.

### 2.3.8 Running

To store (compress) `SRR001666.fastq` file in the `fastq_container` container under the dataset name `SRR001666` using the compiled `cargo_fastqrecord` compressor:

```
./cargo_fastqrecord c -c fastq_container -n SRR001666 -i SRR001666.fastq
```

To retrieve back (decompress) the `SRR001666` dataset from the container and save it as `SRR001666.decomp.fastq` file:

```
./cargo_fastqrecord d -c fastq_container -n SRR001666 -o SRR001666.decomp.fastq
```

# CHAPTER
# THREE

# *CARGO* CONTAINERS

In contrast to the standard file-based approach to compression, *CARGO* uses specially created containers to store, retrieve and query genomic data. This strategy allows to aggregate multiple files, each one having a possibly different file format, into a single container. This allows to store in a compact way sequencing data, analysis intermediates and final results coming from either a single experiment or multiple sequencing projects.

## 3.1 Architecture

The *CARGO* container consists of 3 different areas (or parts). Their conceptual structure is presented in figure *The conceptual structure of a CARGO container*. In order to achieve a better organization and simplify the implementation of backup methods, the container areas are stored on disk as separate files having the common prefix `*.cargo-` (being `*` some arbitrary name for the container specified by the user):

- `*.cargo-meta`: a file storing the container *meta-information* (i.e. the container's internal block configuration)
- `*.cargo-dataset`: a file storing information about the datasets present within the container, including their underlying structure in terms of data streams
- `*.cargo-stream`: a file storing the data streams, distributed into a big number of *large blocks* and *small blocks*.

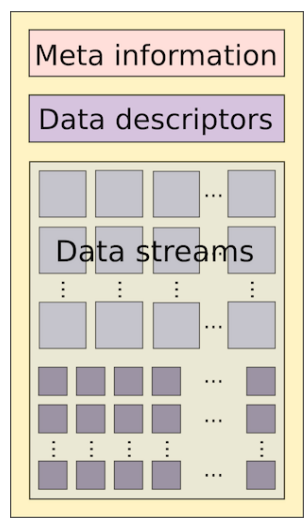

Fig. 3.1: The conceptual structure of a *CARGO* container

> **Warning:** All those three files define the container as a whole, so that a corruption of any may lead to the unrecoverable loss of stored data. In particular, the information contained in the `*.cargo-stream` file cannot be recovered without the other two files. This is why a backup functionality for the *meta-information* and *dataset* areas has been implemented in `cargo_tool` (for more details, see *Container tool*).

### 3.1.1 Meta-information area

The *meta-information area* (file: `*.cargo-meta`) contains information regarding the container blocks configuration and the block allocation table. The container blocks configuration is determined by the sizes and the number of *large blocks* and *small blocks*, which is defined at the moment of container creation. The block allocation table holds the information about the blocks occupancy state, which can be either *free*, *reserved* (being currently written to, but not finalized) or *occupied*. This area is crucial for the proper block allocation mechanisms while operating on the *streams area* of the container.

### 3.1.2 Dataset area

The *dataset area* (file: `*.cargo-dataset`) contains the description of the stored datasets inside the container. Each dataset description contains information about the dataset record type, it's underlying streams hierarchy and blocks it occupies, an optional file header information (if available) and selected data statistics. This area is crucial for 'understanding' the data stored inside the *streams area*, thus allowing for it's retrieval or removal.

### 3.1.3 Streams area

The *streams area* (file: `*.cargo-streams`) is the heart of the container – it holds the genomic data organized in streams determined by the records' type definition. Internally, the data inside each stream is stored as a collection of blocks – the list of the occupied blocks with the other stream information is held in the *dataset area*. The *streams area* is divided into the *large block area* and the *small block area*, which are defined by the number and the size of *large blocks* and *small blocks*. The sizes and the numbers of the blocks are configured at the time of the container creation (for more information, see *Container tool*).

## 3.2 Container tool

The *CARGO container tool* – `cargo_tool` – is a general utility to work with the containers. It provides functionality to create, shrink, remove, backup the containers and to display information about the stored data.

### 3.2.1 Options

When launched from the command line, `cargo_tool` displays the following message:

```
*** CARGO container utility ***
Version: 0.1.0
Date: 28.01.2015
Authors: Lukasz Roguski and Paolo Ribeca

Usage: cargo_tool [options]
options:
  --container-file=<c_file>   - container filename (required for all operations)

  --create-container          - creates a container file with the specified parameters
```

```
--large-block-size=<lb_s>    - large block size in MiB (n = [1 - 256], power of 2)
--large-block-count=<lb_n>   - large blocks count, which will be multiplied by 64
--small-block-size=<sb_s>    - small block size in KiB (n = [64 - 16384], power of 2)
--small-block-count=<sb_n>   - small blocks count, which will be multiplied by 64

--remove-container           - removes a container file
--shrink-container           - shrinks the container size to fit it's content

--snapshot-file=<s_file>     - file name of the snapshot
--create-snapshot            - creates a snapshot of the dataset and meta areas
--restore-snapshot           - restores the meta and dataset areas from the snapshot

--print-blocks               - prints the information about the container blocks
--print-dataset              - prints the information about the specified dataset
--remove-dataset             - removes the specified dataset
--dataset-name=<d_name>      - dataset name
```

where options specify:

- **--container-file=<c_file>**  the container prefix file name (the suffix `.cargo-*` will be added to the file)
- **--create-container**  an action indicator to create container with the specified blocks configuration
- **--large-block-size=<lb_s>**  the size of *large block* (in MiB), must be in range `[1 - 256]` and be the **power of 2**
- **--large-block-count=<lb_n>**  the number of *large blocks* - the actual number of blocks will be multiplied by **64**
- **--small-block-size=<sb_s>**  the size of *small block* (in KiB), must be in range `[64 - 16384]` and the **power of 2**
- **--small-block-count=<sb_n>**  the number of *small blocks* - the actual number of blocks will be multiplied by **64**
- **--remove-container**  an action indicator to remove specified container
- **--shrink-container**  an action indicator to shrink the size of the specified container
- **--snapshot-file=<s_file>**  a file name for the container snapshot of the *meta* and *dataset* areas
- **--create-snapshot**  an action indicator to create a snapshot and save it under the given file name
- **--restore-snapshot**  an action indicator to restore a snapshot from a given file name
- **--print-blocks**  an action indicator to print the blocks information from *meta-information area*
- **--dataset-name=<d_name>**  a name of the queried dataset
- **--print-dataset**  an action indicator to print the information about specified dataset in container
- **--remove-dataset**  an action indicator to remove the specified dataset from the container

**Note:** When creating a new container the actual size of the *streams area* storage is calculated as:

```
storage_size = lb_s * lb_n * 64 MiB + sb_s * sb_n * 64 KiB
```

where:

```
1 KiB = 1024 B
1 MiB = 1024 KiB = 1024 * 1024 B
...
```

For example, when specifying parameters:

- *large block size*: *8* MiB
- *large block count*: *32*
- *small block size*: *512* KiB
- *small block count*: *16*

the total size of the available storage area for the data will be:

```
storage_size = 8 * 32 * 64 MiB + 512 * 16 * 64 KiB
             = 2048 * 8 MiB + 1024 * 512 KiB
             = 16.5 GiB (17716740096 B)
```

with the total number of *2048* large blocks and *1024* small blocks.

---

**Note:** When shrinking the container (to adapt it's size to the size of the stored data), free and non-occupied blocks will be released and the total number of the used blocks will be set to the nearest multiple of **64**.

In the case of previous example, having reserved *2048* large blocks and *1024* small blocks, where only a significant fraction of them is occupied i.e. 500 large blocks and 100 small blocks, the *stream area* size after the container shrinkage is calculated as follows:

```
lb_n' = ((500 / 64 ) + 1) * 64 = 512
sb_n' = ((100 / 64) + 1) * 64 = 128

storage_size = 512 * 8 MiB + 128 * 512 KiB
             = 4.16 GiB (4362076160 B)
```

### 3.2.2 Usage examples

- Creating a container `data_container` with `256 (8 * 64)` *large blocks* of `8` MiB in size and `1024 (16 * 64)` *small blocks* of `256` KiB in size:

    ```
    cargo_tool --create-container --container-file=data_container \
            --large-block-size=8 --large-block-count=8 \
            --small-block-size=256 --small-block-count=16
    ```

- Print the `data_container` block information:

    ```
    cargo_tool --print-blocks --container-file=data_container
    ```

- Print the `HG00380` dataset information from the container `data_container`:

    ```
    cargo_tool --print-dataset --dataset-name=HG00380 --container-file=data_container
    ```

- Shrink the container `data_container` to fit it's content size:

    ```
    cargo_tool --shrink-container --container-file=data_container
    ```

- Create a current snapshot of the `data_container` *meta-information* and *dataset* areas and store it under the `data_container.snap-2015-01-31` file name:

    ```
    cargo_tool --create-snapshot --container-file=data_container \
            --snapshot-file=data_container.snap-2015-01-31
    ```

- Restore the `data_container` *meta-information* and *dataset* areas from the snapshot `data_container.snap-2015-01-31`:

```
cargo_tool --restore-snapshot --container-file=data_container \
        --snapshot-file=data_container.snap-2015-01-31
```

**Note:** When restoring snapshot of the `data_container`, an additional backup files of *meta-information* and *dataset* areas will be created:

```
data_container.cargo-meta.old-*timestamp*
data_container.cargo-dataset.old-*timestamp*
```

where the current time stamp of the operation will be added to the extension of the files. This additional feature is only for security reasons providing the rollback possibility of the snapshot restoring operation, as the `data_container` *meta-information* and *dataset* area files will be overwritten.

# CHAPTER
# FOUR

# THE *CARGO* META-LANGUAGE

With the aim to make genomic data compression prototyping accessible to a wider audience, *CARGO* introduces a flexible meta-language that can be used in order to define record data types – in the spirit of what happens with databases. Subsequently, by running commands from the *CARGO* toolchain one can automatically translate the record type to low-level C++ code, thus achieving both flexibility of implementation and high runtime performance.

In this section we introduce and explain the syntax of the *CARGO* meta-language in Backus-Naur form. The latter represents the formal definition of the language, and hence should be regarded as the authoritative reference to be consulted in case of doubt. However, in order to make the semantics of the language easier to grasp, the material presented in this section is slightly different from the actual implementation: some productions have been rearranged and some replaced with conceptually equivalent ones whenever implementation technical details could be confusing to the reader.

## 4.1 Typesetting conventions

Backus-Naur productions are typeset as follows:

```
example    ::=   | case_one
                 | case_two
                 | "@" case_one COMMA case_two
                 | [case_three]
```

where *example*, `case_one` and `case_two` are nonterminals, `COMMA` is a terminal, and `"@"` is a literal. In addition, constructs in square brackets like `[case_three]` are optional.

## 4.2 Basic meta-programming syntax

### 4.2.1 Meta-programs

Each *CARGO* meta-program file (*cargo_metaprogram*) is a (possibly empty) *sequence of actions* (*one_or_more_actions*):

```
cargo_metaprogram   ::=   | EOF
                          | one_or_more_actions EOF
```

```
one_or_more_actions  ::=   action [one_or_more_actions]
```

where each action (`action`) can be either a *preprocessing directive* (`directive`) or a *sequence of type expressions* (`one_or_more_typeexprs`):

```
action  ::=  | directive
             | one_or_more_typeexprs
```

### 4.2.2 Pre-processing directives

There are two pre-processing directives implemented so far, *inclusions* or *root type declarations*:

```
directive  ::=  | directive_include
                | directive_record
```

Inclusion directives (introduced by the keyword `@include`) allow textual inclusions of other meta-program source files into the current one:

```
directive_include  ::=   "@include" QUOTED_STRING
```

Root type declarations (introduced by the keyword `@record`) have the following form:

```
directive_record  ::=   "@record" UPPER
```

and instruct the parser to translate into C++ the type following the keyword `@record`. For instance, if processed with the `cargo_translate` tool (see *Translator tool*) provided by the current *CARGO* implementation the directive

```
@record Example
```

will emit the 5 following C++ files:

- `Example.h`: C++ definition of the `Example` user record type
- `Example_Parser.h`: C++ parser template for the `Example` C++ user record - will need to be completed by the user
- `Example_Type.h`: C++ *TypeAPI*-based record type specification for the `Example` record type (for subsequent internal use, it does not need to be opened or modified by the user)
- `Example_main.cpp`: template file containing compressor/decompressor applications writing/reading a stream of `Example` records to/from containers
- `Example_Makefile.mk`: *Makefile* template file to build such applications

(see *Examples* for actual examples of use). Other C++ files might be emitted depending on the command-line options given to `cargo_translate` (see *Translator tool*).

### 4.2.3 Type expressions

Consecutive type expressions are separated by semicolons, and can have an optional terminating semicolon:

```
one_or_more_typeexprs   ::=   | typeexpr [";"]
                              | typeexpr ";" one_or_more_typeexprs
```

Each type expression can be either a *type definition* (`typedef`) or a *type extension* (`typeext`).

```
typeexpr   ::=   | typedef [":" one_or_more_annotations]
                 | typeext [":" one_or_more_annotations]
```

Type definitions, which lead to the production of both a C++ type and an associated automatic type interface, can be either bare type declarations, or type declarations followed by modifiers (*annotations*, see `annotation`). Annotations always have the effect of generating a new automatic type interface, even if the type being annotated has been already defined. The generation of new C++ code for the type is not performed when a type is defined in terms of an already defined type. However, it can be forced by using the operator `:=` in lieu of `=` in type definitions (we also say that by doing that the type is being *extended*).

The name of the type being defined can only begin by an uppercase letter (it can only be an *uppercase* identifier):

```
typedef   ::=   UPPER "=" typedef_rhs
```

```
typeext   ::=   UPPER ":=" typedef_rhs
```

The following table summarizes the semantics of type definitions when a new type is derived from a previously defined one:

Table 4.12: Derived type semantics definitions

| Annotations | Operator used in the definition | | | |
|---|---|---|---|---|
| | = | | := | |
| | *New C++ code* | *New auto interface* | *New C++ code* | *New auto interface* |
| Yes | No | Yes | Yes | Yes |
| No | No | No | Yes | No |

### 4.2.4 Assignable types

There are four possible main ways of defining a type:

1. As a *record/product type* (`record_type`, similar in spirit to a C `struct`). Any record type must have two or more named fields.

2. As a *union/variant type* (`union_type`, similar in spirit to a C `union`). Any union type must have two or more named fields.

3. As a *basic type* (`basic_type`), that is a simple combination of predefined/already defined types.

4. As a *subtype* (`referenced_subtype`), i.e. as part of an already defined type.

That is:

```
typedef_rhs   ::=   | record_type
                    | union_type
                    | basic_type
```

```
                     | referenced_subtype
```

**Record and union types**

Essentially, both record and union types are named collections of *member types*, with records being delimited by braces `{ }`, and unions being delimited by square brackets `[ ]`. Both record and union types have two different forms:

1. A *full form*, where record member types are separated by a `&` symbol and union member types are separated by a `|` symbol, respectively. Optional `&` or `|` symbols can also be present at the beginning of the list of members

2. A *simple form*, where record member types are just listed one after the other.

In addition, a final semicolon can be optionally present after the list of member types.

As a result, all the following examples are valid equivalent record type definitions:

```
FullProduct = { a = string; & b = char; & c = enum['F','R'] }
FullProductWithLeadingAnd = {
                            & a = string;
                            & b = char;
                            & c = enum['F','R']
                            }
SimpleProduct = { a = string; b = char; c = enum['F','R'] }
FullProductWithTrailingSemicolon = { a = string; & b = char; & c = enum['F','R']; }
FullProductWithLeadingAndWithTrailingSemicolon = {
                                            & a = string;
                                            & b = char;
                                            & c = enum['F','R'];
                                            }
SimpleProductWithTrailingSemicolon = { a = string; b = char; c = enum['F','R']; }
```

and all the following examples are valid equivalent union type definitions:

```
FullVariant = [ a = string; | b = char; | c = enum['F','R'] ]
FullVariantWithLeadingOr = [
                           | a = string;
                           | b = char;
                           | c = enum['F','R']
                           ]
SimpleVariant = [ a = string; b = char; c = enum['F','R'] ]
FullVariantWithTrailingSemicolon = [ a = string; | b = char; | c = enum['F','R']; ]
FullVariantWithLeadingOrWithTrailingSemicolon = [
                                            | a = string;
                                            | b = char;
                                            | c = enum['F','R'];
                                            ]
SimpleVariantWithTrailingSemicolon = [ a = string; b = char; c = enum['F','R']; ]
```

In programmatic form:

```
 record_type    ::=  | "{" memberdef ";" memberdef_simple_tail "}"
                    | "{" ["&"] memberdef ";" "&" memberdef_product_tail "}"
```

```
 memberdef_product_tail   ::=  | memberdef [";"]
                              | memberdef ";" "&" memberdef_product_tail
```

```
union_type ::= | "[" memberdef ";" memberdef_simple_tail "]"
               | "[" ["|"] memberdef ";" "|" memberdef_variant_tail "]"

memberdef_variant_tail ::= | memberdef [";"]
                           | memberdef ";" "|" memberdef_variant_tail

memberdef_simple_tail ::= | memberdef [";"]
                          | memberdef ";" memberdef_simple_tail
```

Each member definition has the following general form:

```
memberdef ::=    LOWER "=" typedef_rhs
```

where the name of the member can only begin by a lowercase letter (a *lowercase identifier*).

#### Reference to a subtype

Finally, a type definition can be the name of a *subtype* (`subtype`, a subtree of an already defined type) surrounded by parentheses ():

```
referenced_subtype ::=    "(" subtype ")"
```

## 4.3 Basic types

Basic types are the core building blocks of the meta-language. They can be *arrays* (`array_type`), *enumerations* (`enum_type`), qualified or unqualified predefined types (`qual_predef_type` or `unqual_predef_type`, respectively), and references to user-defined types (`user_type`).

```
basic_type ::= | array_type
               | enum_type
               | qual_predef_type
               | unqual_predef_type
               | user_type
```

### 4.3.1 Arrays

There are two kinds of array declarations, *arrays with a known length* (`array_known_length`) and *arrays of unknown (variable) length* (`array_unknown_length`). A special case of arrays are *strings* which can also have either a defined or an undefined length (`string_known_length` or `string_unknown_length`, respectively):

```
array_type ::= | array_known_length
```

```
                       | array_unknown_length
                       | string_known_length
                       | string_unknown_length
```

**General arrays**

```
array_known_length   ::=   basic_type "array" "*" INTEGER
```

```
array_unknown_length ::=   basic_type "array"
```

For instance, the first line declares an array of signed integers of fixed length 4, while the second declares an array of signed integers of unknown (variable) length:

```
int array * 4
int array
```

**Strings**

By definition, the following type equation holds true:

```
string = char array
```

that is, a *string* is an array of characters. Hence the following aliases are provided as a notational shortcut:

```
string_known_length   ::=   "string" "*" INTEGER
```

```
string_unknown_length ::=   "string"
```

### 4.3.2 Enumerations

Enumerations can be *string enumerations* (`string_enum`), *integer enumerations* (`int_enum`) or *character enumerations* (`char_enum`):

```
enum_type   ::=   | string_enum
                  | int_enum
                  | char_enum
```

**String enumerations**

```
string_enum   ::=   "enum" "[" strings "]"
```

```
strings ::=  QUOTED_STRING [","  strings]
```

For instance:

```
enum["a","aa","aaa"]
```

is a string enumerated type whose instances can only be `"a"`, `"aa"` or `"aaa"`.

**Integer enumerations**

```
int_enum   ::=  "enum" "["  int_ranges  "]"
```

```
int_ranges ::=  int_range [","  int_ranges]
```

Ranges of consecutive integers can be written in compact form:

```
int_range  ::=  | INTEGER [":"  INTEGER]
```

and hence for instance the following type equation holds:

```
enum[-1,1,2,3] = enum[-1,1:3]
```

**Character enumerations**

```
char_enum   ::=  "enum" "["  char_ranges  "]"
```

```
char_ranges ::=  char_range [","  char_ranges]
```

Ranges of consecutive characters can be written in compact form:

```
char_range  ::=  QUOTED_CHAR [":"  QUOTED_CHAR]
```

and hence for instance the following type equation holds:

```
enum['a','b','c','z'] = enum['a':'c','z']
```

### 4.3.3 Qualified predefined types

They are characters, signed integers or unsigned integers with a defined bitness:

```
qual_predef_type  ::=  | "char" "^" INTEGER
                      | "int"  "^" INTEGER
                      | "uint" "^" INTEGER
```

```
                  | "bool"
```

At the moment values accepted for the bitness of a *char* are 7, 8 or 16 (the latter being not implemented yet), while *int*-s and *uint*-s can have 8, 16, 32 or 64 bits.

By definition, the following type equation holds true:

```
bool = uint ^ 1
```

that is, a boolean is an unsigned integer with 1 bit – the alias is provided as a notational shortcut.

### 4.3.4 Unqualified predefined types

They are characters, signed integers or unsigned integers without a defined bitness:

```
unqual_predef_type  ::=  | "char"
                         | "int"
                         | "uint"
```

In fact all those definitions are notational shortcuts, as the following type equations hold true:

```
char = char ^ 8
int  = int ^ 64
uint = uint ^ 64
```

i.e. a generic character is assumed to have 8 bits, while a generic integer is assumed to have 64 bits.

### 4.3.5 User-defined types

And finally, a type can also be defined in terms of an already defined type (which must be an identifier starting with an uppercase character, `UPPER`):

```
user_type  ::=  UPPER
```

## 4.4 Subtypes

Subtypes are parts (or more precisely, subtrees) of alredy defined types. Subtypes can be the *empty subtype* (`empty_subtype`), in which case the parser will take as subtype the type defined last; a *subtype previously defined by the user* (`user_subtype`); the *base type of an array type* (`vector_element_subtype`); and the *type of the field of a record type* (`compound_subfield_subtype`):

```
subtype  ::=  | empty_subtype
              | user_subtype
              | vector_element_subtype
              | compound_subfield_subtype
```

```
empty_subtype  ::=
```

```
user_subtype     ::=    UPPER
```

```
vector_element_subtype    ::=    subtype "[" "]"
```

```
compound_subfield_subtype    ::=    subtype "." LOWER
```

For instance, the following type equations will hold true:

```
string [] = char
(bool array) [] = bool
```

and, if

```
SimpleProduct = { a = string; b = char; c = enum['F','R'] };
```

then

```
SimpleProduct.b = char
```

The empty subtype comes handy when annotating a type that has just been defined, as in the following code:

```
SimpleProduct = { a = string; b = char; c = enum['F','R'] } : .a.Block = 8M;
```

(see *Annotations*).

## 4.5 Annotations

Annotations allow the user to provide the *CARGO* framework with more information about one or more (sub)members of record types. A typical example are directives to state that a particular field should be compressed by using some specific compression method. Such information is subsequently gathered and used by the backend in order to generate more efficient C++ code.

Annotations can be one or more:

```
one_or_more_annotations    ::=    annotation ["," one_or_more_annotations]
```

and they can refer to the *block type* (`block_annotation`), the *compression type* (`compression_annotation`) or the *sorting field* (`sorting_field`):

```
annotation    ::=    | block_annotation
                     | compression_annotation
                     | sorting_field
```

### 4.5.1 Block types

The size of the block used when compressing (the content of) a given subtype can be specified by assigning a value to the virtual field `Block` of the subtype:

```
block_annotation  ::=   subtype "."  "Block" = block_type
```

Several block sizes are currently understood by the parser:

```
block_type  ::=  | "64K"
                 | "128K"
                 | "256K"
                 | "512K"
                 | "1M"
                 | "2M"
                 | "4M"
                 | "8M"
                 | "16M"
                 | "32M"
                 | "64M"
                 | "Integer"
                 | "Default"
                 | "Text"
```

where `Integer` and `Text` are predefined values suitable for the compression of the corresponding types (see *The Type API*). `Default` can be used when unsure (see *The Type API*).

### 4.5.2 Compression types

The method used to compress (the content of) a given subtype can be specified by assigning a value to the virtual field `Pack` of the subtype:

```
compression_annotation  ::=   subtype "."  "Pack" = pack_type
```

Several compression methods are currently understood by the parser:

```
pack_type  ::=  | "None"
                | "Integer"
                | "Text"
                | "GzipL1"
                | "GzipL2"
                | "GzipL3"
                | "GzipL4"
                | "Gzip"
                | "Bzip2L1"
                | "Bzip2L2"
                | "Bzip2L3"
                | "Bzip2L4"
                | "Bzip2"
                | "PPMdL1"
                | "PPMdL2"
                | "PPMdL3"
                | "PPMdL4"
```

```
                 | "PPMd"
                 | "LZMAL1"
                 | "LZMAL2"
                 | "LZMAL3"
                 | "LZMAL4"
                 | "LZMA"
```

where `Integer` and `Text` are predefined values suitable for the compression of the corresponding types (see *The Type API*). `None` can be used to turn off compression. For a precise definition of all other methods in terms of their corresponding algorithms, see *The Type API*.

### 4.5.3 Sorting field

The user can optionally annotate (the content of) a subtype as the value by which records should be sorted by assigning a `True` value to the virtual field `Key` of the subtype:

```
sorting_field  ::=    subtype  "."   "Key"  =  "True"
```

The annotation can be used to generate additional C++ code (see *Translator tool*).

## 4.6 Lexical analysis

Prior to parsing, comments and whitespace are removed from the text. In more detail, as in C90 or C++ comments can be either single-line

```
SINGLE_LINE_COMMENT  ::=   "//"  [^"\n"]*
```

or multi-line:

```
MULTI_LINE_COMMENT  ::=   "/*"  ...  "*/"
```

but, different from C/C++, multi-line comments can be nested:

```
/*
  /* I am a valid nested multi-line
     CARGO comment */
*/
```

Tokens can be separated by any amount of whitespace, defined as

```
WHITESPACE  ::=   ["\t" " " "\n"]+
```

And finally, terminals are defined as follows:

```
QUOTED_STRING  ::=   "\""  ...  "\""
```

```
QUOTED_CHAR  ::=   "'" [] "'"

INTEGER  ::=   ["+" "-"]*["0"-"9"]+

UPPER  ::=   ["A"-"Z"]["0"-"9" "A"-"Z" "_" "a"-"z"]*

LOWER  ::=   ["a"-"z"]["0"-"9" "A"-"Z" "_" "a"-"z"]*
```

## 4.7 Translator tool

The `cargo_translate` tool is a utility to translate record data type definition(s) written in *CARGO* metalanguage into all the low-level C++ components needed to produce a working compressor/decompressor tool for the record: a C++ user record definition (see *The Type API*), a *TypeAPI*-based definition (see *The Type API*) and several *CARGO* application template files (see *Application templates*). For complete examples of use, see *Examples*.

### 4.7.1 Command line

`cargo_translate` when launched from the command line (with additional `-h` option for help) displays:

```
*** CARGO translate tool ***
Version: <version>
Date: <rev_date>

Generates C++ TypeAPI record data type definition with CARGO application template files
from the specified high-level CARGO metalanguage type definition
Authors: Lukasz Roguski and Paolo Ribeca

Usage: cargo_translate [-i <input.cargo> -o <output_prefix> -t -k -v -h]
options:
  -i | --input <input.cargo>    - CARGO record data type definition file (default: stdin)
  -o | --output <output_prefix> - output files name prefix (default: OutRecord)
  -t | --transform              - generate record transformation class template
  -k | --keygen                 - generate record key generator class template
  -v | --verbose                - display additional information while parsing
  -h | --help                   - display this message
```

where the available options are:

-i, --input <input.cargo>  file containing CARGO record data type definition (default: *stdin*)

-o, --output <output_prefix>  record name prefix for generating output files: `output_prefix.h`, `output_prefix_Type.h`, ... (default: *OutRecord*)

-t, --transform  generate record transformation class template in order to apply transformations on records while processing data (to file `output_prefix_Transform.h`)

| | |
|---|---|
| **-k, --keygen** | generate key generator class template in order to index sorted records while processing data (to file `output_prefix_KeyGenerator.h`) |
| **-v, --verbose** | display additional information while parsing |
| **-h, --help** | display help message |

As explained in *Pre-processing directives*, `cargo-translate` will generate several C++ files for each *CARGO* meta-language record definition that has been flagged with a `@record` keyword (see *Examples* for examples of use).

# CHAPTER
# FIVE

# THE TYPE API

The main objective of the *CARGO TypeAPI* is to provide an abstract layer separating the low-level *CARGO* data streams representation and the high-level record type definition. In addition, the actual C++ data types and the stream access patterns can be deduced at compile time, resulting in the generation of optimized, high-performance data processing routines. The user only needs to define a record data structure in C++ and provide its definition in terms of the *TypeAPI* – the compiler will then transparently generate all the specialized code that encapsulates the underlying *CARGO* data streams logic.

**Tip:** For ease of use, the *TypeAPI* types definitions and their corresponding C++ record data structures (with some additional helper classes) can be automatically generated from *CARGO* meta-language (see *The CARGO meta-language*) by using the `cargo_translate` tool.

## 5.1 C++ types

From the *CARGO* standpoint, standard C++ types can be divided into two sets – *basic* and *complex* types, which differ in the way they are internally handled by the *TypeAPI* layer.

### 5.1.1 Basic types

*Basic* types are a subset of the *plain old C data types* i.e. integer, character and boolean types. The numeric type names are defined in a manner

```
[u]int(_bits_)
```

where:

- `u` specifies whether the numeric type is `unsigned` – this is optional,
- `(_bits_)` represent the integer width in bits (in powers of two).

In addition to the standard `char` and `bool` types, there are also `uchar` (`unsigned char`) and `byte` (`uint8`) types, which in general case correspond to the same C++ type – all the available C++ *basic* types are presented in the table *Basic C types*.

Table 5.1: Basic C types

| Type name | Bit width | Extra qualifier |
|---|---|---|
| `bool` | 8 | |
| `byte` | 8 | `unsigned` |
| `char` | 8 | |
| `uchar` | 8 | `unsigned` |
| `int8` | 8 | |
| `uint8` | 8 | `unsigned` |
| `int16` | 16 | |
| `uint16` | 16 | `unsigned` |
| `int32` | 32 | |
| `uint32` | 32 | `unsigned` |
| `int64` | 64 | |
| `uint64` | 64 | `unsigned` |

### 5.1.2 Complex types

*Complex* set of C++ data types consist of:

- string types – the C++ *STL* `std::string` type,
- array types – the containers storing *basic* or *complex* types based on C++ `std::vector` from *STL*,
- struct types – the composition or tagged union made of *basic* and *complex* types defined using the standard C++ `struct` type.

## 5.2 Compression options

When defining the record types, *compression method* and *block size* might be specified explicitly to achieve better data compression or performance. By default, when defining types using *TypeAPI*, those parameters are *optional*.

### 5.2.1 Compression methods

*CARGO* currently implements a set of compression methods based on the popular open-source compressors, including gzip, bzip2, PPMd and LZMA, with possible easy extension with other ones as plugins. The compression method names follow a consistent schema:

```
Compression(_method_)L(_level_)
```

where:

- `(_method_)` – defines the compressor,
- `(_level_)` – defines the compression level in range `1-4`.

The available compression methods are presented in tables *Compression methods* and *Default compression methods*.

Table 5.2: Compression methods

| Compressor | Level | Method name | Compressor parameters | |
|---|---|---|---|---|
| gzip | | | *Compression level* | |
| | 1 | `CompressionGzipL1` | 1 | |
| | 2 | `CompressionGzipL2` | 4 | |
| | 3 | `CompressionGzipL3` | 6 | |
| | 4 | `CompressionGzipL4` | 9 | |
| bzip2 | | | *Compression level* | |
| | 1 | `CompressionBzip2L1` | 1 | |
| | 2 | `CompressionBzip2L2` | 4 | |
| | 3 | `CompressionBzip2L3` | 6 | |
| | 4 | `CompressionBzip2L4` | 9 | |
| PPMd | | | *Order* | *Max. memory size* |
| | 1 | `CompressionPPMdL1` | 2 | 32 MB |
| | 2 | `CompressionPPMdL2` | 3 | 32 MB |
| | 3 | `CompressionPPMdL3` | 4 | 32 MB |
| | 4 | `CompressionPPMdL4` | 6 | 32 MB |
| LZMA | | | *Dictionary size* | *Min. match length* |
| | 1 | `CompressionLZMAL1` | 64 kB | 16 |
| | 2 | `CompressionLZMAL2` | 2 MB | 8 |
| | 3 | `CompressionLZMAL3` | 8 MB | 36 |
| | 4 | `CompressionLZMAL4` | 32 MB | 273 |

Table 5.3: Default compression methods

| *Default method* | *Corresponding method* |
|---|---|
| `CompressionGzip` | `CompressionGzipL2` |
| `CompressionBzip2` | `CompressionBzip2L2` |
| `CompressionPPMd` | `CompressionPPMdL2` |
| `CompressionLZMA` | `CompressionLZMAL2` |
| `CompressionText` | `CompressionGzip` |
| `CompressionNumeric` | `CompressionGzip` |
| `CompressionDefault` | `CompressionGzip` |

### 5.2.2 Compression block sizes

Alongside with *compression method*, when defining the record type, the size of the compression block can be selected by specifying the appropriate enumeration – *block size* enumeration is defined in a following way:

```
BlockSize(_size_)(_metric_)
```

where:

- (`_size_`) – specifies a number (power of two),
- (`_metric_`) – is a metric prefix: `K` (Kilo), `M` (Mega), `G` (Giga).

The specified *block size* corresponds to the maximum available size of the work data packet to be compressed (and the size of the internal data buffer) – the data streams are stored in a block-wise manner. The available compression block sizes are presented in the tables *Compression block sizes* and *Default compression block sizes*.

**Note:** The size of the compression block might influence the resulting compression ratio at the higher levels of compression, especially when using *PPMd* or *LZMA* schemes, and, especially when the specified *blocks size* is much smaller than the internal compressor buffer (see *Compression methods*).

Table 5.4: Compression block sizes

| Compression block size name | Block size |
|---|---|
| `BlockSize64K` | 64 KiB |
| `BlockSize128K` | 128 KiB |
| `BlockSize256K` | 256 KiB |
| `BlockSize512K` | 512 KiB |
| `BlockSize1M` | 1 MiB |
| `BlockSize2M` | 2 MiB |
| `BlockSize4M` | 4 MiB |
| `BlockSize8M` | 8 MiB |
| `BlockSize16M` | 16 MiB |
| `BlockSize32M` | 32 MiB |
| `BlockSize64M` | 64 MiB |

Table 5.5: Default compression block sizes

| Compression block size name | Block size |
|---|---|
| `BlockSizeNumeric` | 512 KiB |
| `BlockSizeText` | 2 MiB |
| `BlockSizeDefault` | 2 MiB |

## 5.3 Basic types

*TypeAPI basic types* provide an interface for C++ *basic types* i.e. the numeric, character and boolean types (see: *Basic types*).

### 5.3.1 Type definition

The type definition interface is defined as:

```
TBasicType< _c_basic_type_ [, _compression_method_ , _block_size_ ] >
```

where:

- `_c_basic_type_` – stands for the corresponding *basic* C++ data type,
- `_compression_method_` – stands for *compression method* enumeration (optional), default: `CompressionDefault`,
- `_block_size_` – stands for the underlying *block size* enumeration (optional), default: `BlockSizeDefault`.

### 5.3.2 Integer specialization

In addition to the general type definition, a specialized interface for defining the integer types exists and is defined as follows:

```
TIntegerType< _c_int_type_ >
```

where `_c_int_type_` is *basic* C++ numeric type.

`TIntegerType<>` is a specialized type derived from `TBasicType<>` template class, with specified `CompressionNumeric` as *compression method* and `BlockSizeNumeric` as *block size*. The available predefined types are: `Int8Type`, `UInt8Type`, `Int16Type`, `UInt16Type`, `Int32Type`, `UInt32Type`, `Int64Type`, `UInt64Type`.

### 5.3.3 Character specialization

The specialized type for the character is defined as `CharType`, which uses as a *compression method* `CompressionText` and as a *block size* `BlockSizeText`.

### 5.3.4 Boolean specialization

In a similar way as the character type, the boolean type is defined as `BoolType` using `CompressionDefault` as *compression method* and `BlockSizeDefault` as *block size*.

### 5.3.5 Usage examples

- defining an unsigned 64-bit integer type:

```
typedef TBasicType< uint64 > MyInt64Type;
```

- defining a character type using the *PPMd* level 2 as *compression method* and 32 MiB as *block size*:

```
typedef TBasicType< char, CompressionPPMdL2, BlockSize32M > MyCharType;
```

## 5.4 Basic array types

*TypeAPI* provides also the interface for *array type* containing elements of C++ *basic* types (see: *Basic C types*). The *array type* is based on the standard C++ `std::vector` and `std::string` types with a variable (the standard behavior) or fixed length.

### 5.4.1 Array type definition

The variable- and fixed-length *array types* are defined as follows:

```
TBasicArrayType< _c_basic_type_ [ , _compression_method_ , _block_size_ ] >

TFixedBasicArrayType<  _c_basic_type_, _length_
                    [ , _compression_method_ , _block_size_ ]
                 >
```

where:

- `_c_basic_type_` – defines *basic* C++ data type,
- `_length_` – specifies the length of the fixed array,
- `_compression_method_` – specifies *compression method* enumeration (optional), default: `CompressionDefault`,
- `_block_size_` – specifies *block size* enumeration (optional), default: `BlockSizeDefault`.

Despite using the fixed length in the case of `TFixedVectorType`, both *array types* are based on the standard C++ `std::vector` type for compatibility, ease of use and ease of integration.

### 5.4.2 Character array (string) specialization

The variable– and fixed-length *character array* (or *string*) are the special cases of *array types* – they are defined as follows:

```
TStringType< [ _compression_method_ , _block_size_ ] >
```

```
TFixedStringType< _length_ [ , _compression_method_ , _block_size_ ] >
```

where:

- `_length_` – specifies the length of the fixed array,
- `_compression_method_` – specifies the compression method enumeration (optional), default: `CompressionText`,
- `_block_size_` - specifies the block size enumeration (optional), default: `BlockSizeText`.

Similarly, like in the case of `TFixedArrayType<>`, both *string* types are based on the standard C++ `std::string` type. In addition, *TypeAPI* defines `StringType`, which is a specialized case of `TStringType<>` and which uses `CompressionText` as the *compression method* and `BlockSizeText` for size of the compression block.

### 5.4.3 Integer array type specialization

*TypeAPI* provides also a specialized *integer array types* in both variants: variable– and fixed-length; they are defined as:

```
TIntArrayType< _c_int_type_ >
```

```
TFixedIntArrayType< _c_int_type_, _length_ >
```

where:

- `_c_int_type_` – defines basic plain C++ *basic* numeric data type,
- `_length_` – specifies the length of the fixed array.

These types are the specialized cases respectively of `TArrayType<>` and `TFixedArrayType<>` types using `CompressionNumeric` as *compression method* and `BlockSizeNumeric` as *block size*.

### 5.4.4 Usage examples

- defining a byte array type with the *bzip2* level 4 *compression scheme* and the 512 KiB *block size*:

    ```
    typedef TBasicArrayType< byte,
                             CompressionBzip2L4,
                             BlockSize512k
                           > MyByteArrayType;
    ```

- defining a string type using the *PPMd* level 2 *compression scheme* and the 32 MiB *block size*:

    ```
    typedef TStringType< CompressionPPMdL2, BlockSize32M > MyStringType;
    ```

- defining an unsigned 64-bit integer array type of fixed length of 20 and using the default compression options:

    ```
    typedef TFixedIntArrayType< uint64, 20 > MyFixedInt64ArrayType;
    ```

## 5.5 Enumeration types

*Enumeration types* provide functionality for working with integer, character or string labels enumerations. *Enumeration type* is an exceptional case of the *basic data type* and it's C++ definition differs from the *C plain old data* `enum` data type.

### 5.5.1 Type definition

*Enumeration type* is defined as follows:

```
TEnumType< _enum_key_map_ [ , _compression_method_ , _block_size_ ] >
```

where:

- `_enum_key_map_` – specifies a class implementing the enumeration-to-index translation functionality (see below),
- `_compression_method_` – defines *compression method* enumeration (optional), default: `CompressionNumeric`,
- `_block_size_` – defines *block size* enumeration (optional), default: `BlockSizeNumeric`.

The goal of `_enum_key_map_` class template is to provide the mapping functionality of the user-specified enumeration symbols into the numeric index values, which will be internally used by *TypeAPI*. The idea of such key mapping class is defined as follows:

```cpp
class EnumKeyMapper
{
public:
    typedef _key_type_ KeyType;
    static const uint32 KeyCount = _key_count_ ;

    static KeyType Key(uint32 idx_) { ... }
};
```

where:

- `_key_type_` – specifies one of the C++ *basic* types (numeric or character types) or the C++ `std::string` (string label type),
- `_key_count_` – specifies the total number of available keys i.e. user-specified enumerations,

and the static member function `KeyType Key(uint32 idx_)` returns the enumeration value (i.e. the key) associated with the specified index `idx_` value.

### 5.5.2 Usage examples

A sample DNA enumeration type definition consists of a set of `5` characters representing the possible nucleobases i.e. `ACGTN` and with `char` type as the enumeration value:

```cpp
struct SampleDnaEnumKeyMap
{
    typedef char KeyType;
    static const uint32 KeyCount = 5;

    static KeyType Key(uint32 idx)
    {
        static const KeyType SymbolLookupTable[] = {'A', 'C', 'G', 'T', 'N'};
```

```
9        return SymbolLookupTable[idx];
10    }
11 };
12
13 typedef TEnumType< SampleCharEnumKeyMap > DnaEnumType;
```

## 5.6 Struct type

*TypeAPI struct type* provides an interface for defining the product data types similar to the standard `struct` type available in C/C++; *struct* type can be composed either from *basic* or *complex* types.

### 5.6.1 Type definition

To use *TypeAPI struct type*, firstly a definition of a class containing the description of the C++ *struct type* members and its' access is required. The sample C++ *struct type* definition with it's corresponding description in *TypeAPI* is as follows:

```
1  struct _cpp_struct_
2  {
3    _field_1_cpp_type_ _field_1_;
4    _field_2_cpp_type_ _field_2_;
5
6    ...
7
8    _field_N_cpp_type _field_N_;
9  };
10
11
12 class _cpp_struct_description_
13 {
14 public:
15   typedef _cpp_struct_ UserDataType;
16   static const uint32 FieldCount = _members_count_;
17
18   typedef _field_1_typeapi_type_ Type1;
19   static typename Type1::UserDataType& Get1(UserDataType& data_)
20   {
21     return data_._field_1_;
22   }
23   static const typename Type1::UserDataType& Get1(const UserDataType& data_)
24   {
25     return data_._field_1_;
26   }
27
28   typedef _field_2_typeapi_type_ Type2;
29   static typename Type2::UserDataType& Get2(UserDataType& data_)
30   {
31     return data_._field_2_;
32   }
33   static const typename Type2::UserDataType& Get2(const UserDataType& data_)
34   {
35     return data_._field_2_;
36   }
37
```

```
38     ...
39
40     typedef _field_N_typeapi_type_ TypeN;
41     static typename TypeN::UserDataType& GetN(UserDataType& data_)
42     {
43       return data_._field_N_;
44     }
45     static const typename TypeN::UserDataType& GetN(const UserDataType& data_)
46     {
47       return data_._field_N_;
48     }
49 };
```

where:

- _cpp_struct_ – the C++ *structure* type name,
- _cpp_struct_description_ – a *structure* type description class name,
- _field_*_cpp_type_, _field_*_ – the C++ type names with their corresponding member names,
- _members_count_ – the total number of the data fields defined in C++ *structure type*,
- _field_*_typeapi_type_ – *TypeAPI* type names corresponding to C++ _field_*_cpp_type_.

In addition to describing the used C++ types in _cpp_struct_ (and linking them with their equivalent *TypeAPI* types), in _cpp_struct_description_ needs needs also to be defined the static member functions in order to access the corresponding struct members, implementing the mutable and immutable Get* functions. Having the C++ *struct type* described in *TypeAPI* the actual *struct type* definition is as follows:

```
TStructType< _user_struct_description_ >
```

### 5.6.2 Usage examples

- Defining a sample record type with 3 fields of character and numeric types representing a nucleobase with it's corresponding quality score at a given position:

```
1  // a sample C++ record type definition
2  struct VariantRecord
3  {
4    char nucleobase;
5    byte quality;
6    uint64 position;
7  };
8
9  // a sample definition of VariantRecord member types
10 typedef CharType NucleobaseType;
11 typedef TBasicType< byte, CompressionBzip2L2, BlockSize512k > QualityType;
12 typedef TIntegerType< uint64 > PositionType;
13
14 // the record type description for TypeAPI
15 struct VariantRecordDescription
16 {
17   typedef VariantRecord UserDataType;
18   static const uint32 FieldCount = 3;
19
20   typedef NucleobaseType Type1;
21   static typename Type1::UserDataType& Get1(UserDataType& data_)
22   {
```

```
23       return data_.nucleobase;
24     }
25     static const typename Type1::UserDataType& Get1(const UserDataType& data_)
26     {
27       return data_.nucleobase;
28     }
29
30     typedef QualityType Type2;
31     static typename Type2::UserDataType& Get2(UserDataType& data_)
32     {
33       return data_.quality;
34     }
35     static const typename Type2::UserDataType& Get2(const UserDataType& data_)
36     {
37       return data_.quality;
38     }
39
40     typedef PositionType Type3;
41     static typename Type3::UserDataType& Get3(UserDataType& data_)
42     {
43       return data_.position;
44     }
45     static const typename Type3::UserDataType& Get3(const UserDataType& data_)
46     {
47       return data_.position;
48     }
49   };
50
51   // define TypeAPI struct type
52   typedef TStructType< VariantRecordDescription > VariantRecordType;
```

A sample record type `VariantRecord` consist of 3 fields:

- `nucleobase` of `char` type,
- `quality` of `byte` type,
- `position` of `uint64` type.

The character stream representing the consecutive `nucleobase` values is defined in *TypeAPI* as `NucleobaseType` and is compressed using the default *compression scheme* and using the default *block size* associated with `CharType`. The consecutive `quality` values are defined as `QualityType` and are compressed using *bzip2* compressor with level 4 with 512 KiB *block size* size. Finally, the consecutive `position` values are defined as `PositionType` and are compressed using the default options associated with `TIntegerType`. Those types definitions are then used to describe the C++ record type `VariantRecord` in `VariantRecordDescription`, which is later used to create a final *struct record type* definition in *TypeAPI* - `VariantRecordType`.

## 5.7 Union type

*Union type* provides an interface for defining **tagged** union types, which are based on the C/C++ `struct` structure type and differs from the standard C/C++ `union` type.

### 5.7.1 Type definition

As *union type* and *struct type* interfaces are similar, *union type* either requires firstly it's type description in *TypeAPI*. However, as the *union type* defines a **tagged** union type, which, in usability terms, differs from the standard C/C++

union, it is based on C/C++ `struct` type instead of `union`. As a result, it defines an additional field *kind* (`uint32 __kind`), explicitly holding the information (index) of the currently used member of the union. The sample C++ *union type* definition with it's corresponding description in *TypeAPI* is as follows:

```cpp
struct _cpp_union_
{
  uint32 __kind;

  _field_1_cpp_type_ _field_1_;
  _field_2_cpp_type_ _field_2_;

  ...

  _field_N_cpp_type _field_N_;
};

class _cpp_union_description_
{
public:
  typedef _user_union_ UserDataType;
  static const uint32 FieldCount = _members_count_;

  typedef _field_1_typeapi_type_ Type1;
  static typename Type1::UserDataType& Get1(UserDataType& data_)
  {
    return data_._field_1_;
  }
  static const typename Type1::UserDataType& Get1(const UserDataType& data_)
  {
    return data_._field_1_;
  }

  typedef _field_2_typeapi_type_ Type2;
  static typename Type2::UserDataType& Get2(UserDataType& data_)
  {
    return data_._field_2_;
  }
  static const typename Type2::UserDataType& Get2(const UserDataType& data_)
  {
    return data_._field_2_;
  }

  ...

  typedef _field_N_typeapi_type_ TypeN;
  static typename TypeN::UserDataType& GetN(UserDataType& data_)
  {
    return data_._field_N_;
  }
  static const typename TypeN::UserDataType& GetN(const UserDataType& data_)
  {
    return data_._field_N_;
  }
};
```

where:

- `_cpp_union_` – the C++ *union* type name,

- `_cpp_union_description_` – the C++ *union type* description class name,
- `_field_*_cpp_type_`, `_field_*_` – the C++ type names with their corresponding structure member names,
- `_members_count_` – the total number of fields defined in C++ *union type* (excluding `__kind`),
- `_field_*_typeapi_type_` – *TypeAPI* types corresponding to `_field_*_user_type_` available in `_cpp_union_`.

Finally, using `_cpp_union_description_` the *TypeAPI union type* is then defined as follows:

```
TUnionType < _user_union_description_ >
```

### 5.7.2 Usage examples

Defining a sample *union* record type with 2 fields – string and integer, representing either a sequence or a match length, depending on the usage scenario:

```cpp
// a sample C++ record type definition
struct MatchRecord
{
  // a special and obligatory member explicitly specifying
  // the currently used member/value
  uint32 __kind;

  std::string sequence;
  int64 match;
};

// a sample definition of VariantRecord member types
typedef TStringType< CompressionPPMdL4, BlockSize32M > SequenceType;
typedef TIntegerType< int64 > MatchType;

// the record type description for TypeAPI
struct MatchRecordDescription
{
  typedef MatchRecord UserDataType;
  static const uint32 FieldCount = 2;

  typedef SequenceType Type1;
  static typename Type1::UserDataType& Get1(UserDataType& data_)
  {
    return data_.sequence;
  }
  static const typename Type1::UserDataType& Get1(const UserDataType& data_)
  {
    return data_.sequence;
  }

  typedef MatchType Type2;
  static typename Type2::UserDataType& Get2(UserDataType& data_)
  {
    return data_.match;
  }
  static const typename Type2::UserDataType& Get2(const UserDataType& data_)
  {
    return data_.match;
  }
```

```
41    };
42
43    // define TypeAPI struct type
44    typedef TUnionType< MatchRecordDescription > MatchRecordType;
```

A sample record type `MatchRecord` consist of 2 fields – sequence of `std::string` type and `int64` of `byte` type. The string stream values are defined in *TypeAPI* as `SequenceType` and are compressed using *PPMd* level 4 *compression scheme* with 32 MiB *block size*. The consecutive `match` values are defined in *TypeAPI* as `MatchType` and are compressed using default options associated with `TIntegerType` type. Those types definitions are then used to describe the user record `MatchRecord` in `MatchRecordDescription`, which is later used to defined a final user union record type definition in *TypeAPI* – `MatchRecordType`.

## 5.8 Complex array types

*TypeAPI complex array types* provide interfaces for defining more advanced array types used to store C++ *complex types* (see: *Complex types*).

### 5.8.1 Array type definition

In order to use *complex array type* the underlying *complex* type needs to be previously defined in *TypeAPI* – can be either *enum*, *array*, *struct* or *union* type. A variable– and fixed-length array types are then defined as follows:

```
TComplexArrayType < _t_complex_type_ >

TFixedComplexArrayType <  _t_complex_type_, _length_ >
```

where:

- `_t_complex_type_` – stands for a *TypeAPI complex type* definition name,
- `_length_` – specifies the length of the fixed array.

*Compression method* and *block size* are not specified, as they are already defined in the *TypeAPI complex type* definition. Similarly, like in the case of *basic array types*, despite the provided fixed length for `TFixedComplexArrayType` both array types use the standard C++ `std::vector` type for compatibility, ease of use and ease of integration.

### 5.8.2 Usage examples

- defining a complex array of arrays of byes compressed using *bzip2* level 4 *compression method*:

    ```
    typedef TBasicArrayType< byte, CompressionPPMdL4 > MyByteArrayType;

    typedef TComplexArrayType< MyByteArrayType > MyComplexByteArrayType;
    ```

- defining a fixed-length array of strings with length of `20`:

    ```
    typedef TFixedComplexArrayType< StringType, 20 > MyFixedStringArrayType;
    ```

- defining a complex array of `VariantRecord` types (defined in *Usage examples* in *TypeAPI* as `VariantRecordType`):

    ```
    typedef TComplexArrayType< VariantRecordType > MyStringType;
    ```

## 5.9 Metalanguage to TypeAPI cheatsheet

The available in *CARGO metalanguage* types (see *The CARGO meta-language*) with their corresponding *TypeAPI* types are presented in the tables included in the following subchapter.

The used symbols among descriptions mean:

- `T` – *TypeAPI* type or type definition,
- `t` – the C++ user type,
- `C` – *compression method* enumeration (usually optional),
- `B` – the compression *block size* enumeration (usually optional),
- `n` – the length of the fixed array.

### 5.9.1 Basic types

Table 5.6: Boolean CARGO metalanguage and TypeAPI types

| Metalanguage | TypeAPI type | C++ data type |
|---|---|---|
| bool | BoolType<br>TBasicType< bool [, C , B ] > | bool |

Table 5.7: Character CARGO metalanguage and TypeAPI types

| Metalanguage | TypeAPI type | C++ data type |
|---|---|---|
| char | CharType<br>TBasicType< char [, C , B ] > | char |

Table 5.8: Numeric CARGO metalanguage and TypeAPI types

| Metalanguage | TypeAPI type | C++ data type |
|---|---|---|
| int ^ _n_ | TIntegerType< int_n_ ><br>TBasicType< int_n_ [, C , B ] > | int_n_ |
| uint ^ _n_ | TIntegerType< uint_n_ ><br>TBasicType< uint_n_ [, C , B ]> | uint_n_ |
| byte | TIntegerType< byte ><br>TBasicType< byte [, C , B ]> | byte |

### 5.9.2 Basic array types

Table 5.9: Basic array CARGO metalanguage and TypeAPI types

| Variant | Metalanguage | TypeAPI type | C++ data type |
|---|---|---|---|
| Standard | `T array` | `TBasicArrayType< T [, C , B ]>` | `std::vector < t >` |
| Fixed | `T array * n` | `TFixedBasicArrayType< T , n [, C , B ]>` | `std::vector < t >` |

Table 5.10: Basic string CARGO metalanguage and TypeAPI types

| Variant | Metalanguage | TypeAPI type | C++ data type |
|---|---|---|---|
| Standard | `string`<br>`char array` | `StringType`<br>`TStringType< [ C , B ] >`<br>`TBasicArrayType< char [, C , B ] >` | `std::string` |
| Fixed | `string * n`<br>`char array * n` | `TFixedStringType< n [, C , B ] >`<br>`TFixedBasicArrayType< char , n [, C , B ] >` | `std::string` |

### 5.9.3 Enumeration

Table 5.11: Enumeration CARGO metalanguage and TypeAPI type

| Key type | Metalanguage | TypeAPI type | C++ data type |
|---|---|---|---|
| numeric<br>character<br>string | `enum [E1, E2, ...]` | `TEnumType< T [, C , B ] >` | `[u]int ^ _n_`<br>`char`<br>`std::string` |

### 5.9.4 Sum types

Table 5.12: Product CARGO metalanguage and TypeAPI types

| Type | Metalanguage | TypeAPI type | C++ data type |
|---|---|---|---|
| Struct | `S = { ... }` | `TStructType < T >` | `struct S { ... };` |
| Union | `U = [ ... ]` | `TUnionType < T >` | `struct U { ... };` |

### 5.9.5 Complex array types

Table 5.13: Complex array CARGO metalanguage and TypeAPI types

| Variant | Metalanguage | TypeAPI type | C++ data type |
|---|---|---|---|
| Standard | `T array` | `TComplexArrayType< T >` | `std::vector < t >` |
| Fixed | `T array * n` | `TComplexBasicArrayType< T , n >` | `std::vector < t >` |

### 5.9.6 FASTQ example

Following the *FASTQ* example introduced in the *Quickstart* chapter – to define a simple *FASTQ* record type in *CARGO metalanguage* consisting of 3 string fields with explicit *compression methods* i.e. `Bzip2` for *tag* fields compression, `LZMA` for *sequence* and `PPMd` for *qualities*:

```
1  FastqRecord = {
2    tag = string;
3    seq = string;
4    qua = string
5  }
6    .tag.Pack = Bzip2,
7    .seq.Pack = LZMA,
8    .qua.Pack = PPMd;
9
10 @record FastqRecord
```

The corresponding C++ record type definition in:

```
1  struct FastqRecord
2  {
3    std::string tag;
4    std::string seq;
5    std::string qua;
6  };
```

plus the corresponding record type definition in *TypeAPI*:

```
1  typedef TStringType<CompressionBzip2> TagType;
2  typedef TStringType<CompressionLZMA> SequenceType;
3  typedef TStringType<CompressionPPMd> QualityType;
4
5  // describe the C++ struct record type and its fields
6  struct FastqRecordDescription
7  {
8    typedef FastqRecord UserDataType;
9    static const uint32 FieldCount = 3;
10
11   typedef TagType Type1;
12   static typename Type1::UserDataType& Get1(UserDataType& data_)
13   {
14     return data_.tag;
15   }
16   static const typename Type1::UserDataType& Get1(const UserDataType& data_)
17   {
18     return data_.tag;
19   }
20
21   typedef SequenceType Type2;
22   static typename Type2::UserDataType& Get2(UserDataType& data_)
23   {
24     return data_.seq;
25   }
26   static const typename Type2::UserDataType& Get2(const UserDataType& data_)
27   {
28     return data_.seq;
29   }
30
31   typedef QualityType Type3;
```

```
32    static typename Type3::UserDataType& Get3(UserDataType& data_)
33    {
34      return data_.qua;
35    }
36    static const typename Type3::UserDataType& Get3(const UserDataType& data_)
37    {
38      return data_.qua;
39    }
40  };
41
42  // define TypeAPI struct type
43  typedef TStructType< FastqRecordDescription > FastqRecordType;
```

Note: The translated record type definition from *CARGO metalanguage* to *TypeAPI* performed by `cargo_translate` (see: *Translator tool*) might differ in syntax, especially in using shorter versions of type names and automatic type names generation.

# CHAPTER
# SIX

# RECORD DATA TYPE PROCESSING

In order to convert into semantically meaningful records the input raw data passed to *CARGO* applications, and to further process it, the user needs to implement in C++ a small set of parsing functions. The only mandatory ones are the raw data parsing techniques defined in *Record type parser*. To that end, a toolbox made of a set of several helper classes is provided within the *CARGO* library (see *Helper classes*). Furthermore, in order to perform certain operations on records, the user sometimes needs to define a data transformation function (see *Record type transform*) operating on the parsed record data type. Additionally, to store the records according to some sorting criterion specified by the user, a key generation function useful in the case of data extraction and range querying (see *Extractor*) must occasionally be specified (see *Record type key generator*).

## 6.1 General workflow

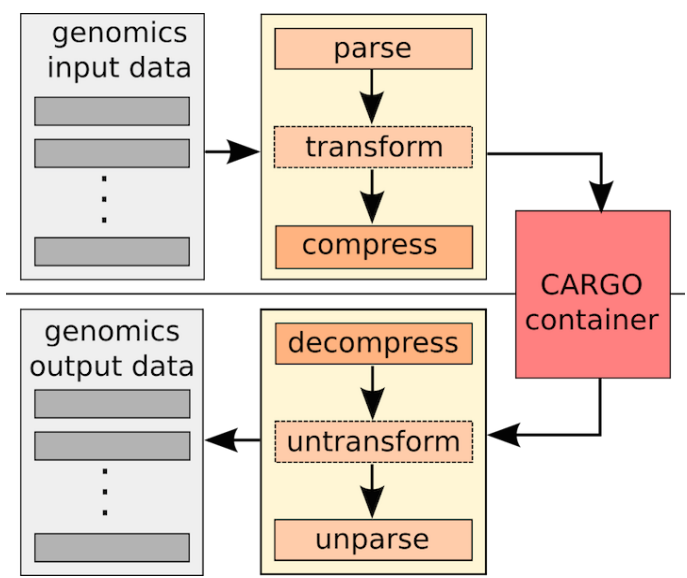

Fig. 6.1: General concepts underlying the CARGO data processing workflow

The general concepts underlying the *CARGO* data processing workflow are presented in *General concepts underlying the CARGO data processing workflow*. More in detail, the data storage pipeline is as follows:

1. First the raw input genomic data in text form is parsed into corresponding C++ data structures by using the parsing function implemented by the the user (see: *Record type parser*)

2. Optionally, in the next step the records are transformed according to the *forward transformation* function specified by the user – see: *Record type transform*

3. Elements of the record are first split into a number of streams, subsequently compressed and finally stored inside the container.

Similarly, the general data access pipeline is as follows:

1. First streams are decompressed from the container and several single stream elements are merged in order to reconstruct the original record

2. Optionally, in the next step the records are un-transformed according to the *backward transformation* function specified by the user

3. Records are unparsed, from the C++ data structure defining the record into raw text, by using the unparsing function implemented by the user.

## 6.2 Helper classes

To quickly and easily implement record data parsers, i.e. in order to write procedures able to read/write raw data from input/to output, *CARGO* provides a set of basic helper classes.

### 6.2.1 Memory stream

`MemoryStream` class (defined in `<cargo/core/MemoryStream.h>`) implements safe and easy to use member functions to read/write the data to/from a specified chunk of memory. The most important member functions are presented in the table *Selected MemoryStream class member functions*.

Table 6.1: Selected MemoryStream class member functions

| `bool ReadByte(uchar& b_)` | |
|---|---|
| Reads the next byte available in the memory, on success advancing the read position by `1` | |
| returns: | `true` – on success |
| | `false` – when reached the end of memory |
| params: | `b_` – reference to the byte to store the read value |
| `bool PeekByte(uchar& b_)` | |
| Reads the next byte available in the memory without advancing the read position | |
| returns: | `true` – on success |
| | `false` – when reached the end of memory |
| params: | `b_` – reference to the byte to store the read value |
| `bool WriteByte(uchar b_)` | |
| Writes the next byte to the memory, on success advancing the read position by `1` | |
| returns: | `true` – on success |
| | `false` – when reached the end of memory |
| params: | `b_` – byte to be stored in memory |

Table 6.2: Selected MemoryStream class member functions - continued

| `int64 Read(uchar* data_, uint64 size_)` |  |
|---|---|
| Reads the next available up to `size_` bytes in the memory, on success advancing the position by the number of bytes read | |
| returns: | number of bytes read, where value less than `size_` means reaching the end of the memory |
| params: | `data_` – pointer to the raw memory region to store the data, containing more than (or equal) *size_*' of available space <br> `size_` – the number of bytes to read |
| `int64 Write(const uchar* data_, uint64 size_)` | |
| Writes the data up to `size_` bytes to the memory, on success advancing the position by the number of bytes written | |
| returns: | number of bytes written - value less than `size_` means reaching the end of the memory |
| params: | `data_` – pointer to the raw memory region to write the data from, containing more than (or equal) `size_` of the stored data <br> `size_` – the number of bytes to write |
| `uint64 Size()` | |
| Returns the total size in bytes of the memory chunk | |
| `uint64 Position()` | |
| Returns the current read/write position (in bytes) | |
| `bool SetPosition(uint64 pos_)` | |
| Sets the read/write position | |
| returns: | `true` – on success <br> `false` – when trying to exceed the total size of the memory chunk |
| params: | `pos_` – position (in bytes) to be set |

**Tip:** When implementing own record data parser code, user can assume that already initialized `MemoryStream` object will be passed to the parsing function.

### 6.2.2 FieldParser

`FieldParser` class (defined in `<cargo/type/UserDataParser.h>`) provides static member functions for parsing the delimiter-separated record fields reading/writing the data directly from/to `MemoryStream`. The most important member functions are presented in the table *Selected FieldParser member functions*.

Table 6.3: Selected FieldParser member functions

| `static bool ReadNextField(MemoryStream& stream_, _T& v_, char separator_)` |
|---|
| Reads and parses the next available field of type `_T` from *memory stream* |
| returns: |
|     `true` – on success |
|     `false` – when reached the end of the *memory stream* or error while parsing |
| params: |
|     `stream_` – *memory stream* object to read from |
|     `v_` – reference to the output value to store the parsed data |
|     `separator_` – records fields separator character |
| `static bool WriteNextField(MemoryStream& stream_, _T v_, char separator_)` |
| Parses the next numeric field of type `_T` and writes to the *memory stream* |
| returns: |
|     `true` – on success |
|     `false` – when reached the end of the *memory stream* or error while parsing |
| params: |
|     `stream_` – *memory stream* object to read from |
|     `v_` – integer value to be parsed and stored |
|     `separator_` – records fields separator character |
| `static bool WriteNextField(MemoryStream& stream_, const std::string& str_, char sep_)` |
| Parses the next string field writing the data to the specified stream |
| returns: |
|     `true` – on success |
|     `false` – when reached the end of the *memory stream* or error while parsing |
| params: |
|     `stream_` – *memory stream* object to read from |
|     `str_` – reference to the string to be parsed and stores |
|     `sep_` – records fields separator character |
| `static bool PeekNextByte(MemoryStream& stream_, uchar& b_)` |
| Peeks the next available byte from specified *memory stream* |
| returns: |
|     `true` – on success |
|     `false` – when reached the end of the *memory stream* |
| params: |
|     `stream_` – *memory stream* object to read from |
|     `b_` – reference to the byte to store the read value |

Table 6.4: Selected FieldParser member functions - continued

| <td colspan="2">`static bool SkipNextField(MemoryStream& stream_, char separator_)`</td> |  |
|---|---|
| <td colspan="2">Skips the next field value reading the from the specified *memory stream*</td> |  |
| returns: | `true` – on success <br> `false` – when reached the end of the *memory stream* or error while parsing |
| params: | `stream_` – *memory stream* object to read from <br> `separator_` – records fields separator character |
| <td colspan="2">`static bool SkipBlock(MemoryStream& stream_, char blockStart_, char blockEnd_)`</td> |  |
| <td colspan="2">Skips the whole block (or line) of text delimited by `blockStart_` and `blockEnd_`</td> |  |
| returns: | `true` – on success <br> `false` – when reached the end of the *memory stream* |
| params: | `stream_` – *memory stream* object to read from <br> `blockStart_` – beginning of the text block character delimiter <br> `blockEnd_` – end of the text block character delimiter |

## 6.3 Record type parser

*Record type parser* is a class providing the static member functions to parse the raw textual data chunks filling the provided C++ record data structures. The class can be seen as a link between the user-understandable text data and *CARGO*-understandable structured data. It's the only one obligatory class which member functions needs to be implemented by user by using previously introduced helper classes i.e. `FieldParser` and/or `MemoryStream`.

### 6.3.1 Parser class template

*Parser class* code template is as follows:

```cpp
class TRecord_Parser
{
public:
  static void SkipToEndOfHeader(MemoryStream& stream_)
  {
    // ...
  }

  static void SkipToEndOfRecord(MemoryStream& stream_)
  {
    // ...
  }

  static bool ReadNextRecord(MemoryStream& stream_, TRecord& record_)
  {
```

```
16       // ...
17   }
18
19   static bool WriteNextRecord(MemoryStream& stream_, TRecord& record_)
20   {
21       // ...
22   }
23 };
```

where:

- `MemoryStream& stream_` – represents the *memory stream* from/to which the record data is going to be read/written,
- `TRecord& record_` – represents a user C++ record data which contains or will contain the data in interest.

### 6.3.2 # skip until the end of the header

The function:

```
void SkipToEndOfHeader(MemoryStream& stream_)
```

implements skipping of the following next bytes in the *memory stream* until the end of the (file) header, which may appear depending on file format. Quite often the genomic file header lines begin with a special symbol e.g. `@` (SAM format) or `##` (VCF format) and they need to be filtered out before the actual records processing. This function can be left empty, as not always the genomic file format uses file header (e.g. FASTA or FASTQ format).

### 6.3.3 # skip until the end of the record

The function:

```
void SkipToEndOfRecord(MemoryStream& stream_)
```

implements skipping the next bytes in the *memory stream* until the end of the (current) record in order to properly position parser at the beginning of the next one. As quite often the records are stored line-by-line (e.g. in *SAM* or *VCF* formats), the function will just implement skipping to the end of the (current) line. In case of *FASTQ* or *FASTA* formats, where one record spans over multiple lines, an analysis of the line ending the record needs to be taken into account (see *Simple FASTQ format compressor*).

### 6.3.4 # read next record

The function:

```
bool ReadNextRecord(MemoryStream& stream_, TRecord& record_)
```

implements filling the translated C++ record data structure of `TRecord` type with the data parsed from *memory stream*. For the records defined in the tab-separated text format using only the helper class `FieldParser` for parsing the data should be sufficient.

### 6.3.5 # write next record

The function:

```
bool WriteNextRecord(MemoryStream& stream_, TRecord& record_)
```

implements parsing the contents of the translated C++ record data structure of `TRecord` type and further writing it to *memory stream*. For the records defined in tab-separated text format using just the helper class `FieldParser` for parsing the data should be sufficient.

### 6.3.6 FASTQ example

Following the *FASTQ* example from *Simple FASTQ format compressor* subchapter, the simple *FASTQ* records parser can be defined as:

```cpp
class FastqRecord_Parser
{
public:
    static void SkipToEndOfHeader(io::MemoryStream& stream_)
    {
        // there's no header in the FASTQ file format -> member function left empty
    }

    static void SkipToEndOfRecord(io::MemoryStream& stream_)
    {
        // skip to the new line
        FieldParser::SkipNextField(stream_, '\n');

        byte nextByte = 0;
        while (FieldParser::PeekNextByte(stream_, nextByte))
        {
            // beginning of next record - found read id line?
            if (nextByte == '@')
                return;
            FieldParser::SkipNextField(stream_, '\n');
        }
    }

    static bool ReadNextRecord(io::MemoryStream& stream_, FastqRecord& record_)
    {
        bool result = FieldParser::ReadNextField(stream_, record_.tag, '\n');
        result &= FieldParser::ReadNextField(stream_, record_.seq, '\n');
        result &= FieldParser::SkipNextField(stream_, '\n');  // skip '+' line
        result &= FieldParser::ReadNextField(stream_, record_.qua, '\n');
        return result;
    }

    static bool WriteNextRecord(io::MemoryStream& stream_, FastqRecord& record_)
    {
        bool result = FieldParser::WriteNextField(stream_, record_.tag, '\n');
        result &= FieldParser::WriteNextField(stream_, record_.seq, '\n');
        result &= FieldParser::WriteNextField(stream_, "+", '\n');
        result &= FieldParser::WriteNextField(stream_, record_.qua, '\n');
        return result;
    }
};
```

On line `4` an empty `SkipToEndOfHeader` function was defined – the FASTQ format does not specify any file header, so the body of the function is left empty.

As *FASTQ* record cannot be uniquely identified just by analyzing single line (is defined by 4 consecutive lines), the function `SkipToEndOfRecord` needs to:

1. skip until the end of the current text field (text line), passing a newline separator `\n`,
2. peek the next symbol from the *memory stream*,
3. check whether the next line starts with `@` symbol and if so – exit, otherwise: (4),
4. skip the next line and go to (2).

Having already the *memory stream* positioned at the beginning of the next *FASTQ* record, reading and parsing the next one goes straightforward (starting from line `26`):

1. read and parse the *read id* field from the *memory stream* saving the string as `record_.tag` field,
2. read and parse the record *sequence*,
3. skip the next line, which will be the *plus* control field,
4. read and parse the read *quality*.

Parsing back and writing the *FASTQ* record to *memory stream* goes similar way like reading (starting from line `35`):

1. parse and write to the *memory stream* the `record_.tag` field adding a newline `\n` separator,
2. parse and write the record *sequence*,
3. write the empty + line,
4. parse and write the read *quality*.

## 6.4 Record type transform

*Record type transform* is a class implementing the operations specified by user which will be applied on every record while processing the data. Providing the *transform class* is not obligatory.

### 6.4.1 Transform class template

The *transform class* code template is as follows:

```cpp
class TRecord_Transform
{
public:
    static void TransformForward(TRecord& record_)
    {
        // ...
    }

    static void TransformBackward(TRecord& record_)
    {
        // ...
    }
};
```

### 6.4.2 # transform forward

The function:

```
void TransformForward(TRecord& record_)
```

implements the transformation operation which will be applied on the records **after** parsing from the raw textual data format (i.e.: *# read next record*) and **before** the actual data compression and storage in container. The practical use-cases may include: trimming or filtering of the sequences, down-sampling or transforming the sequencing quality scores values.

### 6.4.3 # transform backward

The function:

```
void TransformBackward(TRecord& record_)
```

implements the transformation operation which will be applied on the records **after** the data decompression from the container and **before** parsing back (i.e.: *# write next record*) to the raw textual data format.

### 6.4.4 FASTQ example

Following the FASTQ example, as a record transformation Illumina quality scores reduction scheme will be applied, reducing the number of possible Q-scores to 8 values. The Q-scores mapping method implementing the specified scheme is presented in the table *Q-scores mapping using Illumina quality scores reduction scheme.*.

Table 6.5: Q-scores mapping using Illumina quality scores reduction scheme

| Quality score bins | Bin quality score |
|---|---|
| N (no call) | N (no call) |
| 2 – 9 | 6 |
| 10 – 19 | 15 |
| 20 – 24 | 22 |
| 25 – 29 | 27 |
| 30 – 34 | 33 |
| 35 – 39 | 37 |
| >= 40 | 40 |

Using *Illumina quality scores reduction* scheme, the sample *FASTQ* record transform code is as follows:

```cpp
class FastqRecord_Transform
{
    // assume a constant Q-score 33 value offset
    static const uint32 QualityOffset = 33;

public:
    static void TransformForward(FastqRecord& record_)
    {
        for (uint32 i = 0; i < record_.qua.length(); ++i)
        {
            uint32 q = record_.qua[i] - QualityOffset;
            if (q < 2) q = 0;
            else if (q < 10) q = 6;
            else if (q < 20) q = 15;
            else if (q < 25) q = 22;
            else if (q < 30) q = 27;
            else if (q < 35) q = 33;
            else if (q < 40) q = 37;
```

```
19          else q = 40;
20          record_.qua[i] = q + QualityOffset;
21      }
22    }
23
24    static void TransformBackward(FastqRecord& record_)
25    {
26        // left empty - we don't need post-processing
27    }
28 };
```

Starting from the line 8 each value of the qualities will be processed, where the type of the quality is stored using a `std::string` type. As a first step, from each quality value is subtracted an offset – `QualityOffset`, which, for simplicity of this example, has a fixed value of 33 (for more information regarding quality values see: FASTQ format). The value of the absolute quality score is assigned to `q` variable, which is later compared with possible quality ranges and assigned a new value following the *Illumina quality scores reduction* scheme. At the end of the processing (line 19), the new, transformed quality score with the offset is assigned to the current one.

> **Tip:** Performance-wise, to skip the 7 conditional checks for the range of the `q` value, a simple translation lookup table can be used and indexed by `q` value itself – see *Record transform* sub-chapter example for the implementation.

## 6.5 Record type key generator

*Record type key generator* is a class implementing the key generation method providing the sorting order of the records stored inside the container. The user needs to implement a function returning a value of the *key* object of C++ `std::string` type, which will be later used to compare the ordering of records. Providing this class by user is optional.

> **Caution:** *CARGO* assumes that records which will be parsed and read are already sorted externally by the user in the specified order e.g. using standard linux `sort` tool.

### 6.5.1 Key generator class template

*Key generator class* template is presented in the code snippet below:

```
1 struct TRecord_KeyGenerator
2 {
3   static std::string GenerateKey(const TRecord& rec_)
4   {
5     std::string key;
6     // ...
7     return key;
8   }
9 };
```

The function

```
std::string GenerateKey(const TRecord& rec_)
```

is the only obligatory one to be implemented, which returns the specified by user record key value to make the further records comparisons.

### 6.5.2 FASTQ example

Following the *FASTQ* example from the previous subchapters, `tag` field will be used to generate the *key* for the further records' order comparison.

```
struct FastqRecord_KeyGenerator
{
  static std::string GenerateKey(const FastqRecord& rec_)
  {
    return rec_.tag;
  }
};
```

# CHAPTER
# SEVEN

# APPLICATION TEMPLATES

When translating a record data type definition in *CARGO meta-language* with `cargo_translate` (see: *Translator tool*), in addition to C++ and *TypeAPI* record type definitions a command-line application template file is also generated. This file, `<record_type_name>_main.cpp`, will contain the definitions of a set of routines able to compress, decompress and optionally extract and transform, input data formatted as a stream of record defined as per the user's original type specification. Such routines are based on *CARGO* library application template interfaces (available in `<cargo/type/IApp.h>`). They are ready to be compiled and used straight on; they support multi-threaded processing and Unix pipes for easy integration with existing tools.

Given a specified record type data, the generated standalone application contained in `<record_type_name>_main.cpp` can provide the following functionality:

- *compressor* routine: it stores the data inside the container under a specified dataset name
- *decompressor* routine: it retrieves a dataset from the container by name
- *extractor* routine: it extracts from the container a contiguous range of the set of records belonging to a sorted dataset
- *transformator* routine: it applies a user-specified transformation on the records belonging to a dataset.

## 7.1 Application interface class

The hierarchy of *CARGO* application interface classes (defined in `<cargo/type/IApp.h>`) is presented in *CARGO application interfaces hierarchy*.

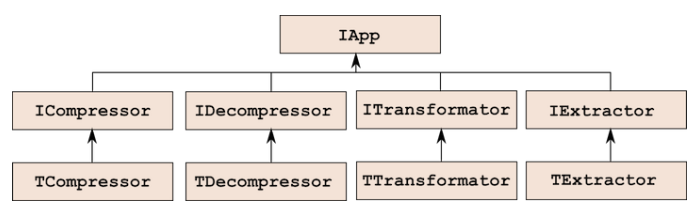

Fig. 7.1: *CARGO* application interfaces hierarchy

### 7.1.1 # IApp

The `IApp` base class provides a simple interface for developing applications based on *CARGO* library. In addition to the virtual destructor, it contains 2 pure virtual methods which need to be implemented by the derived classes:

```
1  class IApp
2  {
3  public:
4    IApp();
5    virtual ~IApp();
6
7    virtual int Run(int argc_, const char* argv_[]) = 0;
8    virtual void Usage() = 0;
9  };
```

The virtual method:

```
int Run(int argc_, const char* argv_[])
```

should contain the implementation of the application logic alongside with the parameters parsing – `argc_` and `argv_` are the passed command-line parameters.

Another virtual method:

```
void Usage()
```

should contain displaying the application usage information or help.

### 7.1.2 # ISpecApp

From `IApp` interface, the further specialized classes – `ICompressorApp`, `IDecompressorApp`, `ITransformatorApp` and `IExtractorApp`, named in general `ISpecApp`, are derived implementing own specific functionality; they share the common concept:

```
1  class ISpecApp : public IApp
2  {
3  public:
4    ISpecApp();
5    ~ISpecApp();
6
7    int Run(int argc_, const char* argv_[]);
8    void Usage();
9
10 protected:
11   struct InputArgs
12   {
13     // ...
14   };
15
16   virtual bool ProcessArgs(InputArgs& args_, int argc_, const char* argv_[]);
17   virtual int RunInternal(const InputArgs& args_);
18
19   virtual IUserDataProcessorProxy* CreateDataProcessor(const InputArgs& args_) = 0;
20 };
```

The specified class interface, derived from `IApp` class, implements the function:

```
int Run(int argc_, const char* argv_[])
```

which functionality is split into two smaller virtual functions for possible future overloading by the derived classes. The first one:

```
bool ProcessArgs(InputArgs& args_, int argc_, const char* argv_[])
```

implements parsing and processing of the command line arguments, saving them into passed `args_` parameter of the specified `InputArgs` type. The second one:

```
int RunInternal(const InputArgs& args_)
```

implements the application-specific logic based on *CARGO* library. In addtion, the function:

```
type::IUserDataProcessorProxy* CreateDataProcessor(const InputArgs& args_) = 0
```

is a pure virtual function, playing a role of a placeholder, which implementation is to provided by the further derived classes i.e. `TCompressor`, `TDecompressor`, `TTransformator` and `TExtractor` or named in general – `TSpecApp`.

### 7.1.3 # TSpecApp

The aim of `TSpecApp` class layer is to provide a record data-type and operation specialized interface, from which the final application can be created. The general interface of the specialized application template is as follows:

```cpp
template < class _TDataRecordType,
           class _TRecordsParser,
           class _TRecordsTransformator,
           class _TRecordsKeyGenerator
         >
class TSpecApp : public ISpecApp
{
protected:
  typedef _TDataRecordType DataRecordType;
  typedef typename DataRecordType::UserDataType DataType;
  typedef TUserDataProcessorProxy< DataRecordType,
                                   _TRecordsParser,
                                   _TRecordsTransformator,
                                   _TRecordsKeyGenerator
                                 > UserDataProcessor;

public:
  TSpecApp();
  ~TSpecApp();

  IUserDataProcessorProxy* CreateDataProcessor(const InputArgs &args_)
  {
      return new UserDataProcessor(args_.generateKey, args_.applyTransform);
  }
};
```

The specified `TSpecApp` template parameters are:

- `class _TDataRecordType` – *TypeAPI* record data type definition (see: *The Type API*),
- `class _TRecordsParser` – the user-specified records data type parser (see: *Record type parser*),
- `class _TRecordsTransformator` – an optional, user-specified record data type transform (see: *Record type transform*),
- `class _TRecordsKeyGenerator` – an optional, user-specified record data type key generator (see: *Record type key generator*).

More information about the usage of the specialized application templates are described below in the corresponding sub chapter *Running the application*.

## 7.2 Application template file

The generated *application template file* – `<record_type_name>_main.cpp`, by `cargo_translate` tool (see: *Translator tool*) contains the implementations of a sample record type *compressor*, *decompressor*, (optionally) *transformator* and (optionally) *extractor* sub-applications. The concept of the such generated standalone application using *CARGO* sub-applications is as follows:

```cpp
typedef app::TCompressorApp<    _record_Type,
                                _record_Parser,
                                _record_Transform,
                                _record_KeyGenerator
                         >      _record_CompressorApp;

typedef app::TDecompressorApp<  _record_Type,
                                _record_Parser,
                                _record_Transform,
                                _record_KeyGenerator
                         >      _record_DecompressorApp;

typedef app::TTransformatorApp< _record_Type,
                                _record_Parser,
                                _record_Transform,
                                _record_Keygenerator
                         >      _record_TransformatorApp;

typedef app::TExtractorApp<     _record_Type,
                                _record_Parser,
                                _record_Transform,
                                _record_KeyGenerator
                         >      _record_ExtractorApp;

int main(int argc_, const char* argv_[])
{
  if (argc_ > 1)
  {
     switch (argv_[1][0])
     {
     case 'c': return _record_CompressorApp().Run(argc_ - 1, argv_ + 1);
     case 'd': return _record_DecompressorApp().Run(argc_ - 1, argv_ + 1);
     case 't': return _record_TransformatorApp().Run(argc_ - 1, argv_ + 1);
     case 'e': return _record_ExtractorApp().Run(argc_ - 1, argv_ + 1);
     }
  }

  std::cout << "CARGO _record_ toolbox" << std::endl;
  std::cout << "usage:" << argv_[0] << " <c|d|t|e> [options]" << std::endl;
  return -1;
}
```

On the lines `1`, `6` and `11` the sample record type *compressor*, *decompressor*, *transformator* and *extraction* sub-application classes are defined, being derived respectively from `TCompressorApp`, `TDecompressorApp`, `TTransformatorApp` and `TExtractor` template class, where the specified template parameters were described in previous sub-chapter *# TSpecApp*.

On the line `26` the application entry point

```
int main(int argc_, const char* argv_[])
```

function is defined, where the number of input arguments and their description depend on the sub-application being selected. The first argument will be always the sub-application selector (`c|d|t|e` - line `40`), which is being checked on line `30` and launching respectively either the records *compressor*, *decompressor*, *transformator* or *extractor*.

## 7.3 Building the application

When a *CARGO* meta-language record definition is translated with `cargo_translate` (see: *Translator tool*), in addition to several C++ files a *Makefile* is also generated. This way, *CARGO* applications can be easily build: the *Makefile* contains all required compilation flags, paths and libraries – which makes it the most convenient and recommended method. Even more advanced examples can be created by tweaking the *Makefile*.

### 7.3.1 Build prerequisites

Before building *CARGO* applications from the generated template application files, the `CARGO_PATH` environment variable needs to be set, pointing to the root of the *CARGO* installation directory:

```
export CARGO_PATH=/path/to/cargo/directory/
```

As *CARGO* relies on several publicly-available compression libraries, the zlib (`libz`) and bzip2 (`libbz2`) libraries need to be present in the system when compiling.

Compiling *CARGO* applications will also require a compiler with *C++11* standard support (for multi–threading support) – by default the *gcc* compiler version *4.8* or above should be used.

### 7.3.2 Using generated Makefile

To build *CARGO* application using the generated *Makefile* file `<record_type_name>_Makefile.mk`, say

```
make -f <record_type_name>_Makefile.mk
```

A successful build will generate an executable named `cargo_<record_type_name>_toolkit`.

## 7.4 Running the application

A compiled application from the provided *CARGO* templates when run from the command line can display following message:

```
CARGO <record_type_name> toolkit
usage:./cargo_<record_type_name>_toolkit <c|d|t|e> [options]
```

where the possible switch launches sub-application related with the user-defined record data type:

- `c` – compressor,
- `d` – decompressor,
- `t` – transformator,

- e – extractor,

and [options] specify the available options depending on the selected sub-application switch (described below).

### 7.4.1 Compressor

The *compressor* sub-application stores the record type input data inside the containers; when launched from command line:

```
./cargo_<record_type_name>_toolkit c
```

displays the following message:

```
CARGO compressor sub-application
available options:
  -c <container>    - container file prefix
  -n <dataset>      - dataset name
 [-i <file>]        - input file, optional (default: stdin)
 [-t <threads>]     - thread count, optional (default: 1)
 [-b <size>]        - input file block size in MiB, optional (default: 64)
 [-a]               - apply records transform (if defined), optional (default: false)
 [-g]               - generate records key (if defined), optional (default: false)
 [-s]               - skip 1st record field as a generated key (used with -a),
                      optional (defaulf: false)
 [-h]               - help - display *this* message
```

The available options are:

> **-c container, --container=container**   *container* file name prefix (e.g. `<container_file>.cargo-`)
>
> **-n name, --dataset_name=name**   *name* of the dataset to be stored
>
> **-i file, --input_file=file**   input *file* name (optional)
>
> **-t n, --threads_num=n**   the number of processing threads (optional)
>
> **-b size, --block-size=size**   the block *size* (in MiB) of the input buffer (optional)
>
> **-a, --apply-transform**   apply records forward transformation (if defined by user, optional)
>
> **-g, --generate-key**   generate a records key for future ranged queries (if defined by user, optional)
>
> **-s, --skip-key**   skip the 1st field as a generated key when parsing records (requires -a, optional)
>
> **-h, --help**   display help message

*Compressor* processes the user defined type input data (either from file or by default – from *standard input*) and stores it in the container under the specific dataset name. To better control the performance, a maximum number of processing threads can be specified. The input block buffer size, depending on the input data type and size, should be picked relatively large (by default 64 MiB), as it influences the compression ratio – in principle, the larger, the better, at the expense of a bigger memory usage. In addition to compressing, the records forward transformations can be applied, but only if the operator has been previously defined by the user in `<record_type>_Transform` class (see: *Record type transform*) and compiled with the final compressor application. The similar concept applies to the records key generation (see: *Record type key generator*), which is used inside the blocks to identify the order of the processed records.

**Important:** *CARGO* applications do not implement sorting of the records internally. Thus, in case of storing the data with requested specified order and using the user provided *key generator*, the input data needs to be previously sorted – this can be done by using a combination of the standard linux tools incl. `sort`.

### 7.4.2 Decompressor

The *decompressor* sub-application retrieves the record type data from the containers; when launched from command line:

```
./cargo_<record_type_name>_toolkit d
```

displays the following message:

```
CARGO decompressor sub-application
available options:
 -c <container>      - container file prefix
 -n <dataset>        - dataset name
 [-o <file>]         - output file, optional (default: stdout)
 [-t <threads>]      - thread count, optional (default: 1)
 [-a]                - apply records transform (if defined), optional
 [-h]                - help - display *this* message
```

The available options are:

> **-c container, --container=container** *container* file name prefix (e.g. `<container_file>.cargo-`)
>
> **-n name, --dataset_name=name** *name* of the dataset to be read
>
> **-o file, --output_file=file** output *file* name (optional)
>
> **-t n, --threads_num=n** the number of processing threads (optional)
>
> **-a, --apply-transform** apply records backward transformation (if defined by user, optional)
>
> **-h, --help** display help message

*Decompressor* retrieves the user defined type data stored in the container outputting it either to a file or *standard output* (by default). In addition to decompression, the record data type backward transformation can be applied if it was previously defined by user in `<record_type>_Transform` class (see: *Record type transform*). The maximum number of processing threads can be also specified.

### 7.4.3 Transformator

The *transformator* sub-application performs the record type data transformations, applying the user specified transform (either forward or backward) on all of the input records and saving result to the output; when launched from command line:

```
./cargo_<record_type_name>_toolkit t
```

displays the following message:

```
CARGO transformator sub-application
available options:
 [-i <file>]         - input file, optional (default: stdin)
 [-o <file>]         - output file, optional (default: stdout)
 [-t <threads>]      - thread count, optional (default: 1)
 [-b <size>]         - input file block size in MiB, optional (default: 64)
```

```
[-a]                    - apply records transform (if defined), optional
[-r]                    - do reverse transform (requires -a), optional
[-g]                    - generate records key (if defined), optional
[-h]                    - help - display *this* message
```

Available options are:

>  **-i file, --input_file=file**   input *file* name, optional (default: stdin)
>
>  **-o file, --output_file=file**   output *file* name, optional (default: stdout)
>
>  **-t n, --threads_num=n**   the number of processing threads, optional (default: `1`)
>
>  **-b size, --block_size=size**   he block *size* (in MiB) of the input buffer, optional (default: `64`)
>
>  **-a, --apply-transform**   apply records transformation (if defined by user, optional)
>
>  **-r, --reverse**   do reverse (backward) transform (requires -a, optional)
>
>  **-g, --generate_key**   generate records key when processing (if defined by user, optional)
>
>  **-h, --help**   display help message

*Transformator* is the only one sub-application not operating directly on *CARGO* containers. It applies the user defined record data type transformation either forward or backward (see: *Record type transform*) and/or key generation (see: *Record type key generator*) on all the records – such generated key will be appended as an extra field at the beginning of the record. The input data is read either from a file or the *standard input* and saved either to a file or the *standard output*. The input buffer block size and number of processing threads control the overall performance.

### 7.4.4 Extractor

The *extractor* sub-application retrieves the record type data from the containers within a specified key range; when launched from command line:

```
./cargo_<record_type_name>_toolkit e
```

displays the following message:

```
CARGO extractor sub-application
available options:
  -c <container>     - container file prefix
  -n <dataset>       - dataset name
  -k <key_0>::<key_n> - records extract range from <key_0> to <key_n>
 [-o <file>]         - output file, optional (default: stdout)
 [-t <threads>]      - thread count, optional (default: 1)
 [-a]                - apply records transform (if defined), optional
 [-h]                - help - display *this* message
```

Available options are:

>  **-c container, --container=container**   *container* file name prefix (e.g. `<container_file>.cargo-`)
>
>  **-n name, --dataset_name=name**   *name* of the dataset to be read
>
>  **-k k_begin-k_end, --key=k_begin-k_end**   records extraction range from `k0` to `kn`
>
>  **-o file, --output_file=file**   output *file* name (optional)
>
>  **-t n, --threads_num=n**   the number of processing threads (optional)
>
>  **-a, --apply-transform**   apply records backward transformation (if defined by user, optional)

| **-h, --help** | display help message |

*Extractor* retrieves the user defined type data stored in the container from the specified dataset and within the specified key range – the key generation function must be earlier defined by the user (see: *Record type key generator*). In addition to extraction, the record data type backward transformation can be applied if it was previously defined by user in `<record_type>_Transform` class (see: *Record type transform*). The sub-application outputs the data either to a file or *standard output* (by default). The maximum number of processing threads can be also specified.

**Important:** The format of specified key in the extraction range from `k_begin` to `k_end` must match the key format defined by user in the `<record_type>_KeyGenerator` class.

### 7.4.5 Example usages

- Using the compiled SAM record type toolkit `cargo_samrecord_toolkit`, store the `HG00306.sam` input SAM file in the `HG00` container as a dataset named `HG00306`:

```
cargo_samrecord_toolkit c -c HG00 -n HG00306 -i HG00306.sam
```

- Retrieve the SAM dataset `HG00306` from container `HG00` using 8 processing threads and saving the output as `HG00306.out.sam`:

```
cargo_samrecord_toolkit d -c HG00 -n HG00306 -o HG00306.out.sam -t 8
```

- Decompress the `HG00306.bam` BAM file using *SAMTools* and stream output SAM stream to *CARGO* container `HG00` under `HG00306` dataset name, using 8 processing threads and applying user-defined records transformation:

```
samtools view -h HG00306.bam | cargo_samrecord_toolkit c -c HG00 \
     -n HG00306 -t 8 -a
```

- Apply user-specified records forward transformation and key generator to the `HG00306.bam` file, sort the records and store the streamed data in the `HG00` container as a dataset named `HG00-sorted`:

```
cargo_samrecord_toolkit t -a -g -i HG00306.bam | sort -S 2G \
     | cargo_samrecord_toolkit c -c HG00 -n HG00306
```

CHAPTER

EIGHT

EXAMPLES

All the examples presented in this chapter can be found in the `cargo/examples/` subdirectory of the standard *CARGO* distribution. Their compiled versions are available in the `cargo/examples/bin` subdirectory.

When creating a specific file format compressor or decompressor, the general workflow is as follows:

1. Define the record data type in *CARGO* metalanguage
2. Translate the definition using `cargo_translate` tool – a set of user files will be generated
3. Fill the generated record data type parser template file
4. Optionally, define the record data type transformation (forward and backward)
5. Optionally, define the record data type key generator (for sorting order)
6. Build the application using the generated *Makefile* file.

**Important:** Before building the examples make sure that build prerequisites described in subchapter *Build prerequisites* are met.

## 8.1 FASTQ

The *FASTQ* example shows a simple proof-of-concept whereby with a few lines of code we create a compressor specialized to the FASTQ format . Results are comparable to those produced by state-of-the-art *FASTQ* format compressors.

The sources for this example are available in the `cargo/examples/fastq/fastq-simple` subdirectory of the standard *CARGO* distribution. A precompiled binary can be found in the directory `cargo/examples/bin/cargo_fastqrecord_toolkit-simple`.

**Tip:** Should one prefer to skip the following steps and build the FASTQ example straight away, a `Makefile` file is provided in the main directory together with a FASTQ record type definition in *CARGO* metalanguage (`FastqRecord.cargo`) and a complete record parser (`FastqRecord_Parser.bak`).

### 8.1.1 General format description

The FASTQ format is an ASCII text-based format useful to store biological sequences together with their quality score values. A sample record looks like the following:

```
1  @SRR001666.1 071112_SLXA-EAS1_s_7:5:1:817:345 length=36
2  GGGTGATGGCCGCTGCCGATGGCGTCAAATCCCACC
3  +
4  IIIIIIIIIIIIIIIIIIIIIIIIIIIIII9IG9IC
```

while a general record contains:

- *read id*: an identifier of the read starting after the `@` symbol
- *sequence*: a sequence of nucleotides encoded using `AGCTN` letters
- *plus*: a control line, optionally containing a repetition of the read identifier
- *quality*: a Phred sequencing quality score of the sequence.

**Note:** In the *sequence* field other IUPAC symbols can also appear and, in special cases, nucleotide sequences using a lowercase notation `agctn`.

### 8.1.2 Record type definition

In our example, the FASTQ record is defined in *CARGO metalanguage* as follows:

```
1  FastqRecord = {
2    tag = string;
3    seq = string;
4    qua = string
5  }
6
7  @record FastqRecord
```

Such defined record type consists only of 3 `string` fields: *read id* `tag`, *sequence* `seq` and *quality* `qua`, omitting the optional *plus* control field. No explicit compression methods were selected, so the default ones (see: *Default compression methods*) will be applied, while storing the data inside container.

### 8.1.3 Translation

Once the record type definition in *CARGO* meta-language has been saved as a `FastqRecord.cargo` file, the definition can be used as an input for the translation step. Running the command

```
cargo_translate -i FastqRecord.cargo
```

will generate the following files:

- `FastqRecord.h`: C++ definition of the `FastqRecord` user record type
- `FastqRecord_Parser.h`: C++ parser template for the `FastqRecord` C++ user record - will need to be completed by the user
- `FastqRecord_Type.h`: C++ *TypeAPI*-based record type specification for the `FastqRecord` record type (for subsequent internal use, it does not need to be opened or modified by the user)
- `FastqRecord_main.cpp`: template file containing compressor/decompressor applications writing/reading a stream of `FastqRecord` records to/from containers
- `FastqRecord_Makefile.mk`: *Makefile* template file to build such applications.

### 8.1.4 Translated record type definition

The translated C++ *FASTQ* record definition, contained in `FastqRecord.h`, is as follows:

```cpp
struct FastqRecord {
  std::string tag;
  std::string seq;
  std::string qua;
};
```

and its corresponding *TypeAPI* definition, contained in `FastqRecord_Type.h`, is

```cpp
struct __Compound1
{
  typedef FastqRecord UserDataType;
  static const uint32 FieldCount = 3;

  typedef type::TStringType<streams::CompressionText, streams::BlockSizeText> Type1;
  static typename Type1::UserDataType& Get1(UserDataType& data_)
  {
    return data_.tag;
  }
  static const typename Type1::UserDataType& Get1(const UserDataType& data_)
  {
    return data_.tag;
  }

  typedef type::TStringType<streams::CompressionText, streams::BlockSizeText> Type2;
  static typename Type2::UserDataType& Get2(UserDataType& data_)
  {
    return data_.seq;
  }
  static const typename Type2::UserDataType& Get2(const UserDataType& data_)
  {
    return data_.seq;
  }

  typedef type::TStringType<streams::CompressionText, streams::BlockSizeText> Type3;
  static typename Type3::UserDataType& Get3(UserDataType& data_)
  {
    return data_.qua;
  }
  static const typename Type3::UserDataType& Get3(const UserDataType& data_)
  {
    return data_.qua;
  }
};
typedef type::TStructType<__Compound1> FastqRecord_Type;
```

### 8.1.5 Records parser

The generated parser file `FastqRecord_Parser.h` contains the skeleton of the record data type parser class. Some of its functions need to be implemented by the user (for more information see: *Record type parser*). After this step has been completed, the code will look as follows:

```cpp
using namespace type;

```

```cpp
3   class FastqRecord_Parser
4   {
5   public:
6     static void SkipToEndOfHeader(io::MemoryStream& stream_)
7     {
8       // there's no header in the FASTQ file format - member function left empty
9     }
10
11    static void SkipToEndOfRecord(io::MemoryStream& stream_)
12    {
13      // skip to the new line
14      FieldParser::SkipNextField(stream_, '\n');
15
16      byte nextByte = 0;
17      while (FieldParser::PeekNextByte(stream_, nextByte))
18      {
19        // beginning of next record - found read id line?
20        if (nextByte == '@')
21          return;
22        FieldParser::SkipNextField(stream_, '\n');
23      }
24    }
25
26    static bool ReadNextRecord(io::MemoryStream& stream_, FastqRecord& record_)
27    {
28      bool result = FieldParser::ReadNextField(stream_, record_.tag, '\n');
29      result &= FieldParser::ReadNextField(stream_, record_.seq, '\n');
30      result &= FieldParser::SkipNextField(stream_, '\n');  // skip '+' line
31      result &= FieldParser::ReadNextField(stream_, record_.qua, '\n');
32      return result;
33    }
34
35    static bool WriteNextRecord(io::MemoryStream& stream_, FastqRecord& record_)
36    {
37      bool result = FieldParser::WriteNextField(stream_, record_.tag, '\n');
38      result &= FieldParser::WriteNextField(stream_, record_.seq, '\n');
39      result &= FieldParser::WriteNextField(stream_, "+", '\n');
40      result &= FieldParser::WriteNextField(stream_, record_.qua, '\n');
41      return result;
42    }
43  };
```

On line 4 an empty `SkipToEndOfHeader` function was defined – the FASTQ format does not specify any file header, so the body of the function has been intentionally left empty.

As the *FASTQ* record cannot be uniquely identified just by analyzing a single line (it's defined by 4 consecutive lines), the function `SkipToEndOfRecord` needs to:

1. Skip until the end of the current text line, passing a newline separator `\n`
2. Peek the next symbol from the *memory stream*
3. Check whether the next line starts with `@` symbol and if so – exit, otherwise: (4)
4. Skip the next line and go to (2).

Once the *memory stream* is positioned at the beginning of the next *FASTQ* record, reading and parsing it can be done straightforward (starting from line 26), while unparsing the *FASTQ* record and writing it to a *memory stream* goes in a similar way (starting from line 35).

### 8.1.6 Building

The generated *Makefile* `FastqRecord_Makefile.mk` contains the information required in order to build a *FASTQ* toolkit taking the options specified by user into account. Running *gnu* `make` in the example's directory as

```
make -f FastqRecord_Makefile.mk
```

will produce the executable `cargo_fastqrecord_toolkit`.

### 8.1.7 Running

The generated application uses the default command line parameters in order to select compressor (`c`) and decompressor (`d`) sub-routines (details are described in *Application templates*).

- Using the compiled *FASTQ* record type toolkit `cargo_fastqrecord_toolkit`, store the `ERR217195_2.fastq` input FASTQ file in the `C_FASTQ` container as a dataset named `ERR217195`:

  ```
  cargo_fastqrecord_toolkit c -c C_FASTQ -n ERR217195 -i ERR217195_2.fastq
  ```

- Retrieve the previously stored *FASTQ* dataset `ERR217195` from container `C_FASTQ` using 8 processing threads and saving the output as `ERR217195_2.out.fastq`:

  ```
  cargo_fastqrecord_toolkit d -c C_FASTQ -n ERR217195 -o ERR217195_2.out.fastq -t 8
  ```

- Print info for the `ERR217195` dataset stored in container `C_FASTQ`:

  ```
  cargo_tool --print-dataset --dataset-name=ERR217195 --container-file=C_FASTQ
  ```

- Remove the `ERR217195` dataset from the `C_FASTQ` container:

  ```
  cargo_tool --remove-dataset --dataset-name=ERR217195 --container-file=C_FASTQ
  ```

## 8.2 SAM-STD

The *SAM-STD* example shows a simple proof-of-concept whereby with a few lines of code we create a compressor specialized to the SAM . The compression ratio and speed achieved by this example outperform the widely used *BAM* format as implemented by *SAMTools*. In addition, with minor parameter tweakings of the record type definition one can achieve a compression ratio comparable to that of current state-of-the-art *SAM* compression tools.

The sources for this example are available in the `cargo/examples/sam/sam-std` subdirectory of the standard *CARGO* distribution. A precompiled binary can be found in the directory `cargo/examples/bin/cargo_samrecord_toolkit-std`.

**Tip:** Should one prefer to skip the following steps and build the *SAM-STD* example straight away, a `Makefile` file is provided in the main directory together with a SAM record type definition in *CARGO* metalanguage (`SamRecord.cargo`), a complete record parser (`SamRecord_Parser.bak`) and a record transformation function (`SamRecord_Transform.bak`).

### 8.2.1 General format description

SAM – Sequence Alignment/Map format is a tab-delimited text format used by a variety of bioinformatics tools to store sequence alignment information in a record-like way. The file consist of an optional header block (having lines starting with the `@` symbol) followed by an arbitrary number of consecutive *SAM* records, each consisting of the fields described in table *SAM format field description*. In addition to the standard 11 *SAM* format fields, there are also optional fields defined as TAG:TYPE:VALUE, where TAG is a two-character string matching /[A-Za-z][A-Za-z0-9]/ expression – they are presented in table *SAM format optional fields*.

Table 8.1: SAM format field description

| Col | Field | Type | Regexp/Range | Brief description |
|---|---|---|---|---|
| 1 | QNAME | String | [!-?A-~]{1,255} | Query template NAME |
| 2 | FLAG | Int | [0, 2^16 - 1] | bitwise FLAG |
| 3 | RNAME | String | \*|[!-()+-<>-~][!-~]* | Reference sequence NAME |
| 4 | POS | Int | [0, 2^31 - 1] | 1-based leftmost mapping POSition |
| 5 | MAPQ | Int | [0, 2^8 - 1] | MAPping Quality |
| 6 | CIGAR | String | \*|([0-9]+[MIDNSHPX=])+ | CIGAR string |
| 7 | RNEXT | String | \*|=|[!-()+-<>-~][!-~]* | Ref. name of the mate/next read |
| 8 | PNEXT | Int | [0, 2^31 - 1] | Position of the mate/next read |
| 9 | TLEN | Int | [-2^31 + 1, 2^31 - 1] | observed Template LENgth |
| 10 | SEQ | String | \*|[A-Za-z=.]+ | segment SEQuence |
| 11 | QUAL | String | [!-~]+ | ASCII of Phred-scaled base QUALity+33 |

Table 8.2: SAM format optional fields

| Type | Regular expression matching "VALUE" | Description |
|---|---|---|
| A | [!-~] | Printable character |
| i | [-+]?[0-9]+ | Singed 32-bit integer |
| f | [-+]?[0-9]*\.?[0-9]+([eE][-+]?[0-9]+)? | Single-precision floating number |
| Z | [ !-~]+ | Printable string, including space |
| H | [0-9A-F]+ | Byte array in the Hex format |
| B | [cCsSiIf](,[-+]?[0-9]*\.?[0-9]+([eE][-+]?[0-9]+)?)+ | Integer or numeric array |

### 8.2.2 Record type definition

The simple version of *SAM* record in *CARGO metalanguage* is defined in the following way:

```
1  SamRecord = {
2    qname = string;
3    flag = uint^16;
4    rname = string;
5    pos = uint^32;
6    mapq = uint^8;
7    cigar = string;
8    next = string;
9    pnext = int^32;
10   tlen = int^32;
11   seq = string;
12   qua = string;
13   opt = string;
14 } :
15   .qname.Pack = PPMdL4,
16   .flag.Pack = PPMdL4,
17   .rname.Pack = GzipL2,
18   .pos.Pack = LZMAL1,
19   .mapq.Pack = PPMdL1,
20   .cigar.Pack = PPMdL4,
21   .next.Pack = GzipL2,
```

```
22      .pnext.Pack = LZMAL1,
23      .tlen.Pack = PPMdL4,
24      .seq.Pack = LZMAL1,
25      .qua.Pack = PPMdL1,
26      .opt.Pack = Bzip2L4;
27
28   @record SamRecord
```

The type names of the presented *SAM* record fields correspond to their definition presented in table *SAM format field description*. However, the optional fields are treated as a one single long string. The specified *compression methods* (using the default *block sizes*) were selected through a set of experiments and provide a satisfactory compression ratio and performance.

### 8.2.3 Translation

The record definition was saved under `SamRecord.cargo` file name; after translating the record definition:

```
cargo_translate -t -i SamRecord.cargo
```

the following files will be created:

- `SamRecord.h`: the C++ record type definition
- `SamRecord_Parser.h`: user record parser template, which is to be modified by user
- `SamRecord_Transform.h`: user record transformation template, which is to be modified by user
- `SamRecord_main.cpp`: application template file
- `SamRecord_Makefile.mk`: the application build script file
- `SamRecord_Type.h`: internal record type definition in *TypeAPI*.

### 8.2.4 Translated record type definition

The translated C++ record definition `SamRecord.h`:

```cpp
1  struct SamRecord {
2    std::string qname;
3    uint16 flag;
4    std::string rname;
5    uint32 pos;
6    uint8 mapq;
7    std::string cigar;
8    std::string next;
9    int32 pnext;
10   int32 tlen;
11   std::string seq;
12   std::string qua;
13   std::string opt;
14 };
```

The corresponding record type definition in *TypeAPI* due to it's length and for clarity is skipped here – it can be found in the generated `SamRecord_Type.h` file in the example subdirectory `cargo/examples/sam/sam-std`.

### 8.2.5 Records parser

The generated parser file `SamRecord_Parser.h` contains the skeleton of the user record data type parser class which functions are to be implemented by the user (for more see: *Record type parser*). The code for the complete parser class is presented on the snippet below:

```
1  using namespace type;
2
3  class SamRecord_Parser
4  {
5    static const char FieldSeparator = '\t';
6    static const char RecordSeparator = '\n';
7    static const char HeaderStartSymbol = '@';
8
9  public:
10   static void SkipToEndOfHeader(io::MemoryStream& stream_)
11   {
12     FieldParser::SkipBlock(stream_, HeaderStartSymbol, RecordSeparator);
13   }
14
15   static void SkipToEndOfRecord(io::MemoryStream& stream_)
16   {
17     FieldParser::SkipNextField(stream_, RecordSeparator);
18   }
19
20   static bool ReadNextRecord(io::MemoryStream& stream_, SamRecord& record_)
21   {
22     bool result = true;
23     result &= FieldParser::ReadNextField(stream_, record_.qname, FieldSeparator);
24     result &= FieldParser::ReadNextField(stream_, record_.flag, FieldSeparator);
25     result &= FieldParser::ReadNextField(stream_, record_.rname, FieldSeparator);
26     result &= FieldParser::ReadNextField(stream_, record_.pos , FieldSeparator);
27     result &= FieldParser::ReadNextField(stream_, record_.mapq , FieldSeparator);
28     result &= FieldParser::ReadNextField(stream_, record_.cigar, FieldSeparator);
29     result &= FieldParser::ReadNextField(stream_, record_.next, FieldSeparator);
30     result &= FieldParser::ReadNextField(stream_, record_.pnext, FieldSeparator);
31     result &= FieldParser::ReadNextField(stream_, record_.tlen , FieldSeparator);
32     result &= FieldParser::ReadNextField(stream_, record_.seq, FieldSeparator);
33     result &= FieldParser::ReadNextField(stream_, record_.qua, FieldSeparator);
34     result &= FieldParser::ReadNextField(stream_, record_.opt, RecordSeparator);
35     return result;
36   }
37
38   static bool WriteNextRecord(io::MemoryStream& stream_, SamRecord& record_)
39   {
40     bool result = true;
41     result &= FieldParser::WriteNextField(stream_, record_.qname, FieldSeparator);
42     result &= FieldParser::WriteNextField(stream_, record_.flag, FieldSeparator);
43     result &= FieldParser::WriteNextField(stream_, record_.rname, FieldSeparator);
44     result &= FieldParser::WriteNextField(stream_, record_.pos, FieldSeparator);
45     result &= FieldParser::WriteNextField(stream_, record_.mapq, FieldSeparator);
46     result &= FieldParser::WriteNextField(stream_, record_.cigar, FieldSeparator);
47     result &= FieldParser::WriteNextField(stream_, record_.next, FieldSeparator);
48     result &= FieldParser::WriteNextField(stream_, record_.pnext, FieldSeparator);
49     result &= FieldParser::WriteNextField(stream_, record_.tlen, FieldSeparator);
50     result &= FieldParser::WriteNextField(stream_, record_.seq, FieldSeparator);
51     result &= FieldParser::WriteNextField(stream_, record_.qua, FieldSeparator);
52     result &= FieldParser::WriteNextField(stream_, record_.opt, RecordSeparator);
```

```
53      return result;
54    }
55 };
```

As in the *SAM* format definition – a sample *SAM* file can contain an optional header, so a function `SkipToEndOfHeader` needs to be defined. The function implements skipping of the blocks of text, where each line starts with @ symbol and finish with a newline \n character (line `12`).

Skipping the position in the *memory stream* until the end of the currently parsed *SAM* record is trivial, as it is only requires to skip until the end of the currently read text line (line `17`).

The actual parsing the next *SAM* record starts in line `20` (and ``40) and it goes straightforward, reading (or writing) the concurrent fields in the specified in the SAM format description.

### 8.2.6 Record transform

In this example, as a transformation function, the Illumina quality scores reduction scheme was implemented based on *Q-scores mapping using Illumina quality scores reduction scheme*.

```
1  class SamRecord_Transform
2  {
3    // assume and use hardcoded quality offset
4    static const uint32 QualityOffset = 33;
5
6  public:
7    static void TransformForward(SamRecord& record_)
8    {
9      static const char qualityTranslationTable[64] = {
10                            0, 0, 6, 6, 6,    6, 6, 6, 6, 6,    // 0x
11                           15,15,15,15,15,   15,15,15,15,15,    // 1x
12                           22,22,22,22,22,   27,27,27,27,27,    // 2x
13                           33,33,33,33,33,   37,37,37,37,37,    // 3x
14                           40,40,40,40,40,   40,40,40,40,40,    // 4x
15                           40,40,40,40,40,   40,40,40,40,40,    // 5x
16                           40,40,40,40   };                     // 6x -- 64
17
18      for (uint32 i = 0; i < record_.qua.size(); ++i)
19      {
20        uint32 q = record_.qua[i] - QualityOffset;
21        record_.qua[i] = qualityTranslationTable[q] + QualityOffset;
22      }
23    }
24
25    static void TransformBackward(SamRecord& record_)
26    {
27
28    }
29 };
```

To implement the quality transformation function, a lookup table with quality values was created in order to skip multiple comparison of quality value range – the simplified version with branches was implemented in *FASTQ example* subchapter.

CARGO Documentation, Release 0.7rc-internal

### 8.2.7 Building

The generated build script file `SamRecord_Makefile.mk` contains the required recipes for building *SAM* toolkit with options specified by user. Running from the example's directory *gnu* `make`:

```
make -f SamRecord_Makefile.mk
```

will produce the executable `cargo_samrecord_toolkit`.

### 8.2.8 Running

The generated application uses the default command line parameters for compressor `c` and decompressor `d` subapplications, which in detail were described in *Application templates* chapter.

- Using the compiled *SAM* record type toolkit `cargo_samrecord_toolkit` store the `C_SAM306.sam` input *SAM* file in `C_SAM` container under dataset name `C_SAM306`:

    ```
    cargo_samrecord_toolkit c -c C_SAM -n C_SAM306 -i C_SAM306.sam
    ```

- Retrieve the *SAM* dataset `C_SAM306` from container `C_SAM` using 8 processing threads saving the output as `C_SAM306.out.sam`:

    ```
    cargo_samrecord_toolkit d -c C_SAM -n C_SAM306 -o C_SAM306.out.sam -t 8
    ```

- Store the `C_SAM306.sam` input SAM file in `C_SAM` container under `C_SAM306` dataset name applying additionally records transformation with 256 MiB as an input block buffer and 8 processing threads:

    ```
    cargo_samrecord_toolkit c -c C_SAM -n C_SAM306 -i C_SAM306.sam -a -t 8 -b 256
    ```

- Decompress the `C_SAM306.bam` using *SAMTools* and apply the user transformation (in this case *Illumina quality scores reduction*) saving the output to `C_SAM306.out` file:

    ```
    samtools view -h C_SAM306.bam | cargo_samrecord_toolkit t -o C_SAM306.out -a
    ```

- Print the `C_SAM306` dataset info from the `C_SAM` container:

    ```
    cargo_tool --print-dataset --dataset-name=C_SAM306 --container-file=C_SAM
    ```

- Remove the `C_SAM306` dataset from the `C_SAM` container:

    ```
    cargo_tool --remove-dataset --dataset-name=C_SAM306 --container-file=C_SAM
    ```

## 8.3 SAM-EXT

The *SAM-EXT* example is an extended version of the previous *SAM-STD*, which implements a tokenization of *SAM* optional fields with an additional Illumina quality scores reduction scheme transformation.

The sources for this example are available in the `cargo/examples/sam/sam-ext` subdirectory of the standard *CARGO* distribution. A precompiled binary can be found in the directory `cargo/examples/bin/cargo_samrecord_toolkit-ext`.

**Tip:** Should one prefer to skip the following steps and build the *SAM-EXT* example straight away, a `Makefile` file is provided in the main directory together with a SAM record type definition in *CARGO* metalanguage (`SamRecord.cargo`), a complete record parser (`SamRecord_Parser.bak`) and a record transformation function (`SamRecord_Transform.bak`).

### 8.3.1 Record type definition

The extended version of *SAM* record in *CARGO metalanguage* differs only in optional fields representation and is defined the following way:

```
1  OptionalValue = [
2          intValue = int^32;
3          charValue = char;
4          stringValue = string;
5  ] :
6  .intValue.Pack = Bzip2L4,
7  .charValue.Pack = GzipL2,
8  .stringValue.Pack = Bzip2L4;
9
10 OptionalField = {
11         tag = string;
12         value = OptionalValue;
13 } :
14 .tag.Pack = PPMdL4;
15
16 SamRecord = {
17         qname = string;
18         flag = uint^16;
19         rname = string;
20         pos = uint^32;
21         mapq = uint^8;
22         cigar = string;
23         next = string;
24         pnext = int^32;
25         tlen = int^32;
26         seq = string;
27         qua = string;
28         opt = OptionalField array;
29 } :
30 .qname.Pack = PPMdL4,   .qname.Block = 8M,
31 .flag.Pack = PPMdL4,    .flag.Block = 4M,
32 .rname.Pack = GzipL2,
33 .pos.Pack = LZMAL1,     .pos.Block = 4M,
34 .mapq.Pack = PPMdL1,    .mapq.Block = 8M,
35 .cigar.Pack = PPMdL4,   .cigar.Block = 8M,
36 .next.Pack = GzipL2,
37 .pnext.Pack = Bzip2L4,  .pnext.Block = 4M,
38 .tlen.Pack = PPMdL4,    .tlen.Block = 4M,
39 .seq.Pack = LZMAL1,     .seq.Block = 8M,
40 .qua.Pack = PPMdL1,     .qua.Block = 8M;
41
42 @record SamRecord
```

The types of the presented *SAM* record fields correspond to their definition presented in table *SAM format field description* with optional fields following the simplified definition presented in *SAM format optional fields*. In this example, the optional fields are defined as an *array* of tokens of *OptionalField* type, where each token contains a *tag* identifying the data with it's value of *OptionalValue* union type.

The fields compression methods with block sizes were selected through a set of experiments and provide a satisfactory compression ratio and performance.

**CARGO Documentation, Release 0.7rc-internal**

### 8.3.2 Translation

The translation process follows a similar path defined in the previous *SAM-STD* example. Translating *SAM* record definition saved in `SamRecord.cargo`:

```
cargo_translate -t -i SamRecord.cargo
```

will generate the following files:

- `SamRecord.h`: a C++ record type definition
- `SamRecord_Parser.h`: the record parser template, which is to be modified by user
- `SamRecord_Transform.h`: the record transformation template, which is to be modified by user
- `SamRecord_main.cpp`: the application template file
- `SamRecord_Makefile.mk`: the application build script file
- `SamRecord_Type.h`: the internal record type definition in *TypeAPI*.

### 8.3.3 Translated record type definition

The translated into C++ record definition `SamRecord.h` is as follows:

```cpp
struct OptionalValue {
  uint32 __kind;
  int32 intValue;
  char charValue;
  std::string stringValue;
};

struct OptionalField {
  std::string tag;
  OptionalValue value;
};

struct SamRecord {
  std::string qname;
  uint16 flag;
  std::string rname;
  uint32 pos;
  uint8 mapq;
  std::string cigar;
  std::string next;
  int32 pnext;
  int32 tlen;
  std::string seq;
  std::string qua;
  std::vector<OptionalField > opt;
};
```

The corresponding record type definition in *TypeAPI*, due to it's length and for clarity, is skipped here and can be found in the generated `SamRecord_Type.h` in the example's subdirectory `cargo/examples/sam/sam-ext`.

### 8.3.4 Records parser

The generated parser file `SamRecord_Parser.h` contains the skeleton of the user record data type parser class which functions are to be implemented by the user (for more see: *Record type parser*). The code for such filled parser class is presented below:

```cpp
using namespace type;

class SamRecord_Parser
{
  static const char FieldSeparator = '\t';
  static const char RecordSeparator = '\n';
  static const char HeaderStartSymbol = '@';

  enum OptionalValueEnum
  {
    IntValueType = 1,
    CharValueType = 2,
    StringValueType = 3
  };

public:
  static void SkipToEndOfRecord(io::MemoryStream& stream_)
  {
    FieldParser::SkipNextField(stream_, RecordSeparator);
  }

  static void SkipToEndOfHeader(io::MemoryStream& stream_)
  {
    FieldParser::SkipBlock(stream_, HeaderStartSymbol, RecordSeparator);
  }

  static bool ReadNextRecord(io::MemoryStream& stream_, SamRecord& record_)
  {
    bool result = true;

    // parse standard fields
    //
    result &= FieldParser::ReadNextField(stream_, record_.qname, FieldSeparator);
    result &= FieldParser::ReadNextField(stream_, record_.flag, FieldSeparator);
    result &= FieldParser::ReadNextField(stream_, record_.rname, FieldSeparator);
    result &= FieldParser::ReadNextField(stream_, record_.pos, FieldSeparator);
    result &= FieldParser::ReadNextField(stream_, record_.mapq, FieldSeparator);
    result &= FieldParser::ReadNextField(stream_, record_.cigar, FieldSeparator);
    result &= FieldParser::ReadNextField(stream_, record_.next, FieldSeparator);
    result &= FieldParser::ReadNextField(stream_, record_.pnext, FieldSeparator);
    result &= FieldParser::ReadNextField(stream_, record_.tlen, FieldSeparator);
    result &= FieldParser::ReadNextField(stream_, record_.seq, FieldSeparator);
    result &= FieldParser::ReadNextField(stream_, record_.qua, FieldSeparator);

    // parse optional fields into tokens
    //
    record_.opt.clear();

    std::string rawOptString;
    result &= FieldParser::ReadNextField(stream_, rawOptString, RecordSeparator);
    if (!result)
      return false;

    uint32 iStart = 0;
    for (uint32 i = 0; i < rawOptString.length() + 1; ++i)
    {
      if (i == rawOptString.length() || rawOptString[i] == '\t')
```

```
58            {
59              ASSERT(i - iStart > 5);
60
61              uint32 sLen = i - iStart;
62              const char* s = &rawOptString[iStart];
63
64              OptionalField o;
65              o.tag.append(s, s + 2);
66
67              switch (s[3])
68              {
69                case 'A':  // printable character
70                  o.value.__kind = CharValueType;
71                  o.value.charValue = s[5];
72                  break;
73
74                case 'i':  // signed 32-bit integer
75                  o.value.__kind = IntValueType;
76                  int64 v;
77                  algo::ParseInt(v , s + 5, sLen - 5);
78                  o.value.intValue = (int32)v;
79                  break;
80
81                case 'Z':  // printable string
82                  {
83                    o.value.__kind = StringValueType;
84                    o.value.stringValue.assign(s + 5, s + sLen);
85                    break;
86                  }
87              }
88
89              record_.opt.push_back(o);
90              iStart = i + 1;
91            }
92        }
93
94        return result;
95      }
96
97      static bool WriteNextRecord(io::MemoryStream& stream_, SamRecord& record_)
98      {
99        bool result = true;
100
101       // write standard fields
102       //
103       result &= FieldParser::WriteNextField(stream_, record_.qname, FieldSeparator);
104       result &= FieldParser::WriteNextField(stream_, record_.flag, FieldSeparator);
105       result &= FieldParser::WriteNextField(stream_, record_.rname, FieldSeparator);
106       result &= FieldParser::WriteNextField(stream_, record_.pos, FieldSeparator);
107       result &= FieldParser::WriteNextField(stream_, record_.mapq, FieldSeparator);
108       result &= FieldParser::WriteNextField(stream_, record_.cigar, FieldSeparator);
109       result &= FieldParser::WriteNextField(stream_, record_.next, FieldSeparator);
110       result &= FieldParser::WriteNextField(stream_, record_.pnext, FieldSeparator);
111       result &= FieldParser::WriteNextField(stream_, record_.tlen, FieldSeparator);
112       result &= FieldParser::WriteNextField(stream_, record_.seq, FieldSeparator);
113       result &= FieldParser::WriteNextField(stream_, record_.qua, FieldSeparator);
114
115       if (!result)
```

```
116         return false;
117
118       // parse and merge optional fields from token to one string
119       //
120       std::string optString;
121       for (uint32 i = 0; i < record_.opt.size(); ++i)
122       {
123         optString.append(record_.opt[i].tag);
124         const OptionalValue& opt = record_.opt[i].value;
125         switch (opt.__kind)
126         {
127           case IntValueType:
128             optString.append(":i:");
129             optString.append(algo::ToString(opt.intValue));
130             break;
131           case CharValueType:
132             optString.append(":A:");
133             optString.push_back(opt.charValue);
134             break;
135           case StringValueType:
136             optString.append(":Z:");
137             optString.append(opt.stringValue);
138             break;
139           default:   ASSERT(0);
140         }
141         optString.push_back('\t');
142       }
143
144       // trim if needed and write
145       //
146       if (optString.length() > 0 && optString.at(optString.length() - 1) == '\t')
147         optString.erase(optString.end() - 1);
148       result &= FieldParser::WriteNextField(stream_, optString, RecordSeparator);
149
150       return result;
151     }
152 };
```

As in the previous example, the SAM file can contain an optional header, so a function `SkipToEndOfHeader` needs to be defined. The function implements skipping of the blocks of text, where each line starts with @ symbol and finish with a newline \n character (line 12).

Skipping the position in the *memory stream* until the end of the currently parsed SAM record is trivial, as it is only requires to skip until the end of the currently read text line (line 17).

The actual parsing the next SAM record starts in line 27 (and ``97) and it goes straightforward, reading (or writing) the concurrent fields in the specified in the SAM format description. From line 45 (and 118) the tokenization of the optional fields i.e. splitting or joining of the sub-fields takes place, parsing and saving the individual sub-fields into an optString array.

### 8.3.5 Record transform

As a transformation function, the *Illumina quality scores reduction* was implemented based on *Q-scores mapping using Illumina quality scores reduction scheme*, with the same code was presented in the previous example *Record transform*.

### 8.3.6 Building

The generated *Makefile* file `SamRecord_Makefile.mk` contains the required recipes for building SAM toolkit with options specified by user. Running from the example's directory gnu `make`:

```
make -f SamRecord_Makefile.mk
```

will produce the executable `cargo_samrecord_toolkit`.

### 8.3.7 Running

- Using the compiled *SAM* record type toolkit `cargo_samrecord_toolkit` store the `C_SAM306.sam` input SAM file in `C_SAM` container under dataset name `C_SAM306`:

```
cargo_samrecord_toolkit c -c C_SAM -n C_SAM306 -i C_SAM306.sam
```

- Retrieve the *SAM* dataset `C_SAM306` from container `C_SAM` using 8 processing threads saving the output as `C_SAM306.out.sam`:

```
cargo_samrecord_toolkit d -c C_SAM -n C_SAM306 -o C_SAM306.out.sam -t 8
```

- Store the `C_SAM306.sam` input *SAM* file in `C_SAM` container under `C_SAM306` dataset name applying additionally records transformation with 256 MiB as an input block buffer and 8 processing threads:

```
cargo_samrecord_toolkit c -c C_SAM -n C_SAM306 -i C_SAM306.sam -a -t 8 -b 256
```

- Decompress the `C_SAM306.bam` using *SAMTools* and apply the user transformation (in this case *Illumina quality scores reduction*) saving the output to `C_SAM306.out` file:

```
samtools view -h C_SAM306.bam | cargo_samrecord_toolkit t -o C_SAM306.out -a
```

- Print the `C_SAM306` dataset info from the `C_SAM` container:

```
cargo_tool --print-dataset --dataset-name=C_SAM306 --container-file=C_SAM
```

- Remove the `C_SAM306` dataset from the `C_SAM` container:

```
cargo_tool --remove-dataset --dataset-name=C_SAM306 --container-file=C_SAM
```

## 8.4 SAM-REF

The *SAM-REF* example is a more advanced version of the *SAM-EXT*, which implements a set of additional features, providing much better compression ratio and performance. Those include:

- an extended tokenization of the *SAM* optional fields
- an internal alignment format defined as a special combination of `SEQ`, `CIGAR` fields and `MD` optional field
- reference-based sequence compression
- transformations of *SAM* numerical fields including `TLEN` and `PNEXT`
- optionally, Illumina quality scores reduction transformation
- optionally, BAM-compatible range queries by chromosome and position.

As this example uses additional functionalities outside the standard *CARGO* library, i.e. reference based compression using BFF (*Binary Fasta File* - see *BFF file format*) files, it was necessary to implement our own modified versions of sub-routines in order to be able to handle advanced application initialization (i.e. to load the compressed reference file) and an extended number of parameters. Thus we are not using the application template files generated by default.

The sources for this example are available in the `cargo/examples/sam/sam-ref` subdirectory of the standard *CARGO* distribution. The precompiled binaries can be found in the `cargo/examples/bin/cargo_samrecord_toolkit-ref`, `cargo/examples/bin/cargo_samrecord_toolkit-ref-q8` and `cargo/examples/bin/cargo_samrecord_too` subdirectories. It this subchapter only the essential information will be presented, omitting the advanced technical details.

### 8.4.1 Example contents

*SAM-REF* project directory consist of the following files:

- `SamRecord.cargo`: *SAM* record type definition in *CARGO* metalanguage
- `SamRecord.h`: the translated C++ *SAM* record type definition
- `SamRecord_Parser.{h, cpp}`: the C++ record type parser files
- `SamRecord_Transform.{h, cpp}`: the C++ record type transform files
- `SamRecord_KeyGenerator.h`: the C++ record type key generator used for querying
- `SamRecord_Application.{h, cpp}`: the extended sub-application classes based on *CARGO Application templates*
- `SamRecord_Type.h`: the translated *TypeAPI* SAM record type definition
- `SamRecord_main.cpp`: the main program file launching appropriate sub-application
- `Makefile`: an automated script to build this example.

In addition to the files mentioned above, *SAM-REF* project uses external modules components available in `cargo/examples/sam/common/`:

- `OptMd.cargo`: SAM optional fields (including special case of '`MD` field) definition in *CARGO* metalanguage
- `OptMd.h`, `OptMd_Type.h`: translated C++ user and *TypeAPI* SAM optional fields definition
- `CigarIterator.h`: a simple module used to iterate over SAM `CIGAR` field items
- `MdProcessor.{h, cpp}`: SAM `MD` optional fields parser into internal mismatch format (and vice-versa)
- `SequenceRetriever.{h, cpp}`: *BFF file format* and sequence reference retrieval module.

### 8.4.2 Record type definition

In addition to parsing of the standard 11 fields available in *SAM* format specification (as in *SAM format field description*), tokenization and parsing of the optional fields has been introduced to improve the compression ratio by storing together fields with common types and/or characteristics. From the *SAM* record type specification (*SAM format optional fields*), the single optional field is defined in a way `TAG:T:VALUE`, where:

- `TAG`: describes a 2 character field tag
- `T`: descries a field value type
- `VALUE`: holds the field value.

Such definition of the SAM record, but for simplicity of this tutorial, skipping the compression configuration is presented below – full definition is available in `cargo/examples/sam/sam-ref/SamRecord.cargo`:

```
1  OptionalValue = [
2    intValue = int^32;
3    charValue = char;
4    stringValue = string;
5  ]
6
7  OptionalField = {
8    tag = string;
9    value = OptionalValue;
10 }
11
12 MdOperationValue = [
13   intValue = int^32;
14   stringValue = string;
15 ]
16
17 MdOperation = {
18   opId = int^8;
19   value = MdOperationValue;
20 }
21
22 SamRecord = {
23   qname = string;
24   flag = uint^16;
25   rname = string;
26   pos = uint^32;
27   mapq = uint^8;
28   cigar = string;
29   next = string;
30   pnext = int^32;
31   tlen = int^32;
32   seq = string;
33   qua = string;
34   opt = OptionalField array;
35   md = MdOperation array;
36 }
37
38 @record SamRecord
```

Using *CARGO metalanguage* such information can be represented by defining 2 compound types – `OptionalField` and `OptionalValue` types. `OptionalField` structure holds the field's `tag` information and it's corresponding `value` which is of `OptionalValue` type. An `OptionalValue` type is a tagged union type, which can hold values of either `int^32`, `char` or `string` types.

To aid the reference-based compression, the reference alignment information is represented in a `MdOperation` structure holding the operation identifier `opId` and the mismatch information, which is of the `MdOperationValue` type. The `MdOperationValue` type, in a similar way like `OptionalValue`, defines a tagged union type, which can hold values of 2 different types – `int^32` and `string`.

The actual SAM record type definition in *CARGO* metalanguage in the major part is the same as in the previous *SAM-STD* example differing only in optional fields representation, altering from single *string* type to two *complex array* types (line `36`).

### 8.4.3 BFF file format

To handle the reference-based sequence compression and to keep the sequences in memory in a compact way, a simple BFF – *Binary Fasta File* – format file was developed, which binary encodes the reference sequences and enables random access for sequence retrieval. The source code of *BFF* file methods, conversion to/from FASTA format and methods to access the sequences are available in `cargo/examples/sam/common/SequenceRetriever.h`.

### 8.4.4 Building

To build the *SAM-REF* example a `Makefile` is available in the example directory:

```
make
```

Invoking *gnu* `make` will produce two executables `cargo_samrecord_toolkit-ref`, `cargo_samrecord_toolkit-ref-q8` and `cargo_samrecord_toolkit-ref-q8-max`, where the former one implements additionally *Illumina quality scores reduction* when performing records transformations.

### 8.4.5 Command line options

The *SAM-REF* application example is based on the standard CARGO application templates (see: *Application templates*), but with an additional set of input parameters used for handling the reference file (`-f`) and BAM-compatible key generation (`-z`). When run directly from the command line displays:

```
CARGO SAM toolkit
usage:
./cargo_samrecord_toolkit-ref <c|d|t|e|r> [options]
```

where a selected subprogram can be either a `c`-ompressor, `d`-ecompressor, `t`-ransformator, `e`-xtractor or `r`-eference *BFF* file generator.

**Compressor**

The *SAM-REF* compressor subroutine processes and stores the input SAM data into container; when launched, displays the following message:

```
CARGO compressor sub-application
available options:
 -c <container>     - container file prefix
 -n <dataset>       - dataset name
 [-i <file>]        - input file, optional (default: stdin)
 [-t <threads>]     - thread count, optional (default: 1)
 [-b <size>]        - input file block size in MiB, optional (default: 64)
 [-a]               - apply records transform, optional (default: false)
 [-g]               - generate records key, optional (default: false)
 [-s]               - skip 1st record field as a generated key (requires -g),
                      optional (default: false)
 [-z]               - use SAM chromosome key order (requires -g),
                      optional (default: false)
 [-h]               - help - display *this* message
```

where the available options are:

-c container, --container=container   *container* file name prefix (e.g. `<container_file>.cargo-`)

-n name, --dataset_name=name   *name* of the dataset to be stored

- **-i file, --input_file=file**   input *file* name (optional)
- **-t n, --threads_num=n**   the number of processing threads (optional)
- **-b size, --block-size=size**   the block *size* (in MiB) of the input buffer (optional)
- **-a, --apply-transform**   apply records forward transformation (optional)
- **-g, --generate-key**   generate a records key for future ranged queries (optional)
- **-s, --skip-key**   skip the 1st field as a generated key when parsing records (requires -a, optional)
- **-z, --sam-order**   uses BAM-compatible chromosome key order when generating key (requires -g, optional)
- **-h, --help**   display help message

### Decompressor

The *SAM-REF* decompressor subroutine retrieves the stored SAM dataset(s) from CARGO container; when launched, displays the following message:

```
CARGO SAM type reference-based decompressor sub-application
available options:
  -c <container>     - container file prefix
  -n <dataset>       - dataset name
  -f <file>          - reference FASTA file in BFF format (see: 'r' sub-application)
 [-o <file>]         - output file, optional (default: stdout)
 [-t <threads>]      - thread count, optional (default: 1)
 [-a]                - apply records transform, optional (default: false)
 [-h]                - display *this* message
```

where the available options are:

- **-c container, --container=container**   *container* file name prefix (e.g. `<container_file>.cargo-`)
- **-n name, --dataset_name=name**   *name* of the dataset to be read
- **-f, --bff_file=file**   reference FASTA file in BFF format
- **-o file, --output_file=file**   output *file* name (optional)
- **-t n, --threads_num=n**   the number of processing threads (optional)
- **-a, --apply-transform**   apply records backward transformation (optional)
- **-h, --help**   display help message

### Transformator

The *SAM-REF* transformator subroutine applies user-defined transformations on the input SAM records, outputting the transformed ones; when launched (with -h option), displays the following message:

```
CARGO SAM type reference-based transformator sub-application
available options:
  -f <file>          - reference FASTA file in BFF format (see: tools/bff_tool)
 [-i <file>]         - input file, optional (default: stdin)
 [-o <file>]         - output file, optional (default: stdout)
 [-t <threads>]      - thread count, optional (default: 1)
```

```
[-b <size>]         - input file block size in MiB, optional (default: 64)
[-a]                - apply records transform, optional (default: false)
[-r]                - do backward (reverse) transform (requires -a) (default: false)
[-g]                - generate key (forward) / skip key (reverse),
                      optional (default: false)
[-z]                - use SAM chromosome key order (requires -g),
                      optional (default: false)
[-h]                - display *this* message
```

where the available options are:

> **-i file, --input_file=file**    input *file* name (optional)
>
> **-o file, --output_file=file**    output *file* name (optional)
>
> **-t n, --threads_num=n**    the number of processing threads (optional)
>
> **-b size, --block_size=size**    he block *size* (in MiB) of the input buffer (optional)
>
> **-a, --apply-transform**    apply records transformation (optional)
>
> **-r, --reverse**    do backward (reverse) transformation (requires -a, optional)
>
> **-g, --generate_key**    applies records key (forward) or skips the key (reverse) as the first field (requires -a, optional)
>
> **-z, --sam-order**    uses BAM-compatible chromosome key order when generating key (requires -g, optional)
>
> **-h, --help**    display help message

**Extractor**

The *SAM-REF* extractor subroutine retrieves from the container records matching the specified genomic region range; when launched, displays the following message:

```
CARGO SAM type reference-based extractor sub-application
available options:
 -c <container>      - container file prefix
 -n <dataset>        - dataset name
 -k <key_0>::<key_n> - records extract range, where <key> is in a form <chr>:<pos>
 -f <file>           - reference FASTA file in BFF format (see: tools/bff_tool)
[-o <file>]          - output file, optional (default: stdout)
[-t <threads>]       - thread count, optional (default: 1)
[-a]                 - apply records transform, optional (default: false)
[-z]                 - use SAM chromosome key order (requires -g),
                       optional (default: false)
[-h]                 - display *this* message
```

where the available options are:

> **-c container, --container=container**    *container* file name prefix (e.g. `<container_file>.cargo-`)
>
> **-n name, --dataset_name=name**    *name* of the dataset to be read
>
> **-k k_begin-k_end, --key=k_begin-k_end**    records extraction range, where key is specified in form `<chr>:<pos>`
>
> **-o file, --output_file=file**    output *file* name (optional)
>
> **-t n, --threads_num=n**    the number of processing threads (optional)

**CARGO Documentation, Release 0.7rc-internal**

- **-a, --apply-transform**  apply records backward transformation (if defined by user, optional)
- **-z, --sam-order**  uses BAM-compatible chromosome key order when generating key (requires -g, optional)
- **-h, --help**  display help message

**Reference sequence generator**

The *SAM-REF* reference sequence generator subroutine creates a compressed FASTA reference sequence in BFF file format (see: *BFF file format*). When launched, the following message is displayed:

```
CARGO SAM type reference sequence BFF generator sub-application
available options:
 -i <file>]          - input FASTA file
 -o <file>]          - output BFF file
 [-h]                - display *this* message
```

where the available options are:

- **-i file, --input_file=file**  input FASTA file name
- **-o file, --output_file=file**  output BFF file name
- **-h, --help**  display help message

### 8.4.6 Running

- Using the compiled *SAM-REF* example `cargo_samrecord_toolkit-ref`, create the *BFF* file from the reference *FASTA* file `hs37d5.fasta`:

  ```
  cargo_samrecord_toolkit-ref r -i hs37d5.fasta -o hs37d5.bff
  ```

- Store the `C_SAM306.sam` input *SAM* file in `C_SAM` container under `C_SAM306` dataset name applying records transformations and using 8 processing threads:

  ```
  cargo_samrecord_toolkit-ref c -c C_SAM -n C_SAM306 -i C_SAM306.sam -a -t 8
  ```

- Store the **sorted** `C_SAM306.sam` input *SAM* file in `C_SAM` container under `C_SAM306` dataset name. Additionally, apply records transformations and store the records in a sorted order (BAM-compatible). While compressing use 256 MiB as an input block buffer and 8 processing threads:

  ```
  cargo_samrecord_toolkit-ref c -c C_SAM -n C_SAM306 -a -g -z -t 8 -b 256 \
          -i C_SAM306.sam
  ```

- Retrieve the SAM dataset `C_SAM306` from container `C_SAM` applying records transformations with `hs37d5.bff` reference file, using 8 processing threads and saving the output as `C_SAM306.out.sam`:

  ```
  cargo_samrecord_toolkit-ref d -c C_SAM -n C_SAM306 -f hs37d5.bff -a -t 8 \
          -o C_SAM306.out.sam
  ```

- Query the *SAM* dataset `C_SAM306` from container `C_SAM` for records satisfying the range 1:1,000,000-2,000,000` (`BAM format`--compatible). Apply the records transformations using ``hs37d5.bff reference file; save the output as `query.sam`:

  ```
  cargo_samrecord_toolkit-ref e -c C_SAM -n C_SAM306 -f hs37d5.bff \
          -a -z -k 1:1000000::1:2000000 > query.sam
  ```

- Print the `C_SAM306` dataset info from the `C_SAM` container:

  ```
  cargo_tool --print-dataset --dataset-name=C_SAM306 --container-file=C_SAM
  ```

- Remove the `C_SAM306` dataset from the `C_SAM` container:

  ```
  cargo_tool --remove-dataset --dataset-name=C_SAM306 --container-file=C_SAM
  ```